\documentclass[conference, usenames,dvipsnames,table]{IEEEtran}
\usepackage{etex}
\usepackage[sort,nocompress,adjust]{cite} % autosort groups of citations
\usepackage{amsmath,amssymb,amsfonts}
\usepackage{algorithmic}
\usepackage{graphicx}
\usepackage{textcomp}
\usepackage{fancyhdr}
\usepackage[hyphens]{url}
\usepackage{tabu}
\usepackage{enumitem}
\usepackage[normalem]{ulem}
\usepackage{inconsolata}
\usepackage[scaled]{beramono}
\usepackage{float}
\usepackage{caption}

\def\BibTeX{{\rm B\kern-.05em{\sc i\kern-.025em b}\kern-.08em
T\kern-.1667em\lower.7ex\hbox{E}\kern-.125emX}}

\fancypagestyle{firstpage}{
\fancyhf{}
}

\usepackage{verbatim}   % allows multi-line comments

\usepackage{wrapfig}
\usepackage{multirow}
\usepackage{balance}
\usepackage{cite}

\usepackage{pifont}

\newcommand{\smallcapital}{\fontsize{9pt}{10pt}\selectfont}

\usepackage[bookmarks=true,breaklinks=true,letterpaper=true,colorlinks,linkcolor=black,citecolor=blue,urlcolor=black]{hyperref}

\usepackage{rotating}

\usepackage{arydshln}

\usepackage{placeins}

\usepackage{gensymb}

\title{HiveMind: A Scalable and Serverless Coordination Control Platform for UAV Swarms}

\begin{document}

%
% The "author" command and its associated commands are used to define the authors and their affiliations.
% Of note is the shared affiliation of the first two authors, and the "authornote" and "authornotemark" commands
% used to denote shared contribution to the research.
\author{Justin Hu, Ariana Bruno, Brian Ritchken, Brendon Jackson, Mateo Espinosa, Aditya Shah, and Christina Delimitrou\\Cornell University}

\date{}

\maketitle

\thispagestyle{firstpage}
\pagestyle{plain}

\begin{abstract}
	{
Swarms of autonomous devices are increasing in ubiquity and size. There are two main trains of thought for controlling 
devices in such swarms; %When it comes to controlling the devices in a swarm, the two main trains of thought are; 
centralized and distributed control. Centralized platforms achieve higher 
output quality but result in high network traffic and limited scalability, while decentralized systems are more scalable, but 
less sophisticated. 
%Centralized control benefits from the vast amount of resources 
%of a large cluster, typically enabling higher-quality decisions, but suffers from high communication latencies, as data is transferred from the edge devices, and 
%can limit scalability+++. 
%Distributed control on the other hand avoids unnecessary data transfer, but must accommodate computation on the limited on-board resources, quickly draining the device's battery, 
%and resulting in lower-quality output. 

In this work we present HiveMind, a centralized coordination control platform for IoT swarms that is both scalable and performant. HiveMind 
leverages a centralized cluster for all resource-intensive computation, deferring lightweight and time-critical operations, 
such as obstacle avoidance to the edge devices to reduce network traffic. HiveMind employs an event-driven serverless framework 
to run tasks on the cluster, guarantees fault tolerance both in the 
edge devices and serverless functions, and handles straggler tasks and underperforming devices. We evaluate HiveMind on 
a swarm of 16 programmable drones on two scenarios; searching for given items, and counting unique people in an area. We show that HiveMind 
achieves better performance and battery efficiency compared to fully centralized and fully decentralized platforms, while 
also handling load imbalances and failures gracefully, and allowing edge devices to leverage the cluster to collectively improve their output quality. 
%the centralized platform 
%achieves faster and better quality execution time than a distributed control framework, especially for scenarios that require significant collaboration across edge devices. 
%We also show that HiveMind further benefits from optimizations including in-memory computation in the cloud, heterogeneity-aware task scheduling, on-board preprocessing, 
%and data compression. 

%TBD... 

%Large swarms of autonomous devices are increasing in size and importance. 
%When it comes to controlling the devices of large-scale swarms there are two 
%main lines of thought. Centralized control, where all decisions - and often compute - 
%happen in a centralized back-end cloud system, and distributed control, where 
%edge devices are responsible for selecting and executing tasks with minimal or 
%zero help from a centralized entity. In this work we aim to quantify the trade-offs 
%between the two approaches with respect to task assignment quality, latency, and reliability. 
%We do so first on a local swarm of 12 programmable drones with a 10-server cluster as 
%the backend cloud, and then using a validated simulator to study the tail at scale effects 
%of swarm coordination control. We conclude that although centralized control almost always 
%outperforms distributed in the quality of its decisions, it faces significant scalability 
%limitations, and we provide a list of system challenges that need to be addressed 
%for centralized control to scale.
 }

\end{abstract}

%
% Keywords. The author(s) should pick words that accurately describe the work being
% presented. Separate the keywords with commas.
%\keywords{cloud computing, datacenter, performance debugging, QoS, deep learning, data mining, tracing, monitoring, microservices, resource management}

\section{Introduction}

Swarms of autonomous edge devices are increasing in number, size, and popularity~\cite{Tong16, Singhvi17, platformLab, Han19, flock18, Floreano18, mit_article, mit_article2, Almeida17, 
Lin18, Lin19, cncf, Markantonakis17, Faticanti18, Alfeo18, Albani17, Yuan18, Campion18, Dogar15, Nasser15, Sanneman15}. 
From UAVs to self-driving cars and supply-chain robots, swarms are enabling new distributed applications, 
which often experience intermittent activity, and are interactive and latency-sensitive~\cite{Gan19,Lin18,Lin19,Delimitrou19,Tong16,Singhvi17,Zhang19}. 

Coordination of large swarms of edge devices usually follows one of two main approaches. The first approach 
argues for \textit{decentralized} control and smart edge devices, which perform most computation \textit{in situ}, only transferring 
the results to the backend system~\cite{Lin18,Lin19,Alfeo18,Faticanti18,Tong16}. This design avoids the high network traffic of a centralized system, 
however, it either requires edge devices to work individually, hence limiting their use cases and missing the potential 
for collective optimizations, or involves peer-to-peer communication between devices, which again increases network traffic. 

The second approach are \textit{centralized} coordination 
systems~\cite{platformLab, flock18, Almeida17,cncf,xcamera}, where the edge devices are merely a way to collect sensor data, 
while all computation and state management happens in a backend cluster. This approach benefits from the ample cloud resources, 
hence it can explore more sophisticated techniques than what is possible on edge devices, but also experiences 
high overheads from network communication, as data is transferred to and from the cloud. 

In response to the emergence of this new class of systems and services, cloud computing operators have developed new programming frameworks to 
make it easier for edge devices to leverage cloud resources, and express the event-driven, interactive nature of their computation~\cite{lambda, azure_functions, google_functions,fission,openlambda,openwhisk,openlambda2}. 
Serverless compute frameworks for example, specifically target highly-parallel, intermittent computation, 
where maintaining long-running instances is not cost efficient~\cite{xcamera,Hellerstein18,vaneyk18,Lloyd18,Klimovic18,Baresi17,Lynn17,sand18,Sock18,Spock18}. 
Serverless frameworks additionally simplify cloud management by letting the cloud provider handle application placement, resource provisioning, and state management, 
with users being charged on a per-request basis~\cite{vaneyk18,Spock18,Klimovic18}. Serverless functions are instantiated in short-lived containers to improve portability, resource isolation, and security, 
and containers are terminated shortly after the process they host completes, freeing up resources for other workloads. 

From the cloud operator's perspective, serverless has two benefits: first, it gives the cloud provider better visibility 
into the characteristics of external workloads, allowing them 
to better optimize for performance, resource efficiency, and future growth. Second, it reduces the long-term resource overprovisioning current 
cloud systems suffer from~\cite{GoogleTrace,Delimitrou13,Delimitrou14,Delimitrou16,Lo15,BarrosoBook,barroso_keynote,Delimitrou15,Delimitrou17,Delimitrou19,Delimitrou14b,Delimitrou13d}, 
by not requiring the end user to handle provisioning, allocating resources at fine granularity and for short periods of time instead. %only allocating resources in fine granularity

While serverless is not exclusively applicable to swarm applications~\cite{Gan18,Gan19,xcamera,gg,salsify,Horn17,Klimovic18}, it is well-suited for their requirements. 
{\smallcapital AWS} Greengrass, for example, is a variant of {\smallcapital AWS}'s general serverless framework 
tailored to the requirements and characteristics of IoT applications, which allows devices to both train their ML models in the cloud, 
and also launch serverless functions for inference~\cite{aws_greengrass}. 
Finally, serverless frameworks offer a centralized persistent storage system for sensor data that the swarm can use to improve its decision quality over time. 
Despite the increasing prevalence of both IoT swarms and serverless compute, there are currently no systems that compare the advantages and issues of centralized and %highlight the trade-offs between centralized and 
decentralized platforms, and highlight the potential of serverless in addressing their performance requirements. 

%present hivemind: key inisghts

%explore tradeoffs between centralized and distributed coordination control

%Microservices are often used in the context of serverless programming frameworks, i.e., frameworks where the application and data are managed by the
%cloud provider, and the user simply launches short-lived ``functions'', and is charged on a per-request basis~\cite{lambda}. Serverless is well-suited for applications with intermittent activity, where
%maintaining long-running instances is cost inefficient. Serverless additionally targets embarrassingly parallel services, which benefit from a massive amount of resources for a brief period of time.

%Cloud computing is making advances to address their performance and resource requirements. For example, serverless... event-driven computation, leverage fine-grained parallelism and get charged only for the time used. 

%- Swarms of autonomous devices are becoming more popular. Example scenarios. 

%- centralized vs. distributed (computation and coordination)

%- advantages/disadvantages

\begin{figure*}
	\centering
	\begin{tabular}{ccc}
		\includegraphics[scale=0.234, viewport=120 20 640 420]{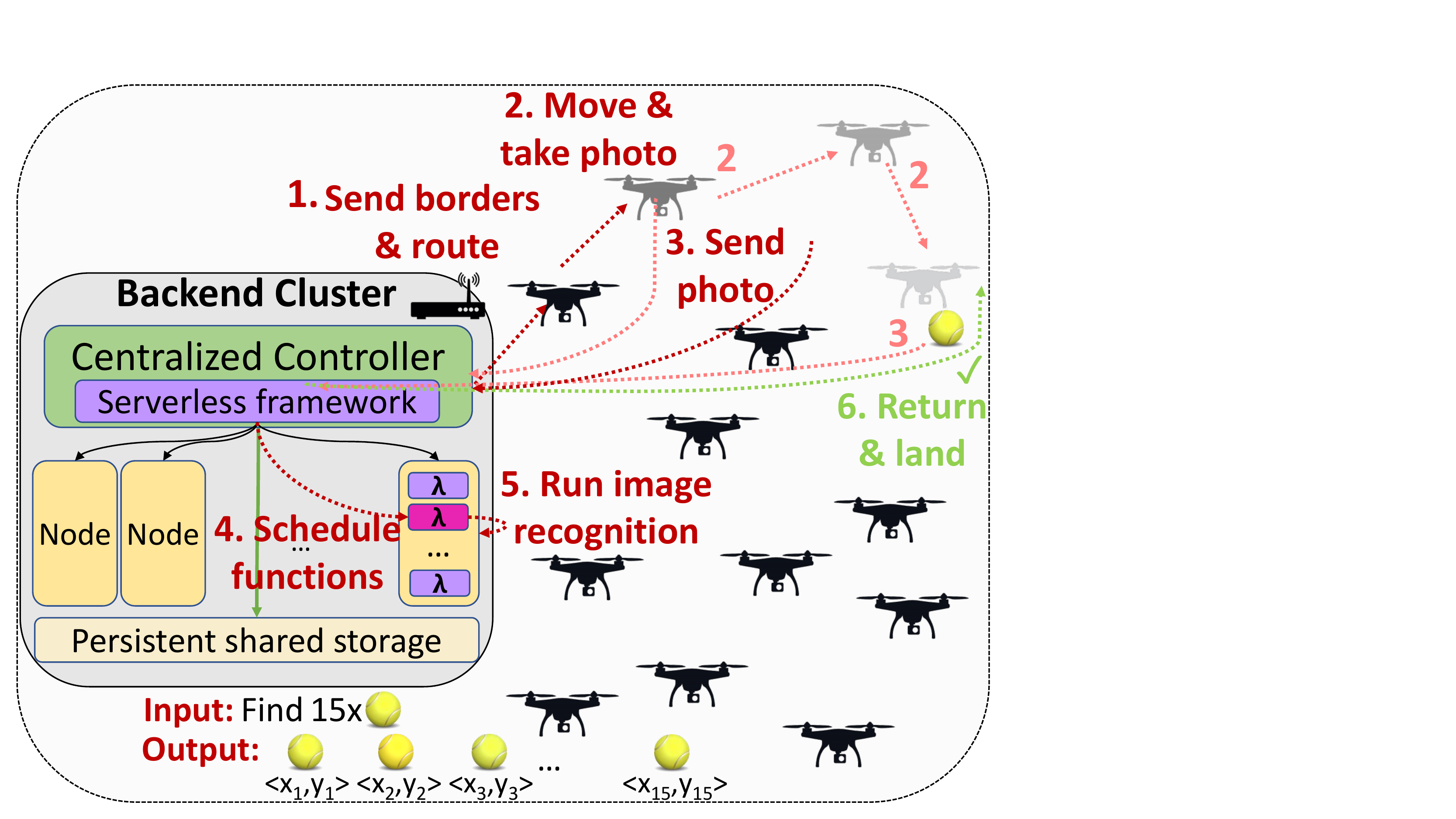} &
		        \includegraphics[scale=0.24, viewport=25 10 700 440]{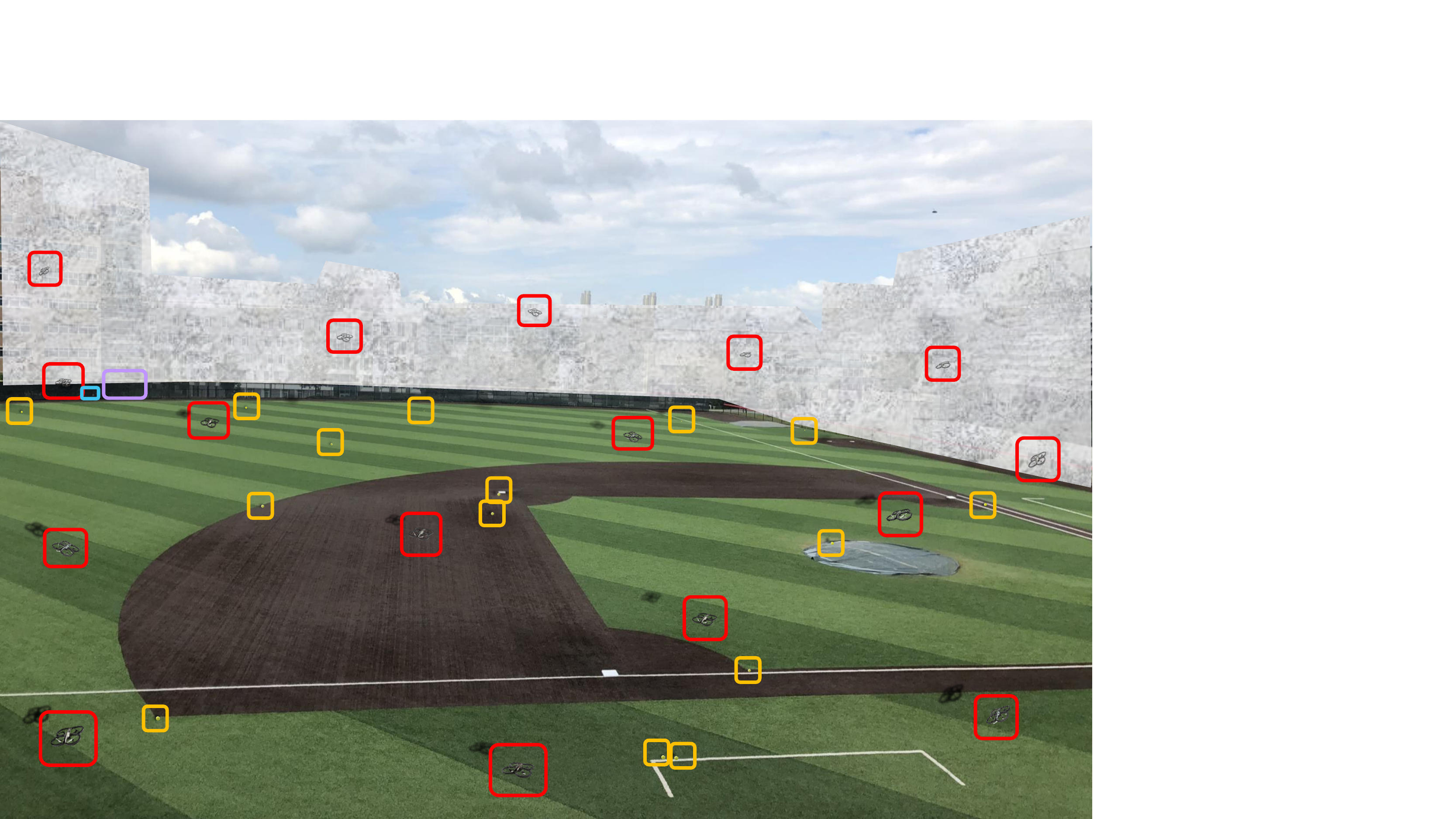} &
		\includegraphics[scale=0.206, viewport=20 10 700 440]{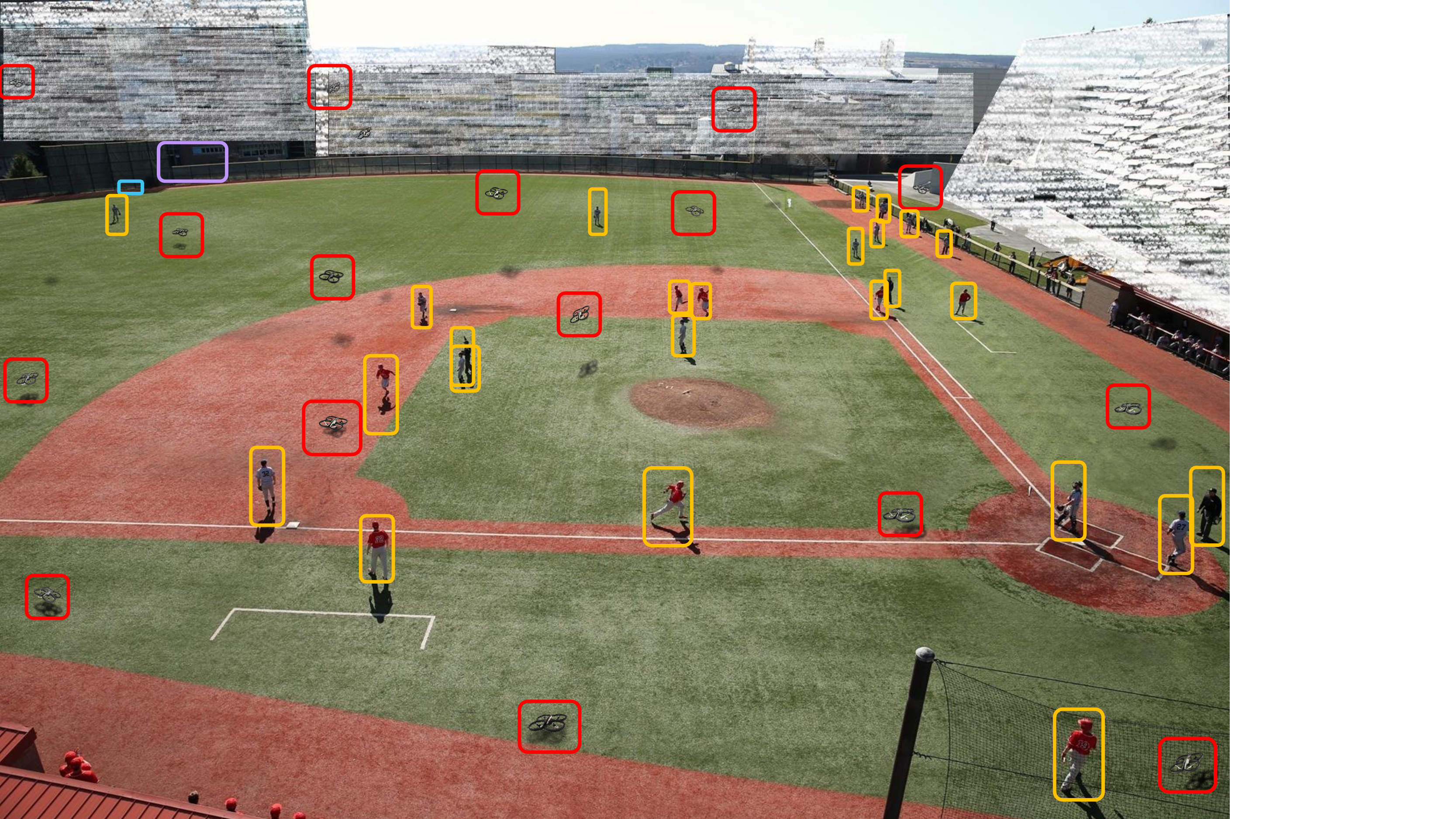}

	\end{tabular}
	\caption{\label{fig:scenario1} {(a) High-level operations for the first scenario in the centralized platform. %Sequence of operations for the scenario of locating 15 tennis balls with HiveMind.
			                (b) The drone swarm executing the first scenario, and (c) the second scenario.
					Red bounding boxes show each of the 16 drones in the swarm, yellow boxes show the 15 tennis balls and 25 people respectively,
					the blue box shows the router the drones use to communicate with the backend cluster, and the purple box
			shows the approximate location of the cluster. Buildings are blurred for double blind review. }}
		\end{figure*}

In this work we first explore the performance and efficiency trade-offs between centralized and decentralized swarm coordination control. % for IoT swarms. 
Based on the findings of this comparison, we present \textit{HiveMind}, a centralized and scalable coordination control platform for swarms of edge devices. 
HiveMind is designed to optimize task latency, battery efficiency, and fault tolerance across the platform's cloud and edge components. 
It leverages event-driven computation using \textit{serverless compute} to 
expose the fine-grained parallelism in operations triggered by edge devices, and improve their performance and efficiency. %predictability. % collected by the edge swarm
HiveMind keeps network traffic low by tasking the edge devices with filtering sensor data, and 
only transferring the most meaningful information to the cloud for further computation. It additionally exploits the centralized system to 
continuously improve the decision quality of edge devices, by letting them learn from each other's mistakes. Finally, HiveMind 
implements fault tolerance, load rebalancing, and straggler mitigation techniques to further improve performance predictability. 
We evaluate HiveMind using a 16-drone swarm, however, the platform's design and programming framework are not drone-specific, and can be used to port 
applications on other IoT swarms, such as self-driving vehicles. 

We explore two application scenarios; locating stationary items, and counting the unique people in a bounded area. We show 
that HiveMind achieves better and more predictable performance and battery efficiency compared to fully centralized and fully decentralized systems, while also 
handling cloud and edge failures gracefully. HiveMind also achieves higher decision quality than a fully decentralized system, and better resource efficiency than a fully 
centralized system, enabling the swarm to efficiently scale to large numbers of edge devices. 

\vspace{-0.05in}

\section{Application Scenarios}
\label{sec:scenarios}

\vspace{-0.07in}
\subsection{Methodology}
\vspace{-0.07in}

\noindent{\bf{Drones: }} The swarm consists of 16 programmable Parrot {\smallcapital{AR.}} Drones 2.0~\cite{parrot}. 
Each drone is equipped with an {\smallcapital ARM} Cortex {\smallcapital A8} 1GHz,
32-bit processor running Linux 2.6.32. There is 1GB of on-board {\smallcapital RAM},
which we complement with a 32{\smallcapital GB} {\smallcapital USB} flash drive 
to store the image recognition model and sensor data. Each drone also has by default
a vertical 720p front-camera used for obstacle avoidance, and the following sensors: 
gyroscope, accelerometer, magnetometer, pressure sensor, and altitude ultrasound sensor.
We additionally fit each drone with a 12{\smallcapital MB} camera connected to the underside 
of the device over {\smallcapital USB}, which is used for high definition photos. 
%While both cameras support video recording as well, for the scenarios we explore, processing video streams is not necessary, and would increase the network bandwidth usage.

\vspace{0.05in}
\noindent{\bf{Server cluster: }}We use a dedicated local cluster with 12, 2-socket, 40-core 
servers with 128-256{\smallcapital GB} of {\smallcapital RAM} each, running Ubuntu 16.04.
Each server is connected to a 40Gbps ToR switch over 10Gbe NICs. Servers communicate 
with the drone swarm over a 867Mbps LinkSys AC2200 MU-MIMO wireless router~\cite{linksys_router} using {\smallcapital TCP}. %check: this is not enough for 120MBps.
Finally, we deploy OpenWhisk~\cite{openwhisk} on the cluster to launch serverless functions, 
and instantiate new jobs inside single-concerned Docker containers. 

%need to explain decentralized in scenarios

\vspace{-0.05in}
\subsection{Stationary Item Detection}
\vspace{-0.05in}

We first explore %the scenario introduced in Section~\ref{sec:example_scenario}, where the drone swarm needs to locate a number of given items (15 tennis balls) in a bounded area. 
a scenario where the swarm needs to locate 15 tennis balls placed 
within the {\smallcapital 2D} borders of a baseball field. Fig.~\ref{fig:scenario1}a shows
the high-level order of operations, and Fig.~\ref{fig:scenario1}b shows 
the swarm executing the scenario. 
%The drones fly at an altitude of 4-6m at 4m/s
%and each receive a partition of the area to cover from the centralized HiveMind controller. Once they receive their assignments, drones begin their routes,
%taking photos every 1s. This data is then analyzed to check if it contains one or more tennis balls. When all drones complete coverage of their assigned areas,
%they return to the base station (take-off location) and land.

\begin{wrapfigure}[10]{l}{0.264\textwidth}
\centering
\vspace{-0.18in}
\includegraphics[scale=0.295, trim=0.9cm 0cm 6cm 8cm, clip=true]{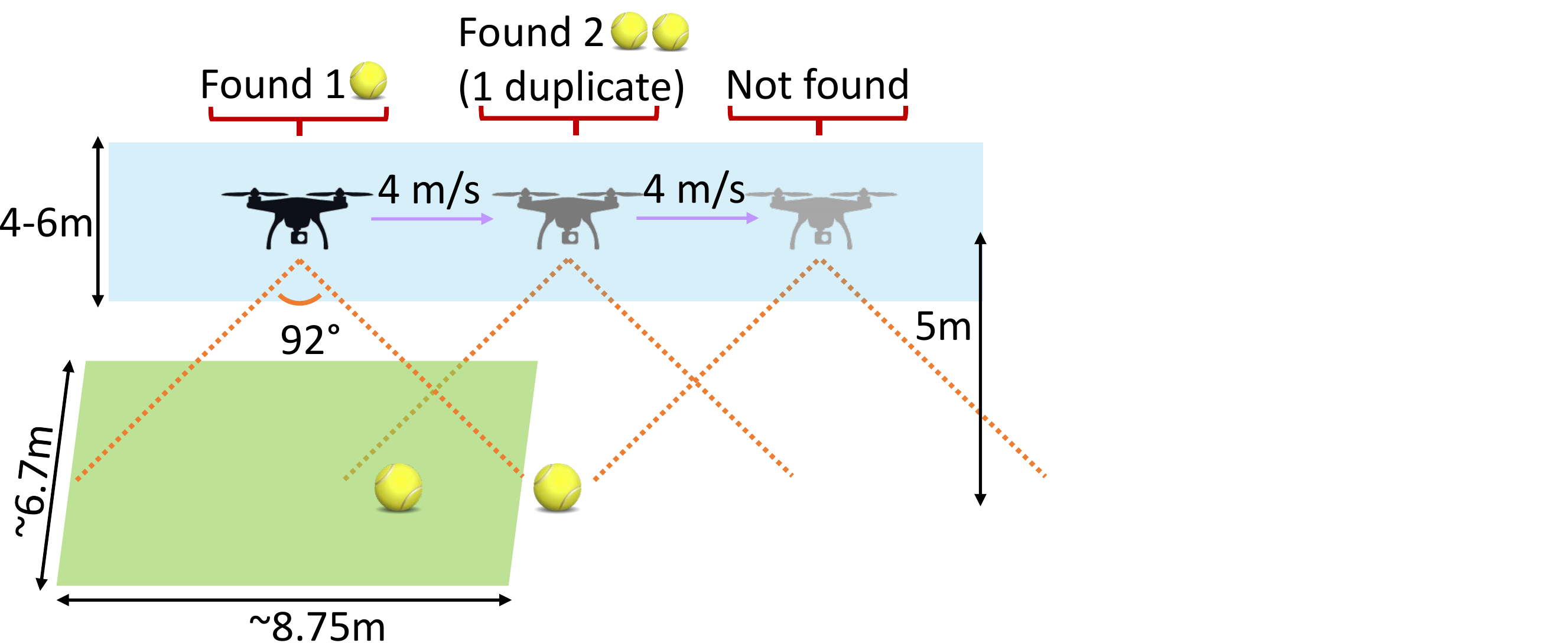}
\vspace{-0.22in}
\caption{\label{fig:fov} {Field of view (FoV), altitude, and speed of a Parrot drone. }}
\end{wrapfigure}

%The specific items are 15 tennis balls, placed within the borders of the baseball field, shown in Fig.~\ref{fig:scenario1}a. Red bounding boxes represent the drones executing the scenario, 
%yellow boxes the tennis balls, while the blue and purple boxes represent the location of the wireless router and backend cluster respectively. 
%changed from 2 m/s every 2s
Drones fly at a height of $4$-$6m$, move at $4m/s$, and take photos of the ground every $1s$ 
to ensure full coverage of the terrain without excessive photo overlap. %minimal overlap in the photos. %3 m/s  
Fig.~\ref{fig:fov} shows an example of consecutive photo taking intervals 
for a drone, flying at an average $5m$ altitude. The camera has a 
92\degree field of view (FoV), which results in an approximate coverage of $6.7m\times8.75m$. 
At $4m/s$, this ensures full coverage of the assigned space with some overlap 
between photos to improve detection accuracy for items close to 
the drone's FoV's borders. Duplicate items are disambiguated using their $<x,y>$ coordinates. 

While Parrot AR 2.0 drones can move at a maximum speed of $12m/s$, 
which would allow faster space coverage, 
we have found that for speeds over $7m/s$ control becomes difficult, 
and flight becomes severely unstable, leading to crashes and equipment damages. 
This is especially the case when computation, such as obstacle avoidance, 
happens on-board. The photo taking interval is also dictated by the fact 
that collecting photos more frequently than every 0.5s can lead to long network 
queueing delays for swarms larger than 15 drones. %collecting 
In Sec.~\ref{sec:evaluation} we also explore photo intervals of $0.5s$ at $6m/s$ speeds 
for cases where not all sensor data are transferred to the cluster. We show 
that while this allows faster space coverage, it can also lead to flight instability when the on-board 
resources are highly-utilized. %compute and memory resources are already highly-utilized. 
Finally, to avoid all devices transferring data to the cloud at the exact same time, 
%and saturating the network bandwidth 
we also insert an initial $0.1s$ delay between the time the drones start their missions. 
%For the cases where image recognition happens in a decentralized fashion on the drones, as well as 
%in the hybrid coordination platform discussed in Section~\ref{sec:design}, 
%we also explore photo intervals of $0.5s$ at speeds of 6 m/s, since not all sensor data are transferred to the backend cluster. In Section~\ref{sec:evaluation} we show that 
%this allows the swarm to cover the required area faster, but can also lead to instability, especially when the on-board compute and memory resources are already highly-utilized. 

%Fig.~\ref{fig:scenario1}b shows the high-level sequence of operations for this scenario. All coordination control platforms evaluated have a centralized controller that runs 
%in the backend cluster. Depending on the type of coordination platform used, the centralized controller is responsible for a varying degree of operations. %varying degree of responsibility. 

We build and deploy a centralized and a decentralized coordination control platform on the swarm, and 
explore their performance, reliability, and efficiency trade-offs in the next section. % between centralized and decentralized coordination control. 
In both systems, there is a centralized controller that performs the initial work assignment between drones, 
and stores the final output in persistent storage. 
The controller evenly divides the area across all drones, and sends them the assigned border coordinates. % to each drone. 
It also communicates with each drone the routing strategy they should follow, and the interval at which 
photos should be collected. The route is derived using A$^*$~\cite{astar}, where each drone tries to minimize 
the total distance traveled by photographing neighboring points in sequence. %, while also exploring its full assigned region, subject to some tennis balls not having been located yet. 
Once the drones receive their assignment, they move to their starting points; 
the corner border point for each of their assigned regions that is closest to their take off point. All drones 
take off from the same location. Once each drone reaches its starting point, they start collecting photos of the ground every $1s$. 
%it moves at the predefined speed of 4m/s (unless otherwise specified) and collects photos of the ground beneath it every second. 
Depending on the structure of the coordination control platform deployed, the drone 
either transfers the data over wifi to the backend cluster, or performs image recognition on-board. 
In both the centralized and decentralized platforms, obstacle avoidance happens on-board to avoid 
catastrophic failures caused by delays in network transfers. Obstacle avoidance leverages the drone's 
med-resolution front camera (non-pivotable) to detect solid 
objects in the drone's vicinity and adjust its route to avoid them. We use the obstacle 
avoidance {\smallcapital SVM} classifier in the \texttt{ardrone-autonomy} library~\cite{autonomy}, and train it, 
in addition to generic solid objects, for trees, people, other drones, and walls as well. 

\noindent{\bf{Centralized platform: }} In this case all sensor data are transferred to the cloud. % for recognition. 
Once the backend cluster receives a new image, the controller triggers the serverless framework to launch 
the recognition task. The OpenWhisk master finds and allocates cluster resources to the new job, and 
launches the serverless functions. 
Image recognition uses an {\smallcapital SVM} classifier, implemented in \texttt{OpenCV} based 
on the \textit{cylon} framework~\cite{cylon}, and trained on a dataset of various balls used in sports.
Once the job completes, OpenWhisk informs the controller whether the image contained a unique tennis ball 
by comparing its coordinates to previously-identified balls. 
In the meantime, the drones continue their routes, and send new photos. % to the cloud. 
%If at any point one or more drones fail, or are close to running out of battery, the centralized controller reassigns their 
%region equally to neighboring drones. %their regions are redistributed among neighboring drones, as described in the controller's design. 
%We further discuss the scheduling and rebalancing policies %and rebalancing strategy 
%%%%in detail 
%in Sec.~\ref{sec:serverless}, and~\ref{sec:load_balancing}. 

%Obstacle avoidance happens at the edge device in all coordination setups, leveraging the drone's front camera (non-pivotable) and accelerometer to detect solid objects
%in the drone's vicinity and avoid them by adjusting their motion. Obstacle avoidance is controlled via the \texttt{ardrone-autonomy} library~\cite{autonomy}.

\noindent{\bf{Decentralized platform: }}In the decentralized system, image recognition happens exclusively at the edge using 
the same {\smallcapital SVM} classifier as in the centralized system, adjusted to account for the drone's different OS and hardware stack. 
Since there is no centralized state where the locations of identified balls are stored, 
the drones need to disambiguate their findings and discard any duplicate balls. 
Once a drone covers its assigned region, it shares all coordinates containing 
tennis balls with its neighboring drones (any drones it is sharing a border with). The recipients 
check for duplicates and only retain uniquely-identified balls. The process continues across 
the swarm, ensuring that disambiguation between a pair of drones is unidirectional 
to avoid discarding the same balls twice. %, i.e., if drone {\smallcapital A} receives 
%the output of drone B and performs disambiguation, it does not also share its output with drone {\smallcapital B}. 
%As drones complete the image disambiguation, they return to their take-off location and land. 

In both platforms, if at any point one or more drones fail, 
or are close to running out of battery, the centralized controller repartitions their 
region equally among neighboring drones. %their regions are redistributed among neighboring drones, as described in the controller's design. 
Rebalancing policies %and rebalancing strategy 
are discussed in detail in Sec.~\ref{sec:load_balancing}. 
Once all drones complete their routes 
they return to their take-off location and land. 
The final output is stored in the cluster, 
and includes photos and coordinates for each tennis ball. 
%When comparing the behavior of different platforms 
%tennis balls are always placed in the same locations. 

%TODO: move to HiveMind section
%In HiveMind, drones perform an initial on-board recognition for circular objects in their FoV, only transferring images with such objects to the backend cluster, and storing all other images 
%in their on-board flash drive for a posteriori accuracy validation. 
%The on-board recognition is using the circular bounding box technique in \texttt{cylon}, implemented in OpenCV~\cite{cylon}. 

%\begin{figure*}
%	\centering
%	\begin{tabular}{cc}
%		\includegraphics[scale=0.32, viewport=100 20 820 490]{swarm_scenario2_5.pdf} & 
%	\includegraphics[scale=0.28, viewport=-20 10 700 480]{swarm_scenario2_photo2.pdf}
%\end{tabular}
%\caption{\label{fig:face_recognition} {Scenario of recognizing unique people within a given region. (a) Shows the order 
%		of operations, and (b) shows the scenario being executed with the 16-drone swarm 
%and people in a baseball field. The red border boxes show each of the drones, the yellow boxes the people to be recognized, 
%the blue box shows the router the drones use with communicate to the backend cluster, 
%and the purple box shows the approximate location of the cluster. Buildings are blurred for double blind review. }}
%\end{figure*}

\begin{figure*}
	\centering
	\begin{minipage}{0.62\textwidth}
		\begin{tabular}{cc}
			\includegraphics[scale=0.24, viewport=20 -6 840 80]{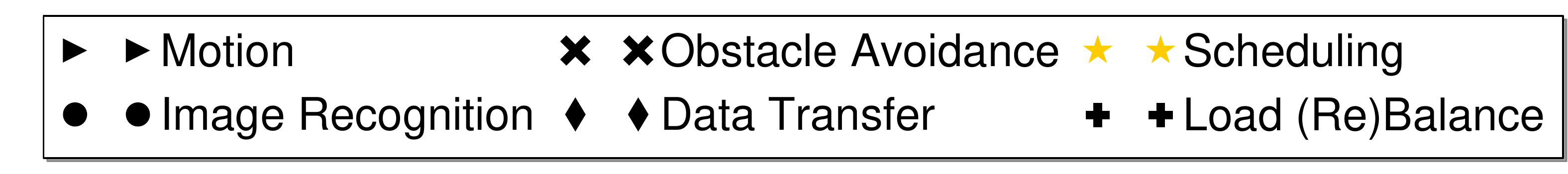} &
			\includegraphics[scale=0.238, viewport=-22 25 1260 85]{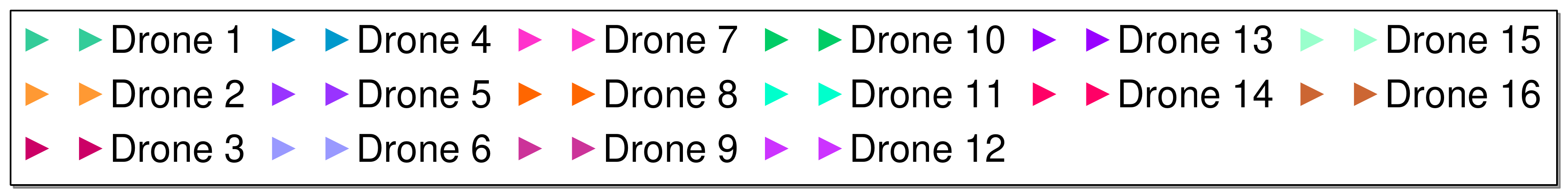} \\
			\includegraphics[scale=0.19, viewport=200 10 1300 370]{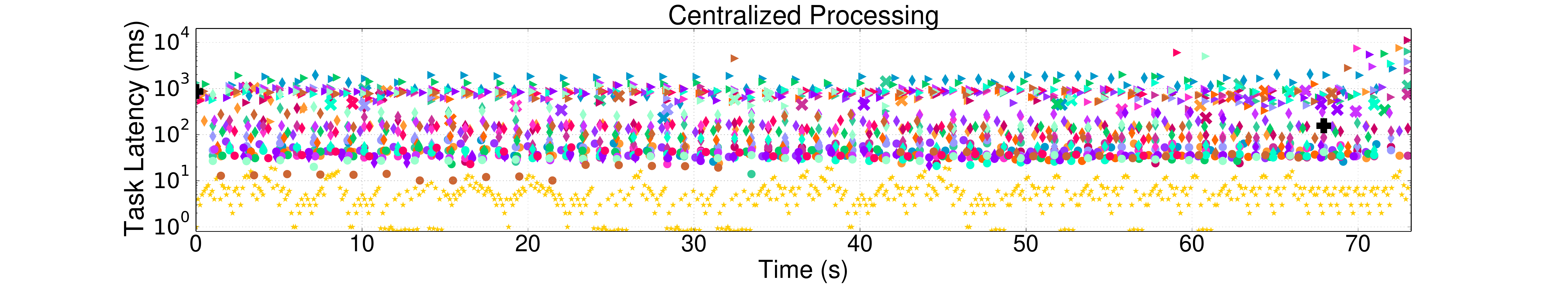} & \\
			\includegraphics[scale=0.19, viewport=200 45 1300 340]{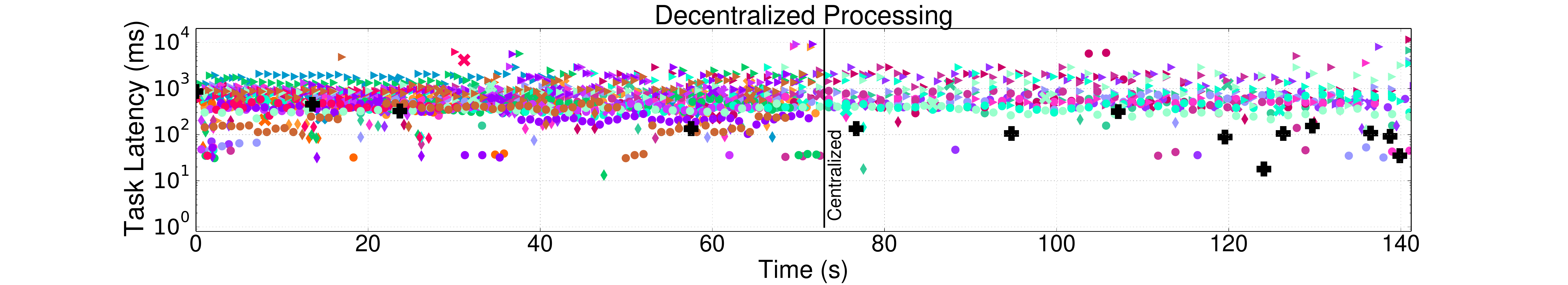} & \\
		\end{tabular}
		\caption{\label{fig:latency_balls} {Latency per task type for centralized (top) and decentralized processing (bottom) in the first scenario (locating tennis balls). The vertical line in the bottom graph signifies
		when the centralized platform completes the scenario. }}
	\end{minipage}
	\hspace{0.3cm}
	\begin{minipage}{0.34\textwidth}
		\begin{tabular}{cc}
			\includegraphics[scale=0.16, viewport=200 50 1400 605]{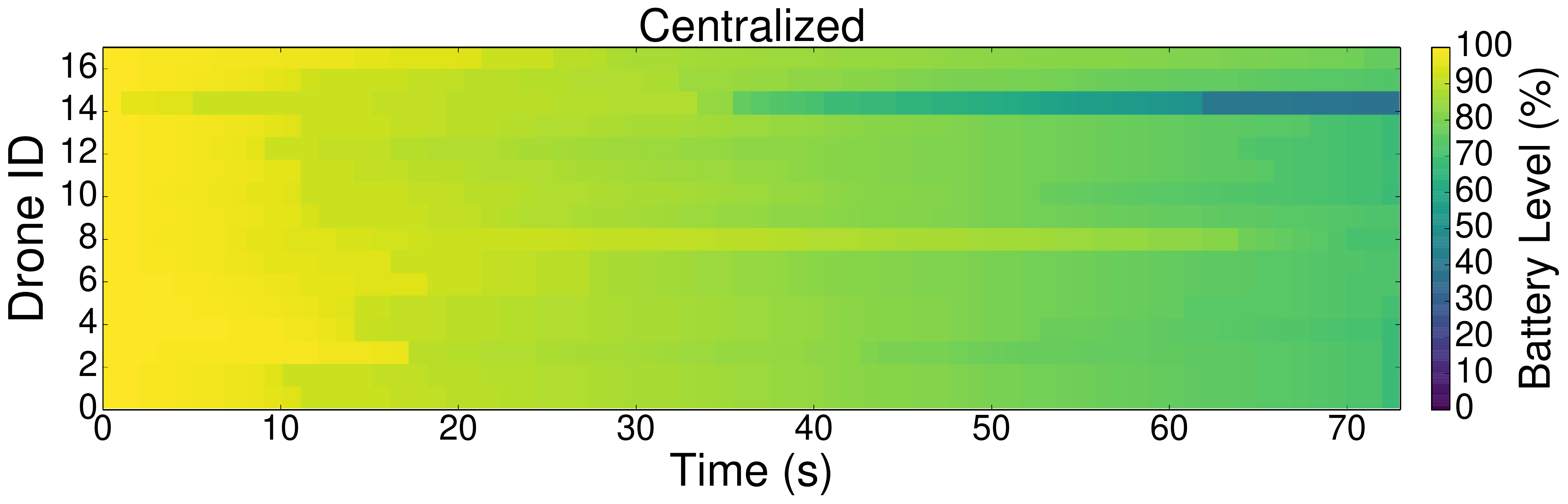} & \\
			\includegraphics[scale=0.16, viewport=200 55 1400 440]{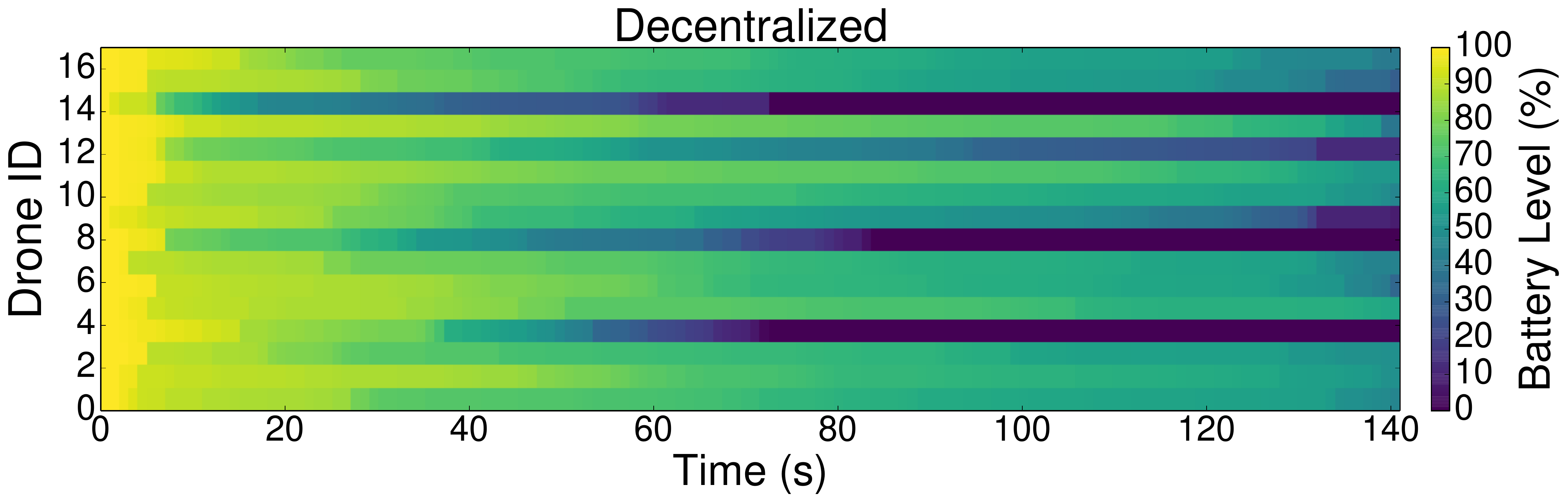} & \\
		\end{tabular}
		\vspace{0.02in}
		\caption{\label{fig:battery_balls} {Per-drone battery level for the centralized and decentralized platforms in the first scenario. }}
	\end{minipage}
\end{figure*}
%unless all tennis balls have been locatee. 
%subject to some tennis balls remaining unlocated. 
%For this scenario, drones explore the space while trying to mini
%\noindent{\bf{Decentralized processing: }}

%- remove target image from figure, add route and interval for photos

%\begin{figure}
%\centering
%\includegraphics[scale=0.32, viewport=220 10 600 420]{swarm_scenario1_detailed2.pdf}
%\caption{\label{fig:image_recognition} {Treasure hunt scenario. }}
%\end{figure}
\vspace{-0.07in}
\subsection{Mobile People Recognition}
\vspace{-0.07in}

We now require the swarm to recognize a total of 25 people present on the baseball field, and count their number; 
the number of people is not known to the drones or cluster in advance.
%~\footnote{The scenario is part of a program to record aerial data of athletic practices for coaching purposes.} 
People are allowed to move within the borders of the field while the 
scenario takes place. This introduces additional challenges as, unlike in the previous scenario, %case of tennis balls, 
the number of people in a region can change over time, resulting in 
counting the same person multiple times as they cross between regions assigned 
to different drones. People can also be double-counted by standing 
close to the borders between regions. Finally, if a person knows a drone's route, 
%a drone will follow, 
they can move between regions, such that they remain outside the FoV 
of nearby drones. We assume that people do not know the drones' routes, and do not 
actively try to avoid being photographed. 
%We also instruct people to follow the exact 
%same route when we compare different coordination platforms. 
%the behavior %starting, ending location and route 
%of different coordination platforms. 

Figure~\ref{fig:scenario1}c shows the swarm executing the scenario. 
%high-level operations for this scenario in the case of the centralized coordination control platform. 
As before, the controller assigns regions to each drone, %in the previous scenario, 
and communicates the coordinates and route with them. 
%Each drone reaches its starting point and commences collecting photos. 

\noindent{\bf{Centralized platform: }} Once the cloud receives a new image, 
it invokes OpenWhisk to launch the new face recognition task. Once the job completes, the OpenWhisk 
master informs the controller whether there was a person in the given photograph. 
%While the serverless tasks are running, drones continue 
%to execute their routes and transfer new photos to the cloud. %Once a new person 
Human recognition is based on the Tensorflow Detection Model Zoo~\cite{tensorflow,tensorflowzoo}, a set 
of pre-trained models compatible with Tensorflow's Object Detection {\smallcapital API}, 
an open-source library used for training and testing object detection models. 
Tensorflow Detection Model Zoo consists of 16 object detection models, 
trained on the {\smallcapital COCO} dataset~\cite{coco_dataset}. 
The models provide bounding boxes around target objects as output, 
and are capable of detecting 80 types of objects, including humans.

Given that a person can move within a given area, we need to disambiguate 
between detected people. Tensorflow's Detection Model Zoo achieves high accuracy in full-body 
object detection, but is not designed for face recognition. Therefore, to disambiguate between people
we use a face recognition framework in OpenCV based on FaceNet~\cite{facenet}, 
a CNN-based system that directly learns a mapping between face images and a compact Euclidean space 
where distances correspond to a measure of face similarity. Having this mapping makes it 
easy to compute similarities between faces with FaceNet embeddings as feature vectors. Disambiguation 
uses the serverless framework, it starts as soon as the first people are identified, 
and usually finishes after the drones have completed their routes and transferred all data. 
To avoid cases where the same person is photographed from the front and back by one or more drones, 
in which case disambiguation based on face recognition is impossible, we limit TensorFlow's Detection Model Zoo
to people where the face is at least partially visible. 
%people participating in the scenario 
%also wear shirts with a unique number on their front and back
%- face upwards (doesn't fix people being photographed from the side or the back)
%- use numbers (complex but ok, still problem with side photos, and if you're going to do that why have the face recognition)
%- only use people recognition model that requires faces (it can miss people if they're always facing the other way -- not really because the drone will eventually get on the other side of the person)
%we request that people 
%face somewhat upwards during the scenario, so their face can at least partially be in the picture. %this doesn't solve pictures taken of someone's side or back

%In the centralized coordination platform, disambiguation also runs on the backend cluster using OpenWhisk to exploit fine-grained parallelism. 
%Disambiguation starts as soon as the first images with people are identified, 

\noindent{\bf{Decentralized platform: }}As with the previous scenario, recognition and disambiguation happen 
at the edge, using the same models as in the centralized system, rewritten in OpenCV, since TensorFlow's dependencies could 
not run on the drones. Disambiguation follows a similar procedure as before, %to the first scenario, 
with the difference that now drones exchange photos to differentiate people instead of coordinates, 
since people may have moved between pictures, which increases network traffic. 

%In the decentralized control platform, both the human recognition and disambiguation happen on the edge devices, with disambiguation following a similar procedure to the one described in the previous scenario. 

%In HiveMind, the drones perform an initial item recognition on-board, placing bounding boxes around shapes that resemble humans, using the simple oval item recognition model of \texttt{cylon} in OpenCV~\cite{cylon}. 
%They then only transfer images with such items to the cloud. %, storing all others in the on-board flash drive for a posteriori validation. 
%The backend cluster then uses the two step process above to verify that the item 
%detected is a human, and disambiguate them against previously-detected people. In Section~\ref{sec:evaluation} we evaluate the accuracy of the on-board detection, and explore the potential of continuous training for it. 

Once all drones complete their missions they return to their take-off location and land. 
The output is stored in the cluster, and contains the photos and total number of identified people. 

\vspace{-0.05in}

%- disambiguate from the numbers in their shirts?? from the faces

%shirts, colors, faces... 

%\noindent{\bf{Decentralized processing: }}

\section{Centralized vs. Decentralized Coordination}
\label{sec:motivation}

We now examine the trade-offs between centralized and distributed coordination platforms for the scenarios of Sec.~\ref{sec:scenarios}. 

\vspace{0.03in}
\noindent{\bf{Performance: }} Fig.~\ref{fig:latency_balls} shows the latency of different types of tasks 
throughout the duration of the first scenario. The top figure shows task latencies for the centralized 
platform and the bottom for the decentralized system. The placement of tennis balls in the field is identical 
in both cases. In the top figure there are six task types, with three executing 
on the drones; motion control, obstacle avoidance, and data transfer, and three on the cluster; 
image recognition, scheduling of serverless tasks, and load (re)balancing. In the decentralized platform, 
there is no need for the serverless framework, since image recognition happens at the edge. 
The cluster is only used for work assignment and rebalancing. % and data storage. 

\begin{figure*}
\centering
\begin{minipage}{0.62\textwidth}
\begin{tabular}{cc}
	\includegraphics[scale=0.24, viewport=50 37 840 117]{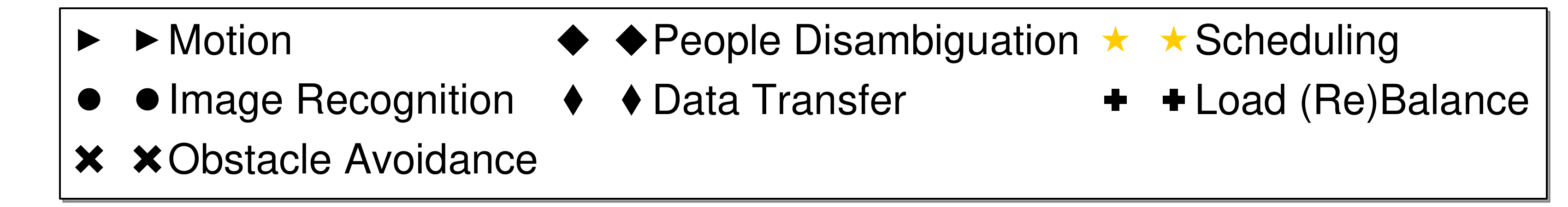} &  
	\includegraphics[scale=0.242, viewport=-52 40 1260 120]{CentrLegendDrone.pdf} \\ 
	\includegraphics[scale=0.19, viewport=200 10 1300 374]{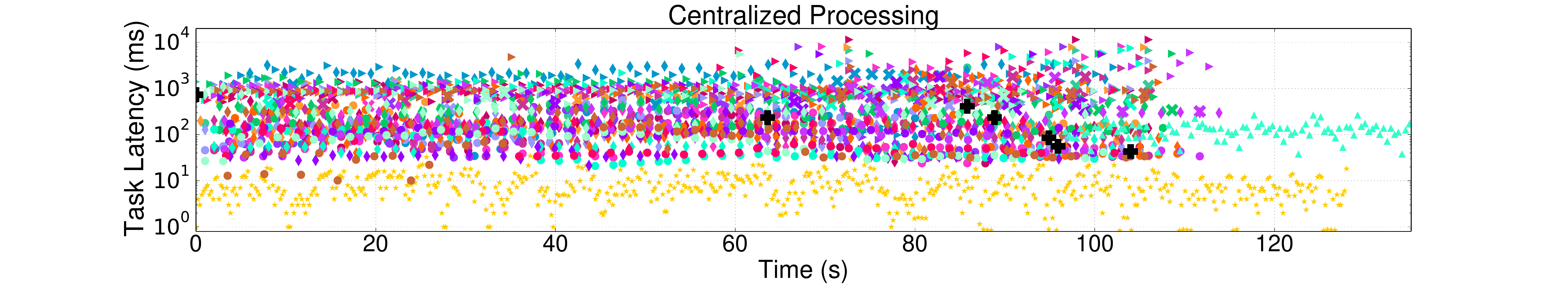} & \\ 
	\includegraphics[scale=0.19, viewport=200 45 1300 340]{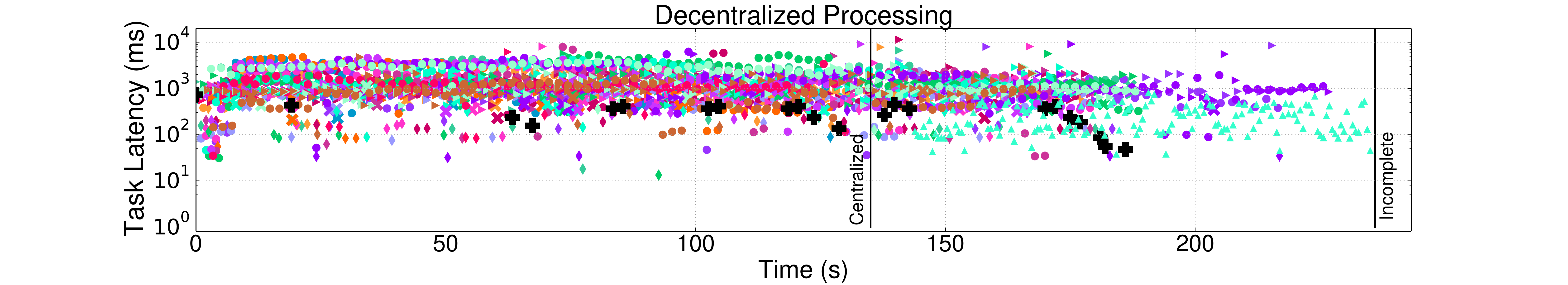} & \\
\end{tabular}
\caption{\label{fig:latency_people} {Latency per task type for centralized (top) and decentralized platforms (bottom) 
in the scenario where we want to count the unique people in an area. We show when the decentralized platform is unable to complete execution. }} 
%We also show when the centralized platform completes the scenario, and when the decentralized platform is unable to continue execution. }}
\end{minipage}
\hspace{0.3cm}
\begin{minipage}{0.34\textwidth}
\begin{tabular}{cc}
	\includegraphics[scale=0.16, viewport=200 46 1400 600]{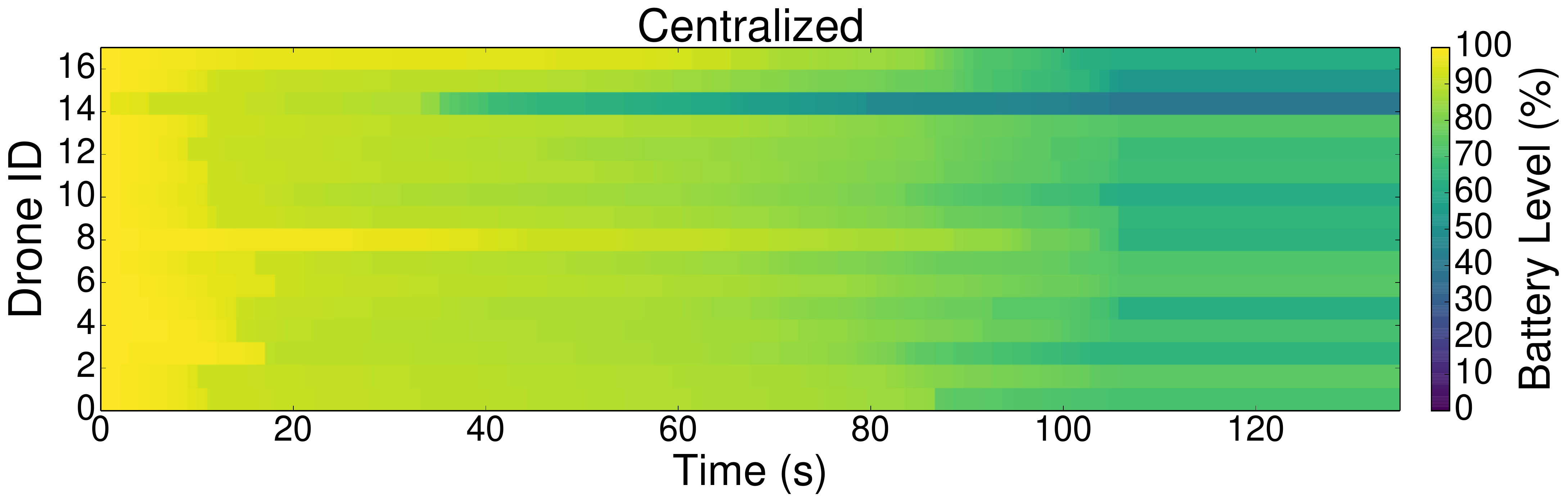} & \\ 
	\includegraphics[scale=0.16, viewport=200 55 1400 435]{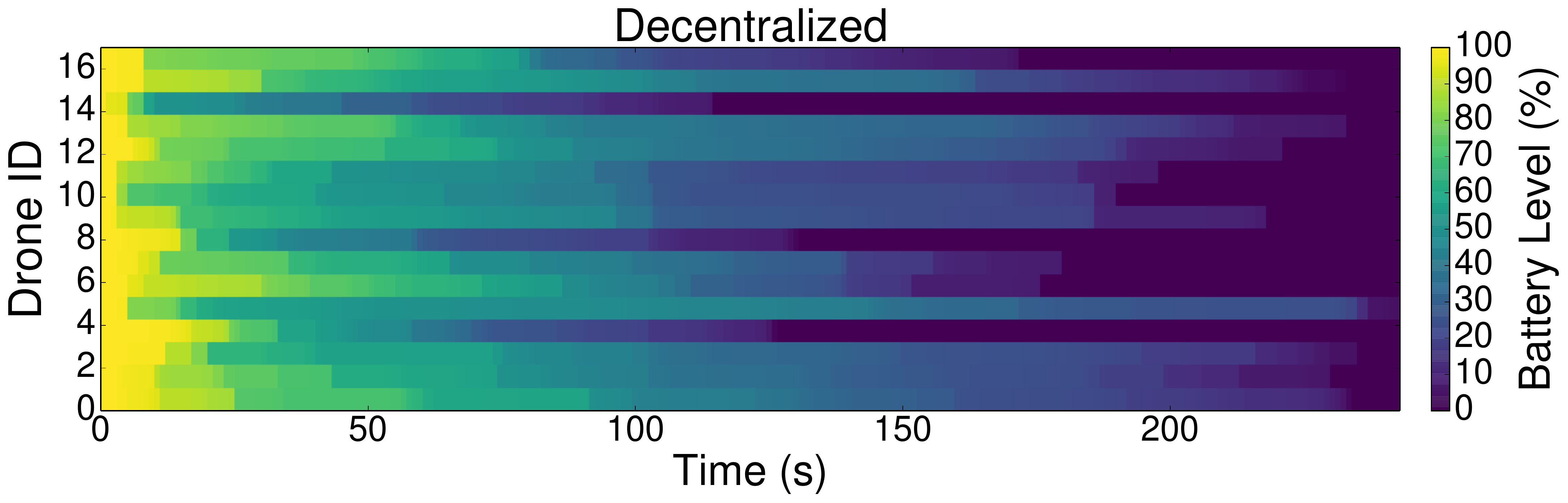} & \\
\end{tabular}
\vspace{0.02in}
\caption{\label{fig:battery_people} {Per-drone battery level for the centralized and decentralized platforms in the second scenario.  }}
\end{minipage}
\end{figure*}

The centralized platform completes the scenario in almost half the time ($74s$) required by the decentralized system ($141s$). %First, 
The most time-consuming tasks involve motion control, as drones move between locations, 
%where photos are taken (approximately 1s per data transfer, although speeds vary slightly in practice), 
followed by tasks transferring data to the cloud. 
%that transfer sensor data to the cloud. 
Inserting a small amount of delay between drones avoids high spikes in data transfer, 
and reduces network latencies, although queueing latencies can still 
occur when all drones are active. Once data are transferred to the cluster, 
image recognition happens fast, taking $23ms$ on average and $49ms$ for the 99$^{th}$ 
inference percentile. Scheduling serverless tasks is almost instantaneous, %especially when resources are plentiful, 
%low and predictable latency
taking $3.2ms$ on average and $5.4ms$ in the worst-case. 
Finally, work assignment happens at time $0$ and only needs to be revisited once towards the end of execution, 
when drone 14's battery starts draining disproportionately fast, 
and its work is assigned to its neighboring drones to avoid it running out of battery. 
This also results in more obstacle avoidance tasks, as more drones are 
congregating in the same area. 
%There are two major types of obstacles the drones encountered; other drones, 
%especially when moving close to their regions' borders, and other objects, including trees and the field's scoreboard's poles. 
The higher movement latencies in the end of the scenario correspond 
to drones returning to their take-off point, which can be far from their current location. 

The main difference in the decentralized platform is that image recognition 
at the edge takes 1-2 orders of magnitude longer compared to serverless. 
This is not surprising, given the limited on-board resources, and the fact that serverless 
can leverage fine-grained parallelism even within a single recognition task. However, 
it has severe implications in execution time, and in the drones' reliability. 
Occupying the drone's resources with image recognition means that the device 
is less agile and less able to adjust its course in a timely manner when needed, e.g., to avoid an obstacle. 
This makes motion control slower and more unpredictable, 
seen by a number of slow motion operations in the middle of the scenario, 
and in the case of drones \texttt{4} and \texttt{14}, it also resulted in them 
crashing at $t=74s$, and being unable to continue their missions. 
This forces the backend controller to rebalance the load several times %an increasing number of times 
as the scenario progresses, both due to failures and low battery reserves for several 
drones. The benefit of the decentralized platform is reducing network traffic, as only photos with 
the target object are shared with the backend cluster. %, considerably reducing the amount of bandwidth used. 
Both coordination platforms correctly identified all 15 tennis balls. 
%the centralized and decentralized systems 

Fig.~\ref{fig:latency_people} shows a similar comparison for the second scenario. 
%where we want to recognize and count the number of unique people on the field. 
This scenario is more computationally-intensive, given the diversity 
in people's anatomy compared to uniform tennis balls, and the need to disambiguate 
people using their faces as opposed to coordinates. As with the first scenario, 
movement is the most time consuming operation for the centralized platform, 
followed by data transfer and image recognition. 
Compared to the first scenario, recognizing people takes longer, $59ms$ on average 
and $159ms$ for the 99$^{th}$ percentile. 
%Disambiguating people starts once the first people are identified by discarding  %139
%those detected two or more times by the same or different drones. 
Disambiguation is the last set of tasks to finish after all drones have completed 
their missions, and incurs similar latencies to image recognition. 
Scheduling serverless tasks incurs similar latencies to the first 
scenario, despite the higher intra-job parallelism, because cluster resources 
never become oversubscribed. 
%As the ratio of edge devices to cluster servers increases, keeping scheduling latency low will become more challenging. 
Between $72$-$100s$ the controller has to rebalance the work assignment to account 
for a subset of drones whose battery is draining quickly, and to counteract the fact that 
drones \texttt{4}, \texttt{5}, and \texttt{14} are moving slower than the rest, 
and hence would degrade overall execution time. %take longer to cover the same area, degrading execution time. 
This causes their neighboring drones to move further away to 
cover the additional areas, seen by the higher movement latencies after $t=72s$. 

In contrast, the decentralized platform progresses at a slower pace, despite avoiding most data transfers, 
and is eventually unable to complete the scenario due to several drones 
running out of battery. The centralized controller rebalances the load several times, 
however, the remaining devices ultimately do not have sufficient battery to accommodate the extra work. 
Both people recognition and face disambiguation take 1-3 orders of magnitude 
longer than in the centralized platform, affecting the drones' flight stability, 
and draining their batteries. Drone \texttt{5} is the last to run out of battery 
at $t=243s$. Most of the drones return to their take-off location before completely 
running out of battery, with the exception of drones \texttt{4} and \texttt{14} which ran out of battery before 
returning to the base station. %was depleted at a faster rate not allowing them to return to the base station before powering off. 

Unlike the centralized platform, the decentralized system is also penalized by the sensor data 
remaining isolated across drones, thus not benefiting from each other's decisions. In Sec.~\ref{sec:evaluation} 
we study the impact of a centralized data repository on a swarm's ability to continue learning online. %during the execution of a mission. 
Finally, the centralized system correctly identified all 25 people in the field, 
while the decentralized system missed 7 people due to drones running out of battery. 
When comparing the two systems, we instruct people to move in exactly the same way. % across tests. 

\vspace{0.03in}
\noindent{\bf{Battery efficiency: }} Fig.~\ref{fig:battery_balls} shows the per-drone 
battery level for the first scenario across the two platforms. All drones start with 100\% 
battery charge. In the case of the centralized platform, battery depletion is mostly uniform, 
with the exception of drone \texttt{14} whose battery is draining at a faster rate, 
due to a fault in the power controller's firmware that kept the core always at the highest frequency. 
To address this, the centralized controller rebalances the load between drone \texttt{14} 
and its neighboring drones to avoid completely draining its battery. 
The average battery level across all drones at the end of the scenario is 73.2\%. 

Battery depletion is less uniform in the decentralized platform, 
with some drones losing charge at higher rates, depending on how quickly they 
perform the on-board image recognition. The increased resource 
load also results in less reliable motion control, causing drones \texttt{4} and 
\texttt{14} to crash and power off. The average battery level in the end is 29.8\%. 

Fig.~\ref{fig:battery_people} shows the per-drone battery level 
for the second scenario. The trade-offs are clearer here %than before %similar %here too 
%but more extreme 
due to the increased computational requirements 
people recognition has, compared to recognizing a single 
stationary item. The average battery level at the end of the scenario 
for the centralized platform is 65\%, lower than before, but much %significantly 
higher than the decentralized system, where almost all drones ran out of 
battery. % either on the field, or have to interrupt their mission and return to the base station. 

\begin{figure}
\centering
\begin{tabular}{cc}
	\multicolumn{2}{c}{\includegraphics[scale=0.22, viewport=240 -40 700 70]{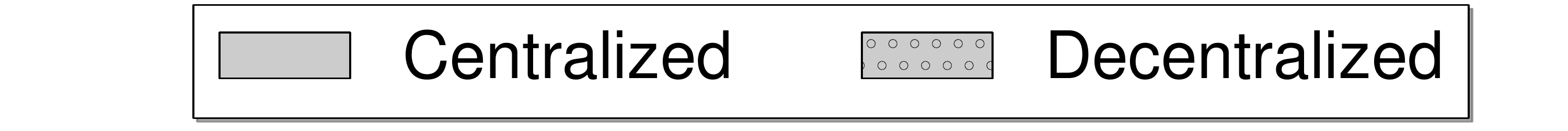}}\\ 
	\includegraphics[scale=0.214, viewport=40 30 600 380]{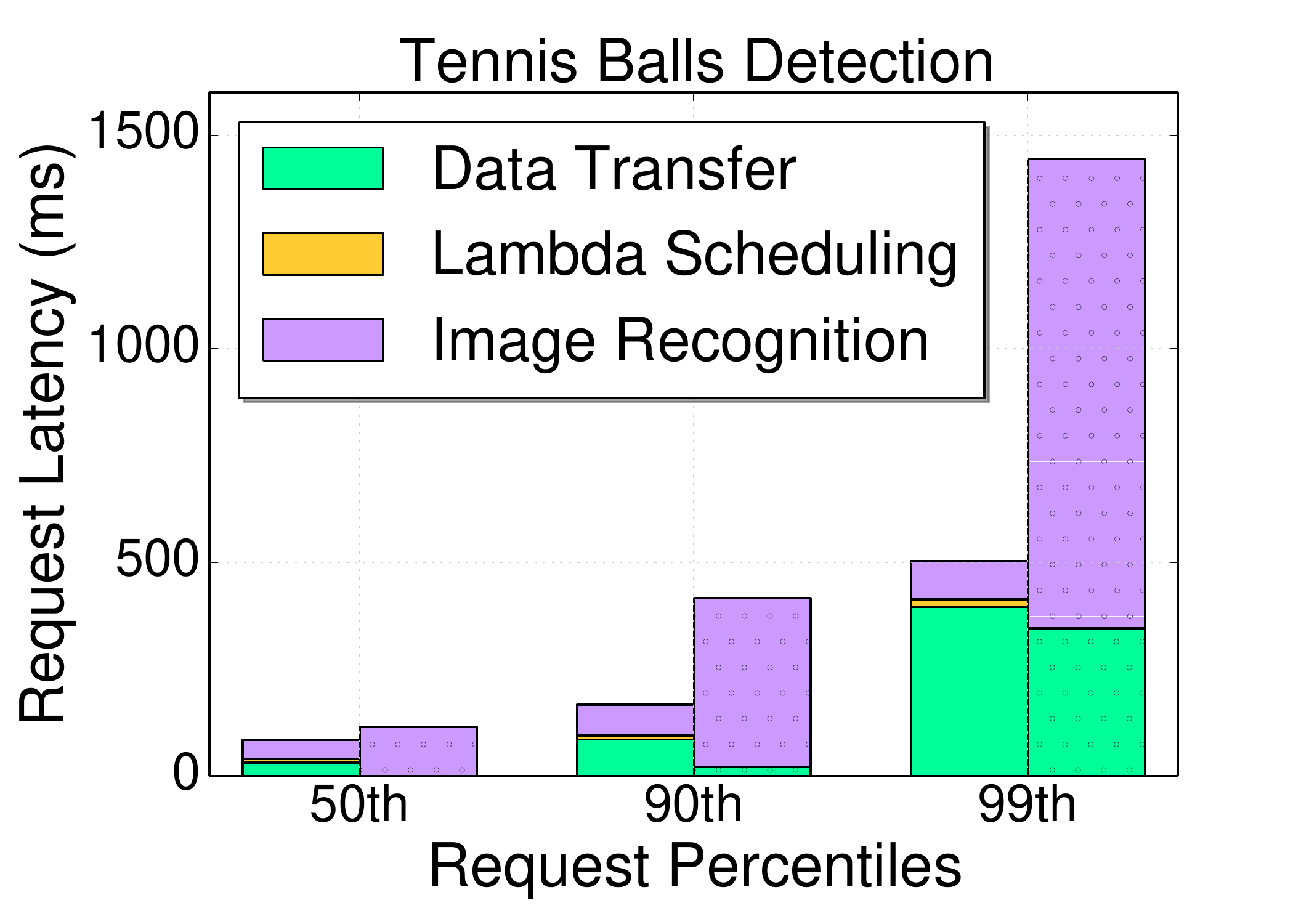} &
	\includegraphics[scale=0.214, viewport=70 30 600 380]{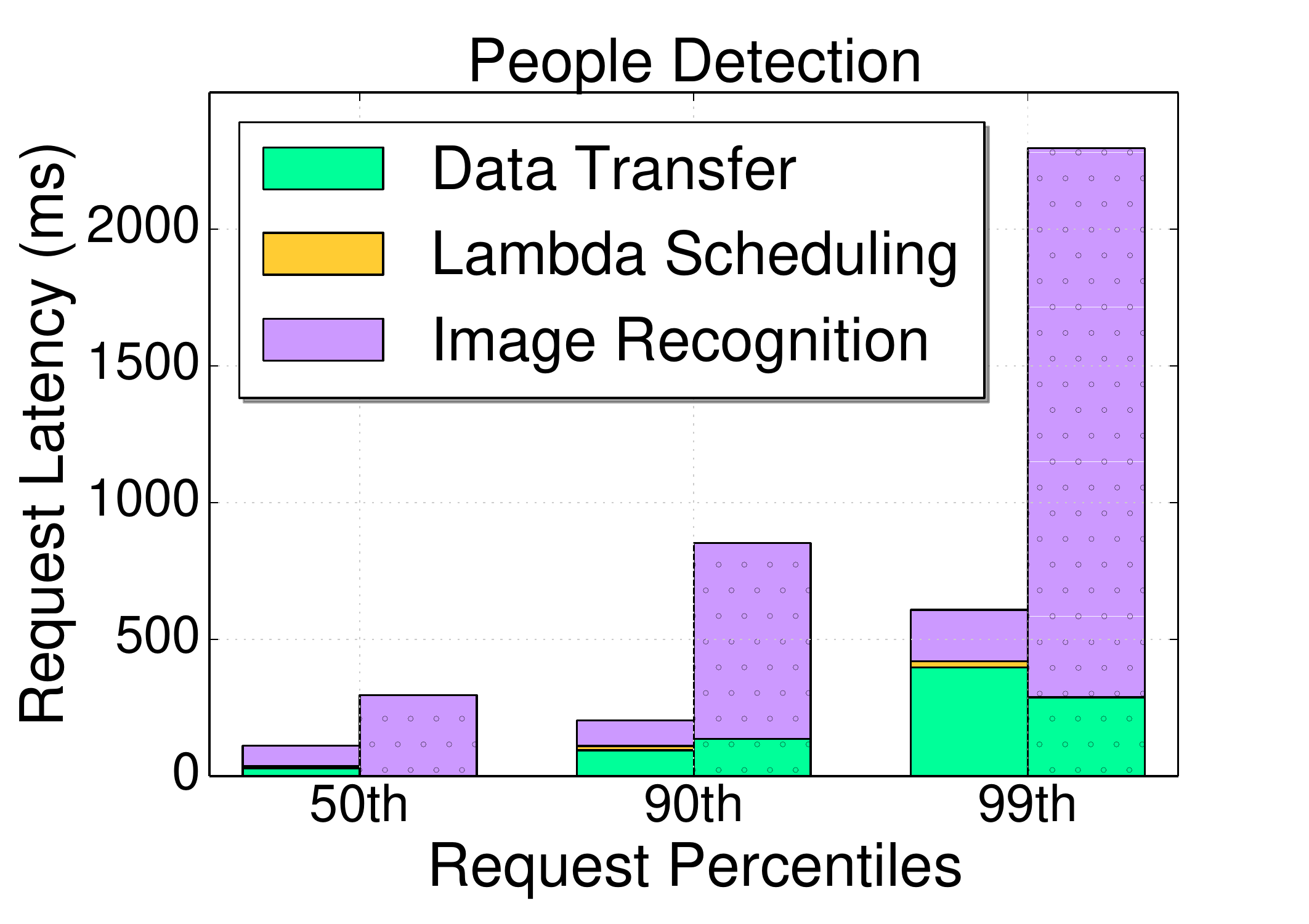} \\
\end{tabular}
\caption{\label{fig:network_proc} {Task latency breakdown for the centralized and decentralized platforms across the two scenarios.}} % Much more predictable latency on cloud, than on drones. }}
\end{figure}

\vspace{0.03in}
\noindent{\bf{Computation vs. communication: }} Fig.~\ref{fig:network_proc} 
shows the breakdown to network communication, computation, and management operations 
for different latency percentiles. Communication includes 
data transfer to the cloud/edge devices, computation includes 
the image recognition and disambiguation tasks, and management 
includes the overhead of serverless task scheduling. We omit 
the load (re)balancing tasks as they are very infrequent. 
In the centralized platform data transfer accounts for the largest latency fraction, 
especially in high percentiles. In comparison, the decentralized platform 
incurs lower latencies for data transfer, which would allow the swarm to scale to larger device numbers. 
%which mostly involve data shared across edge devices, and output results transferred to the cluster. 
On the other hand, computation is much more costly in the decentralized platform, especially in the second scenario, 
with the latency of recognition tasks being additionally highly variant. %, compared to the centralized system. 
For example, the 99$^{th}$ \%ile of image recognition 
for the second scenario is $159ms$ for the centralized platform 
compared to $2006ms$ for the decentralized, 
hurting performance predictability. %hurting predictability
Finally, management tasks introduce negligible overheads, less than $5.5ms$ in all cases. 

%\section{HiveMind Platform Design}
\section{HiveMind Design}
\label{sec:design}

\subsection{Overview}
\label{sec:overview}

The analysis of Sec.~\ref{sec:motivation} showed that centralized platforms achieve better performance, battery efficiency, and 
output quality compared to decentralized coordination control systems, but at the cost of much higher bandwidth usage, which limits their 
scalability. %add something on how the analysis before affected the design of the platform. 
In this section we present HiveMind, a coordination control platform for large IoT swarms designed to achieve the best of centralized 
%(performance, energy efficiency, output quality) 
and decentralized %(scalability) 
platforms, and optimize for performance, battery efficiency, %reliability, 
and fault tolerance. To evaluate HiveMind, we use the same 16-drone swarm as before, however, HiveMind's 
design principles are not drone-specific, and the platform can be used for %to port applications 
diverse types of IoT swarms, including autonomous vehicles. 
%terrestrial robots and 
%We also use HiveMind to highlight the trade-offs between fully centralized and fully decentralized 
%swarm control, and demonstrate the potential serverless compute has in exposing ample cloud resources to the swarm, while keeping cost and latency low. %reducing the cost?? make it event-driven, responsive? 

HiveMind is designed with the following principles, each of which is detailed below: $\bullet$ \textit{centralized control}, 
$\bullet$ \textit{event-driven, fine-grained cloud computation}, $\bullet$ \textit{scalable scheduling and resource allocation}, $\bullet$ \textit{on-board preprocessing}, 
$\bullet$ \textit{fault tolerance}, $\bullet$ \textit{dynamic load balancing}, $\bullet$ \textit{continuous learning}, and a $\bullet$ \textit{generalizable programming framework}. 

\subsection{Centralized Controller}
\label{sec:controller}

HiveMind uses a centralized controller to obtain % with 
global visibility on the state and sensor data of all devices in a swarm. 
This allows the platform to ensure higher quality routing, better fault tolerance (Sec.~\ref{sec:fault_tolerance}) and 
load balancing (Sec.~\ref{sec:load_balancing}), as well 
as to leverage the entire swarm's data to continuously improve its decision quality (Sec.~\ref{sec:continuous_learning}). 

The controller consists of a load balancer, which for the examined scenarios also performs 
the swarm's route mapping, an interface to the serverless framework responsible for cloud-side computation, 
an interface to communicate with the edge devices, and a monitoring system that collects 
tracing information from the edge devices and cloud servers. The controller also has visibility over the 
sequence of operations in a scenario, and is %is also 
responsible for training and deploying the initial models to the edge devices, 
and for retraining them later, if necessary. The controller is implemented as a centralized process in C++, %and 
with two hot standby copies of the master that can take over quickly, in the case of a failure. % in case of a failure. 

\vspace{-0.05in}
\subsection{On-Board Preprocessing}
\label{sec:preprocessing}
\vspace{-0.05in}

The analysis of Sec.~\ref{sec:motivation} showed that offloading all computation 
to the cloud can lead to network link saturation, limiting the swarm's scalability. 
Additionally, there are mission-critical tasks, such as obstacle avoidance, that cannot afford the latency 
of sending data to/from the cloud. 
HiveMind leverages the on-board resources to preprocess sensor data and 
filter those important enough to be sent to the backend cluster for further processing. In the case of the first scenario we examine, 
drones perform an initial image recognition on-board to find objects with an approximate circular shape 
using a simple and lightweight detection library, and only offload to the cloud images that contain such objects. 
Although this approach is prone to false positives, and in some cases false negatives too, it greatly reduces the 
bandwidth usage, allowing HiveMind to support a larger number of IoT devices. 
Similarly, HiveMind uses the on-board compute resources and low-power front-camera to detect obstacles in the drone's vicinity, 
and adjusts its route to avoid them. Offloading this operation to the backend cluster can be subject to %lead to 
high latencies that often result in crashes and catastrophic equipment failures. 

\vspace{-0.05in}
\subsection{Serverless Cloud Framework}
\label{sec:serverless}
\vspace{-0.05in}

HiveMind uses OpenWhisk to launch and execute tasks on the backend cluster. Serverless allows exploiting fine-grained 
parallelism within a single job, decreasing task latency. %the total scenario execution time. 
Once a new photo arrives from a drone, the centralized controller invokes the OpenWhisk scheduler to launch the new job. Each 
job is divided to several serverless functions, either determined empirically by the OpenWhisk master based on the amount 
of data processed, or defined by the user. Each function is then spawned in a Docker container and allocated one core, 
2{\smallcapital GB} of memory, and 512{\smallcapital MB} of disk storage by default, 
consistent with resource allocations on AWS Lambda~\cite{lambda}, Google Functions~\cite{google_functions}, 
and Azure Functions~\cite{azure_functions}. Each function can also access the remote shared persistent storage system holding 
all the training datasets, and sensor data transferred by the edge devices, similar to {\smallcapital AWS} Lambdas accessing {\smallcapital S3} storage. 

Each worker node runs a \textit{worker monitor}, which tracks and reports resource utilization to the OpenWhisk master, 
allowing it to adjust the amount of allocated resources via the Docker resource interface, if necessary. 
A user can also bypass the scheduler's resource allocation policy and customize the amount of resources per serverless function. 
Similarly, users can express priorities for different types of jobs, or different edge devices. In our scenarios 
we assume that all drones have equal priority. 
When cluster resources are plentiful, {\smallcapital CPU}s are dedicated to a single container to avoid contention. Given that each 
function lasts at most a few hundred milliseconds in our scenarios, this does not result in new jobs being queued waiting for resource 
allocations. We plan to explore more resource-efficient allocation strategies in future work. 

We have also implemented a task monitoring infrastructure in OpenWhisk that checks the progress of active serverless functions, and flags
potential \textit{stragglers} that can degrade execution time. If a serverless function takes longer than the 90$^{th}$ percentile 
of functions in that job, OpenWhisk respawns the misbehaving tasks on new physical servers, and uses the results of whichever tasks finish first~\cite{Ousterhout13,tailatscale}. 
The exact percentile that signals a straggler can be tuned depending on the importance of a job. If several underperforming tasks all come from the 
same physical node, that server is put on probation for a few minutes until its behavior recovers. OpenWhisk also respawns any failed tasks 
by default. 

Once a job completes, the OpenWhisk master informs the centralized controller and passes the job's output to it. By default, a container is terminated 
when its function completes execution. This can introduce significant instantiation overheads if job churn is high. In Sec.~\ref{sec:evaluation} 
we explore different container keep-alive policies to reduce start-up overheads. 

\subsection{Fault Tolerance at the Edge}
\label{sec:fault_tolerance}
\begin{wrapfigure}[10]{l}{0.25\textwidth}
	\centering
	\vspace{0.05in}
	        \includegraphics[scale=0.19, viewport=200 100 580 430]{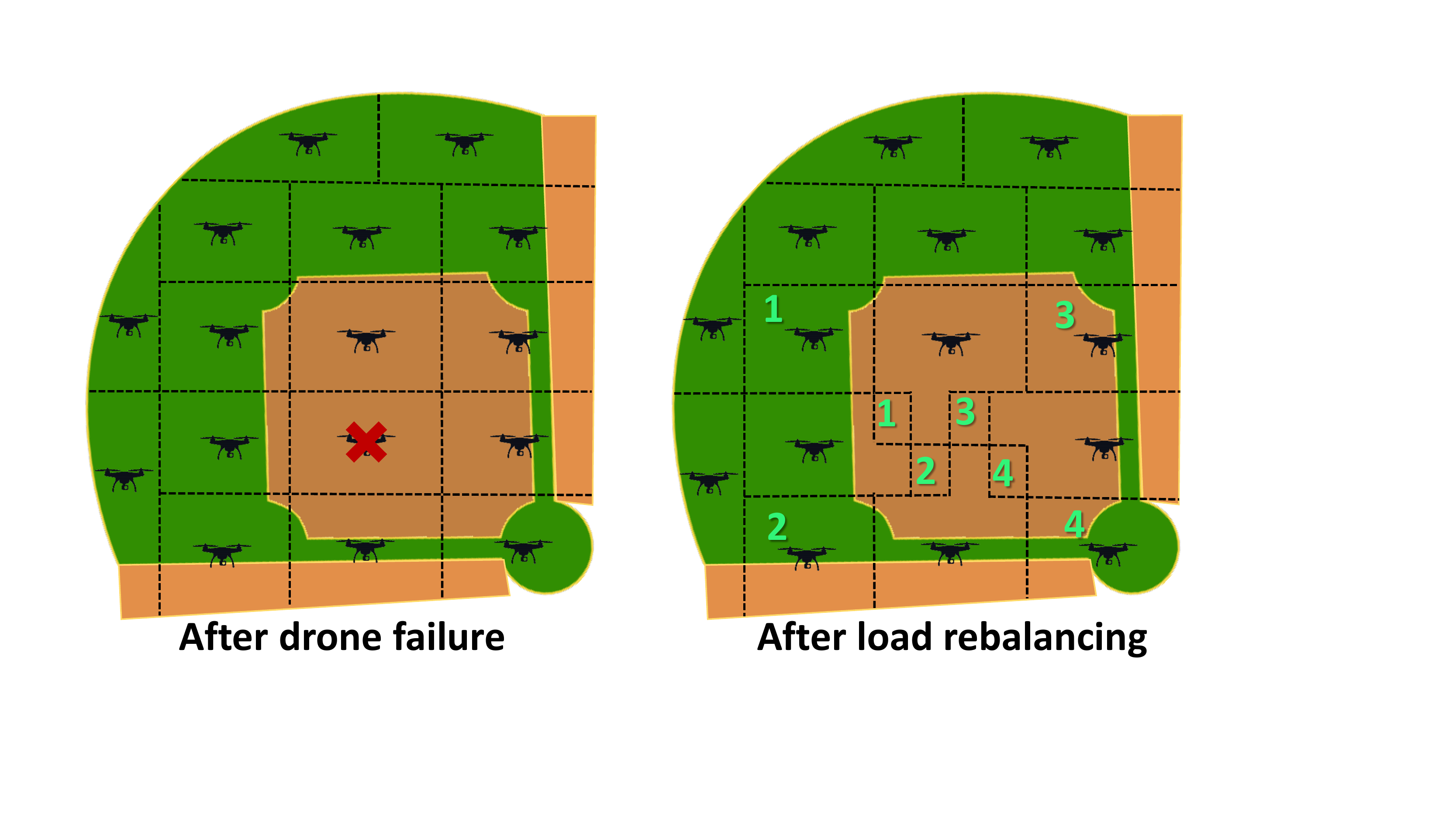}
\caption{\label{fig:fault_tolerance_partitioning} {Load repartitioning to handle a drone failure. }}
\end{wrapfigure}
Edge devices are prone to failures and unpredictable behavior. In the scenarios we examine, all drones 
send a periodic heartbeat to HiveMind (once per sec). If the controller does not receive a heartbeat for 
more than 3s, it assumes that the drone has failed. HiveMind handles such failures by repartitioning 
the load among the remaining drones. Fig.~\ref{fig:fault_tolerance} shows such an example for our scenarios. 
Immediately after HiveMind realizes that the red-marked drone has failed, it repartitions its assigned area equally 
among its neighboring drones assuming they have sufficient battery, and updates their routing information. 
Depending on which drone has failed, this involves reassigning work to 3-8 drones for this example. If the failed 
drone had already executed part of its route, HiveMind only repartitions the remaining area. 
If the non-responsive drone makes contact before the other drones start its work, the adjustment is reverted. 

\vspace{-0.05in}
\subsection{Dynamic Load Rebalancing}
\label{sec:load_balancing}
\vspace{-0.05in}

Failure is not the only reason why load may need to be repartitioned among edge devices. Often some devices 
deplete their battery reserve faster, either due to hardware/software bugs, or due to performing more resource-intensive 
computation or movement. For any of these cases, HiveMind repartitions the work assigned to edge devices. When 
repartitioning work, HiveMind tries to accommodate the extra work using neighboring devices only, if possible, to 
avoid long travel times. 
%In Section~\ref{sec:evaluation} we explore load rebalancing cases caused by both drone failures and underperforming drones. 
We plan to extend load rebalancing to account for heterogeneous edge devices as part of future work. 

\vspace{-0.05in}
\subsection{Continuous Learning}
\label{sec:continuous_learning}
\vspace{-0.05in}

A benefit of centralized coordination is that data from all devices can be collectively used to improve the learning ability of the swarm. 
%to allow edge devices to benefit %is stored in a single, shared location, allowing the edge devices to benefit 
%from each other's actions to improve their learning ability. 
Once HiveMind receives the first few images that edge devices have tagged as containing the target object, it verifies whether 
their detection was accurate. In the case of a false positive, HiveMind penalizes the incorrect decision, periodically retrains the 
on-board detection engines, and redeploys the new model to the edge devices. In Sec.~\ref{sec:evaluation} we show that leveraging 
decisions from all devices improves decision quality much more quickly than retraining a device only based on its own decisions. To handle 
false negatives, after the end of a scenario's execution, HiveMind verifies that any images not sent %also retrains the on-board models after the end of a scenario's execution after verifying that images not sent 
to the cluster did indeed not contain the target object. These images are stored on the drone's local flash drive for validation. 
In case of undetected objects, HiveMind retrains the on-board models, and redeploys them for the next execution. %time the scenario executes. 

\vspace{-0.05in}
\subsection{Programming Framework}
\vspace{-0.05in}

HiveMind uses a high-level, event-driven programming framework to allow users to express new application scenarios. The framework 
supports applications in node.js and Python, and we are currently expanding it to Scala. Users have to express the sequence of 
operations in their scenario, as well as define which operations run on the edge devices, and which on the serverless framework. %determine the routes?? 
Users are also responsible with providing any ML models and training datasets needed for their applications. Finally, 
users can optionally express serverless scheduling policies that deviate from what HiveMind supports by default, as well as 
priorities between edge devices. They can also customize the fault tolerance, load rebalancing, and continuous learning techniques. 
HiveMind automatically handles the communication interfaces between OpenWhisk, the controller, and the edge devices, as %and controller and OpenWhisk, as 
well as the monitoring and tracing frameworks in the cloud and edge. 

\vspace{-0.05in}
\subsection{Putting It All Together}
\vspace{-0.05in}

\begin{figure}
       \centering
               \includegraphics[scale=0.279, viewport=280 10 600 480]{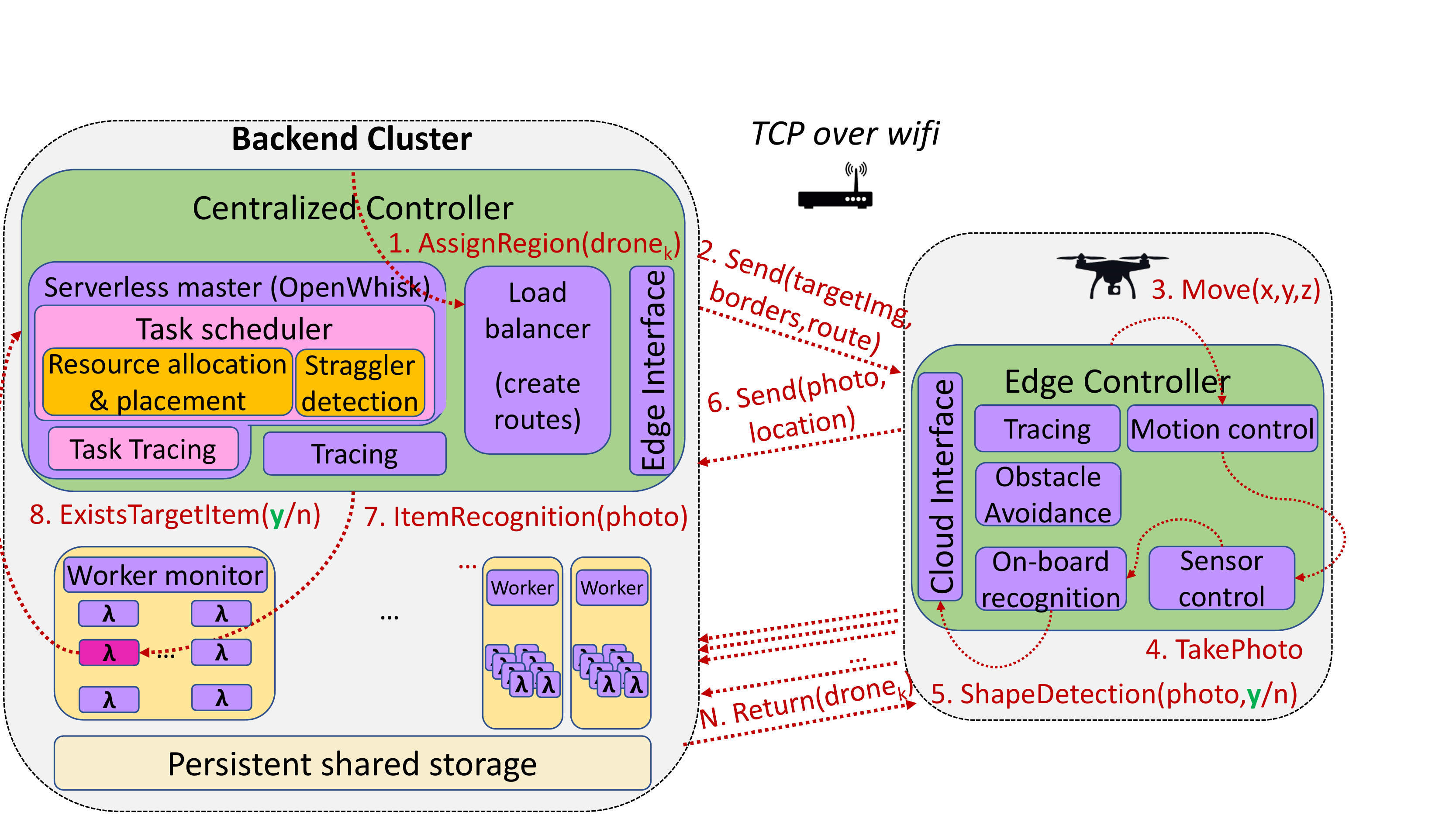}
\caption{\label{fig:hivemind_overview} {HiveMind platform overview, and sequence of operations for the first scenario. }}
\end{figure}

Fig.~\ref{fig:hivemind_overview} shows an overview of the platform's cloud and edge components, and the sequence of operations when executing 
the first scenario. The controller first communicates with the drones their work assignment and route, after which, drones start their 
mission, collecting photographs and analyzing them on-board. 
%Talk about the steps for the two apps in HiveMind and go over the figure here. 
In the first scenario, this involves drones using a simple SVM classifier implemented in OpenCV based on \texttt{cylon}~\cite{cylon} to 
put bounding boxes around any circular objects. Images with such objects are transferred to the cloud for detailed image recognition. After the serverless functions 
complete, OpenWhisk informs the HiveMind controller whether a unique tennis ball indeed existed in the image. The scenario continues until all drones cover their assigned regions. 

In the second scenario, the drones perform an initial human detection, 
placing bounding boxes around shapes that resemble humans, using the rectangular and oval item 
recognition model of \texttt{cylon} in OpenCV~\cite{cylon}.
They then only transfer images with such shapes to the cloud. %, storing all others in the on-board flash drive for a posteriori validation.
The backend cluster uses the detailed TensorFlow model to verify that the image indeed contained a human, 
and disambiguate them against previously-detected people. In Sec.~\ref{sec:evaluation} we evaluate the accuracy of on-board detection for both scenarios. %, and explore the potential of continuous training for it.

\section{Implementation}
\label{sec:implementation}

\noindent{\bf{HiveMind controller: }}The HiveMind controller is written in approximately \texttt{10,000} lines of code in C++ and currently 
supports Ubuntu 14.04 and newer versions. It includes the load balancer, route planner, and the interface to the edge devices via the wireless router, 
as well as the interface to invoke and receive information from the serverless framework. The controller communicates with edge devices over TCP, although 
UDP is also supported. 
We have also implemented a monitoring system in the controller that tracks application progress, 
errors, and device status, and verified that the tracing system has no meaningful impact on performance; 
less than 0.1\% impact on task tail latency, % of scheduling and task execution, 
and less than 0.2\% on  throughput. % scheduling and execution. 

\noindent{\bf{Serverless framework: }}We use OpenWhisk v.0.10.0 and extend the OpenWhisk master by \texttt{4,000} LoC in Go 
to support the task scheduling and straggler detection policies described in Sec.~\ref{sec:design}, 
as well as implement the interface with the HiveMind controller. We also implement a fine-grained monitoring system 
both in the OpenWhisk master and each worker node to track task latency, resource utilization, and function errors. 

\noindent{\bf{Edge devices: }} The platform on the edge devices includes the network interface to talk to the HiveMind controller over {\smallcapital TCP} using the 
wireless router, the motion controller that executes the drone's route, the obstacle avoidance engine, the on-board object detection engines, 
and a logger to track task performance, errors, and battery levels. Finally, the code on edge devices also handles state management by storing in the on-board flash drive 
any images not sent to the cloud. 

\noindent{\bf{Applications: }}HiveMind currently supports applications in node.js and Python, and is being extended to also support Scala. The 
two application scenarios we evaluate use primarily OpenCV libraries for item and people recognition at the edge, and OpenCV and TensorFlow models 
when recognition happens in the backend serverless framework. %We discuss the models in more detail in Section~\ref{sec:scenarios}. 

\section{Evaluation}
\label{sec:evaluation}

\begin{figure*}
	\begin{minipage}{0.48\textwidth}
\centering
\begin{tabular}{ccc}
	\multicolumn{3}{c}{\includegraphics[scale=0.212, viewport=590 -64 1210 60]{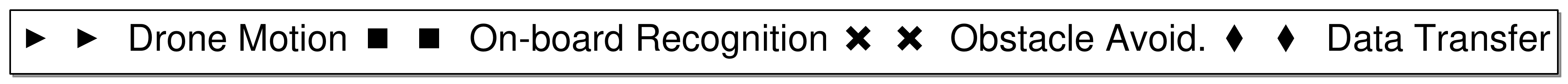}} \\
	\includegraphics[scale=0.156, viewport=100 10 1400 270]{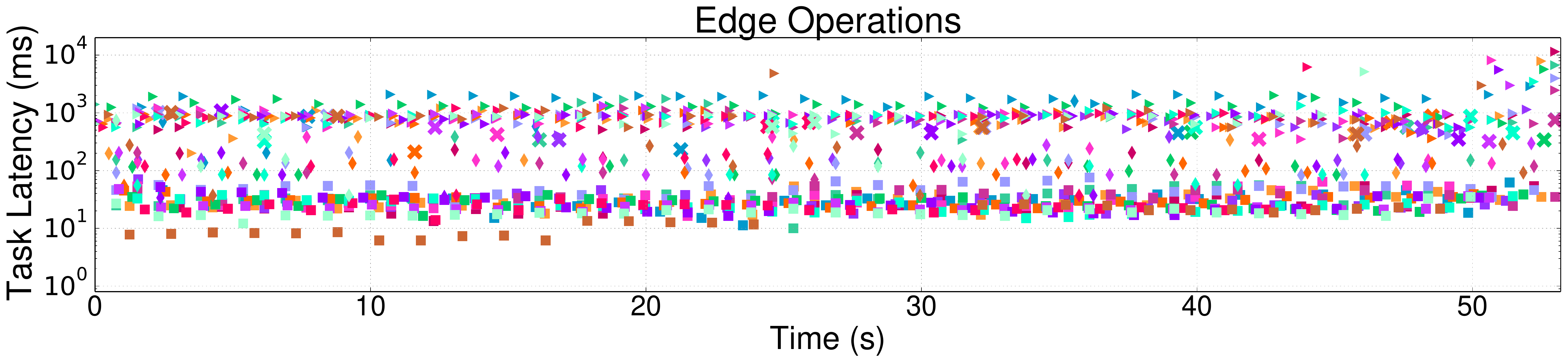} & & \\ 
	\includegraphics[scale=0.156, viewport=70 10 1400 330]{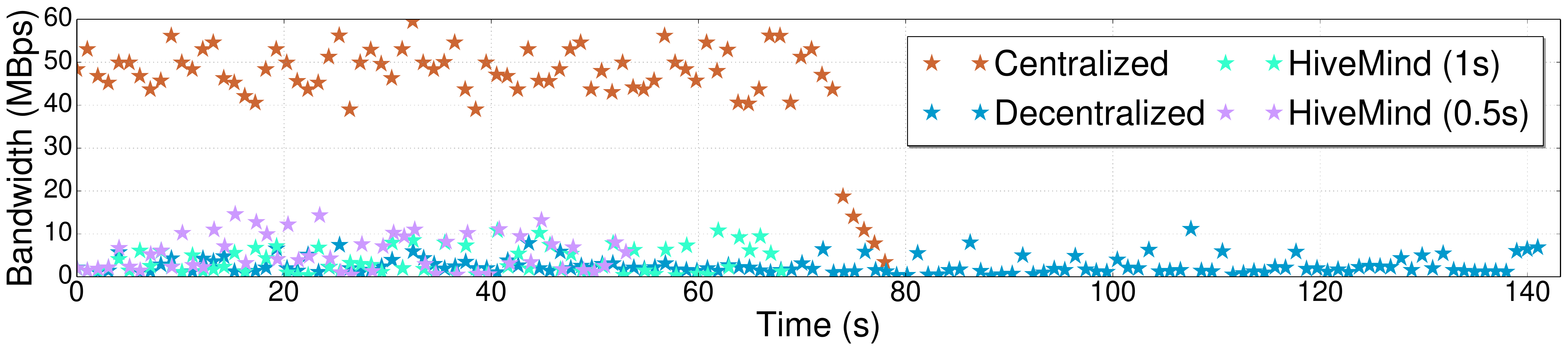} & & \\ 
	\includegraphics[scale=0.144, viewport=350 10 1205 350]{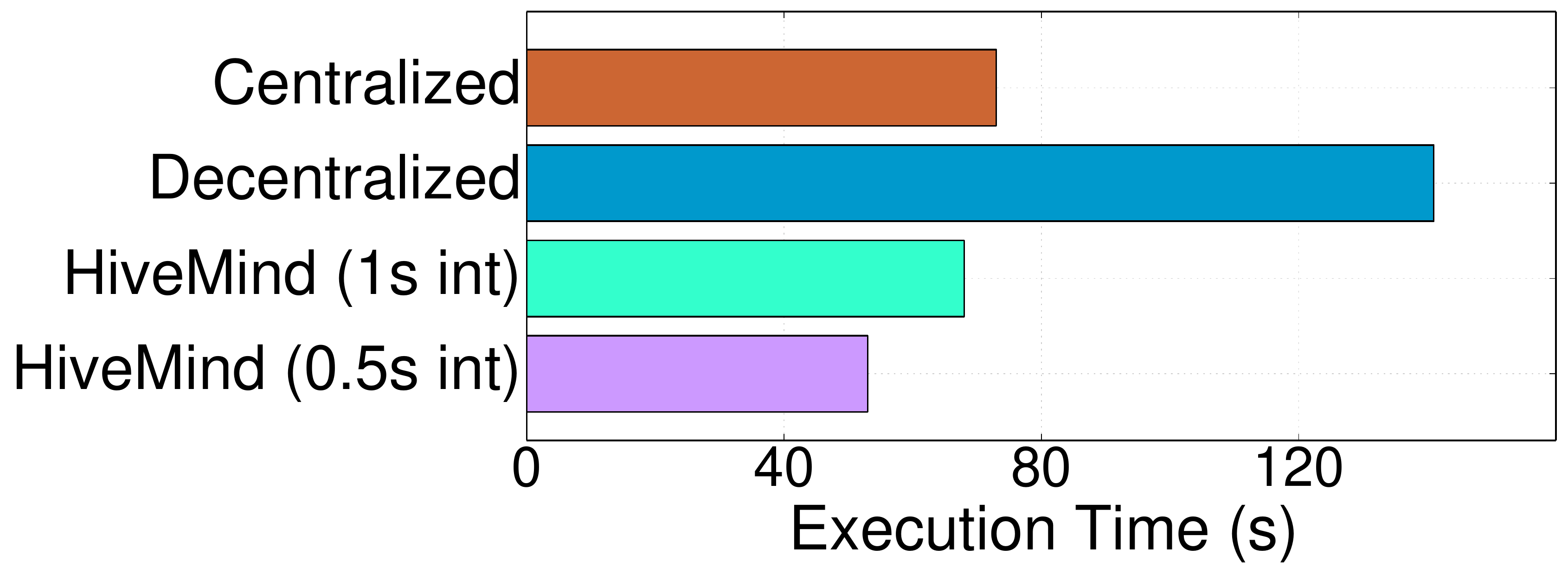} & 
	\includegraphics[scale=0.144, viewport=604 10 1200 330]{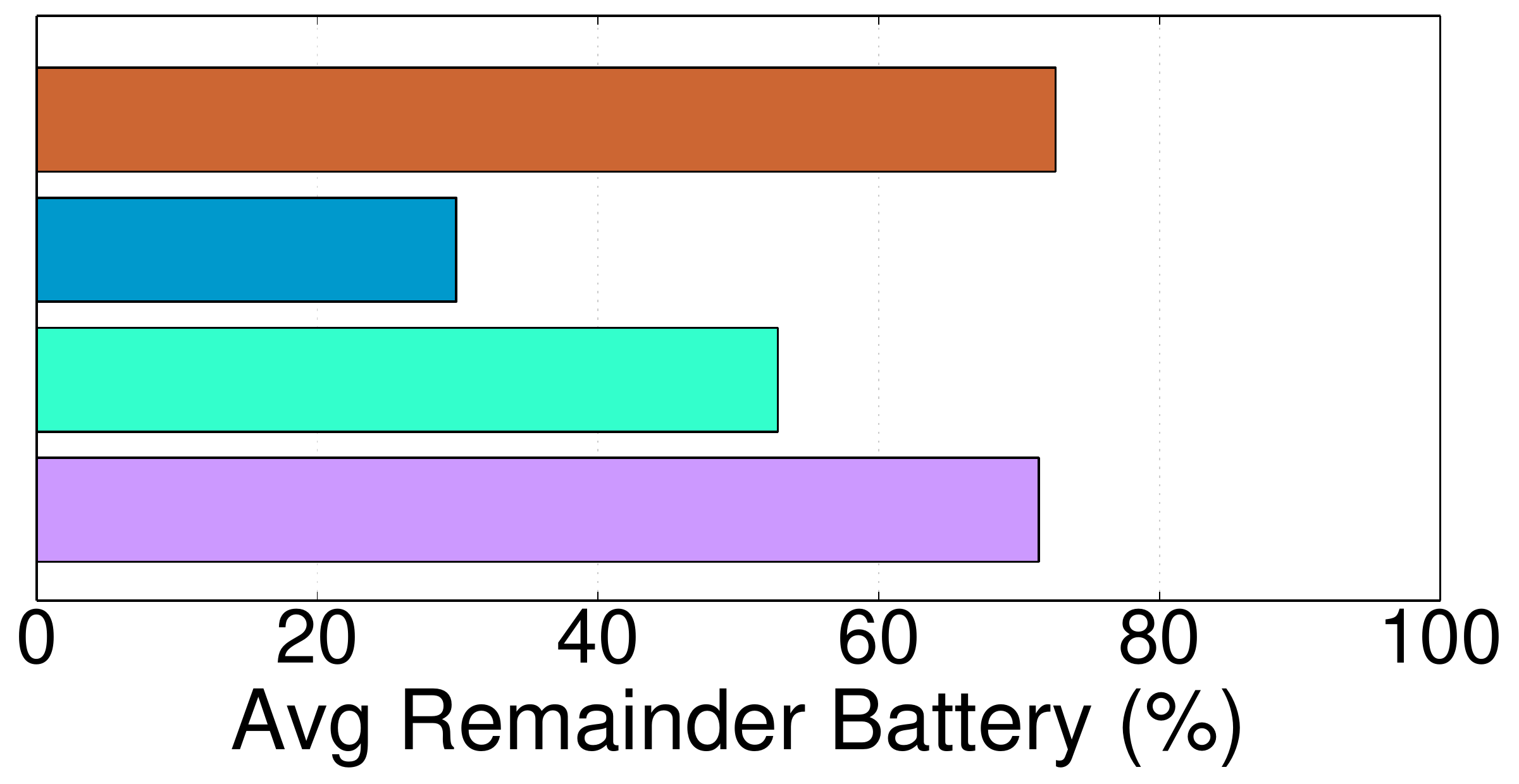} & \\ 
	\includegraphics[scale=0.15, viewport=580 40 900 350]{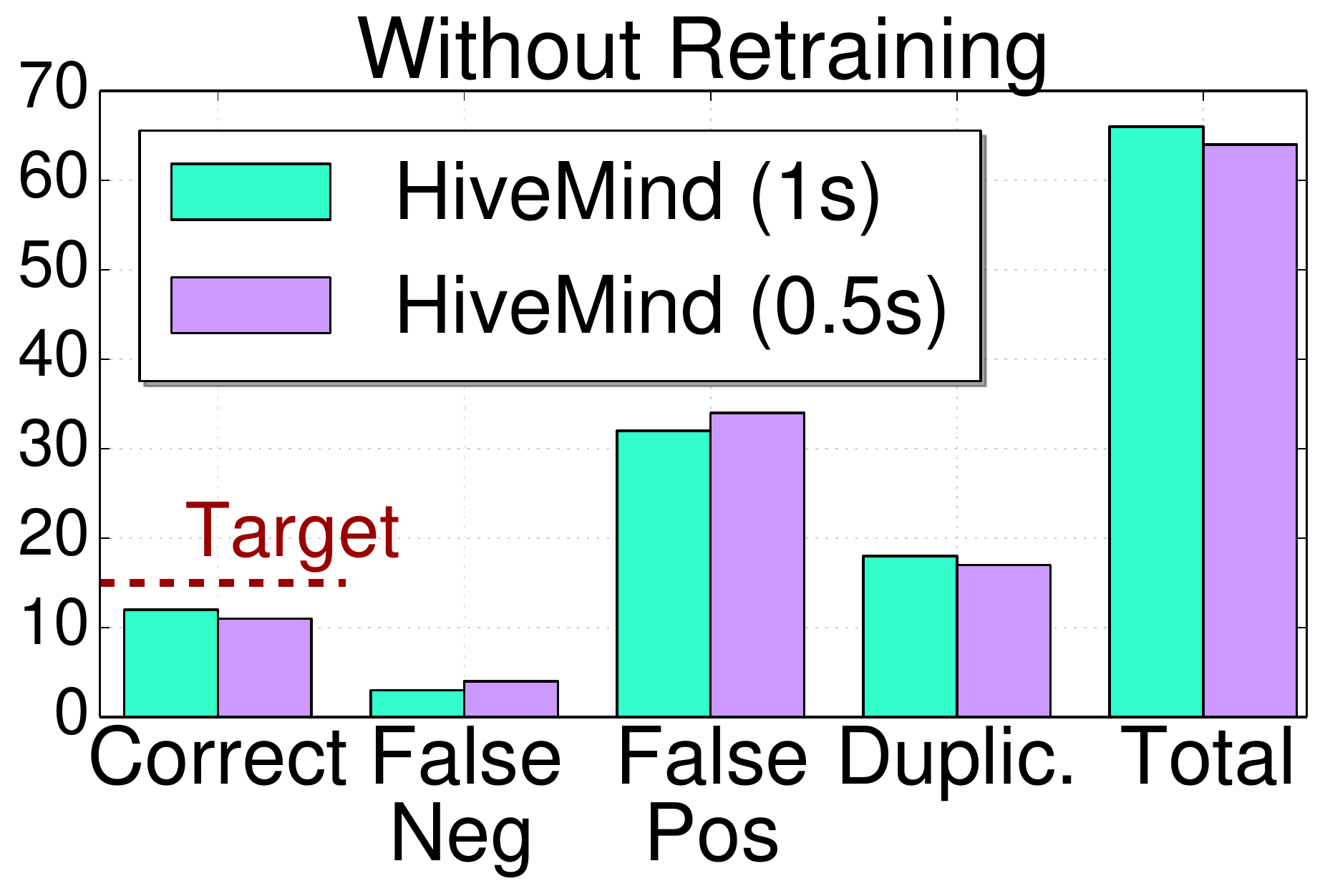} & %accuracy with/without reinforcement learning
	\includegraphics[scale=0.15, viewport=1010 40 1500 350]{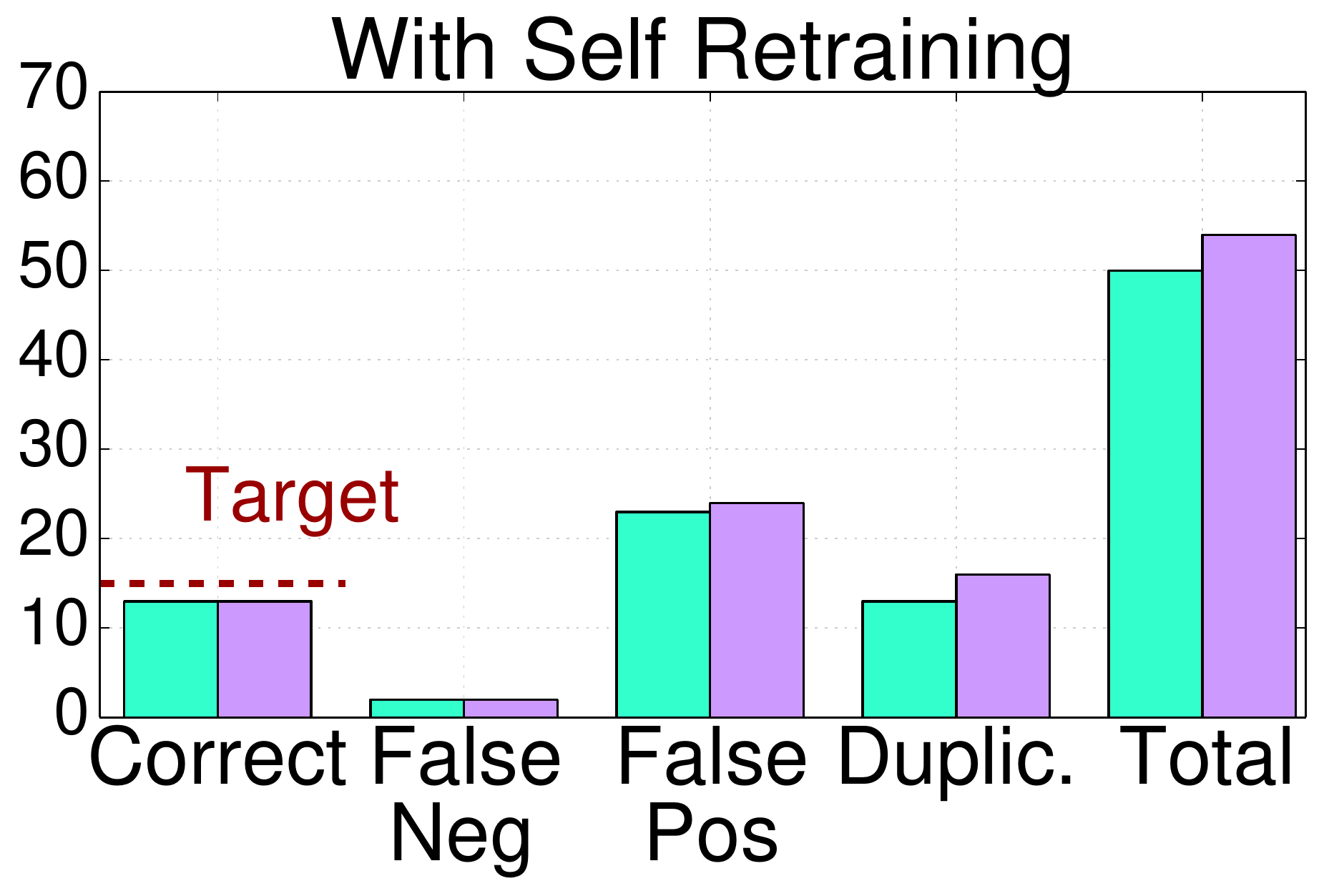} & %accuracy with/without reinforcement learning
	\includegraphics[scale=0.15, viewport=1074 40 1600 330]{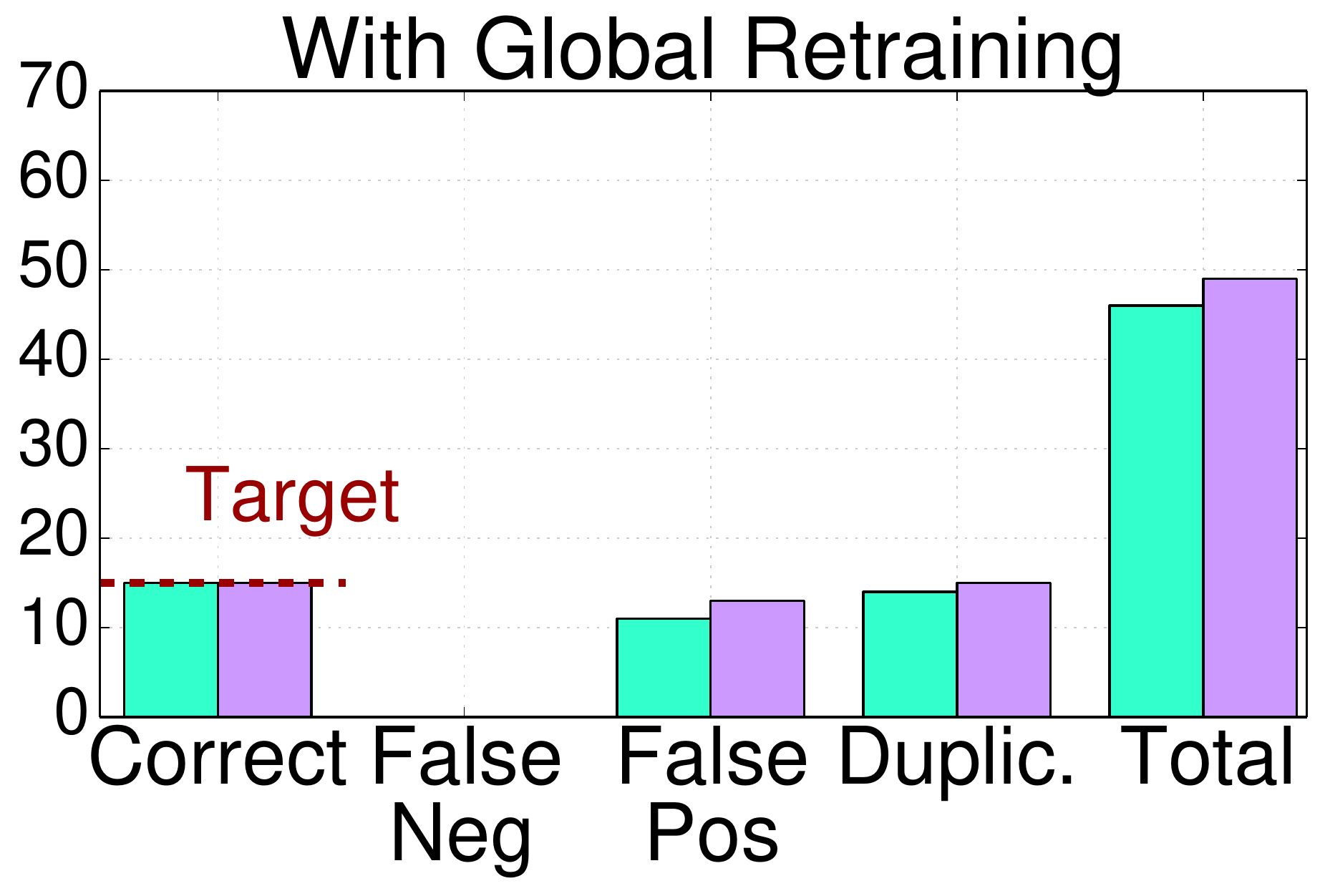} \\ 
\end{tabular}
\caption{\label{fig:hybrid_balls_edge} {From top to bottom for the first scenario: (a) latency of different tasks running on the drones with HiveMind, %the hybrid platform, 
		(b) bandwidth usage across platforms, (c) performance and battery efficiency comparison, and (d) the impact of online 
retraining to the accuracy of on-board recognition. }}
\end{minipage}
\hspace{0.4cm}
\begin{minipage}{0.48\textwidth}
\centering
\begin{tabular}{ccc}
	\multicolumn{3}{c}{\includegraphics[scale=0.17, viewport=952 -70 1540 80]{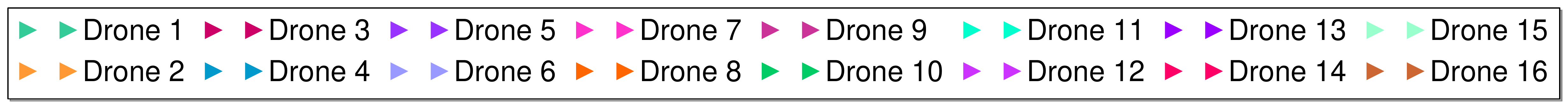}} \\
	\includegraphics[scale=0.156, viewport=80 10 1400 270]{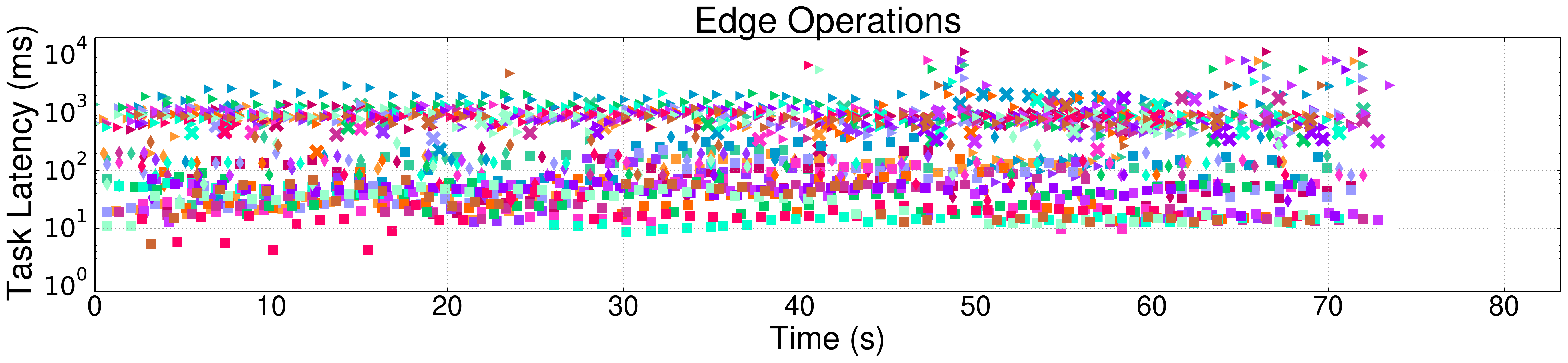} & & \\ 
	\includegraphics[scale=0.156, viewport=50 10 1400 330]{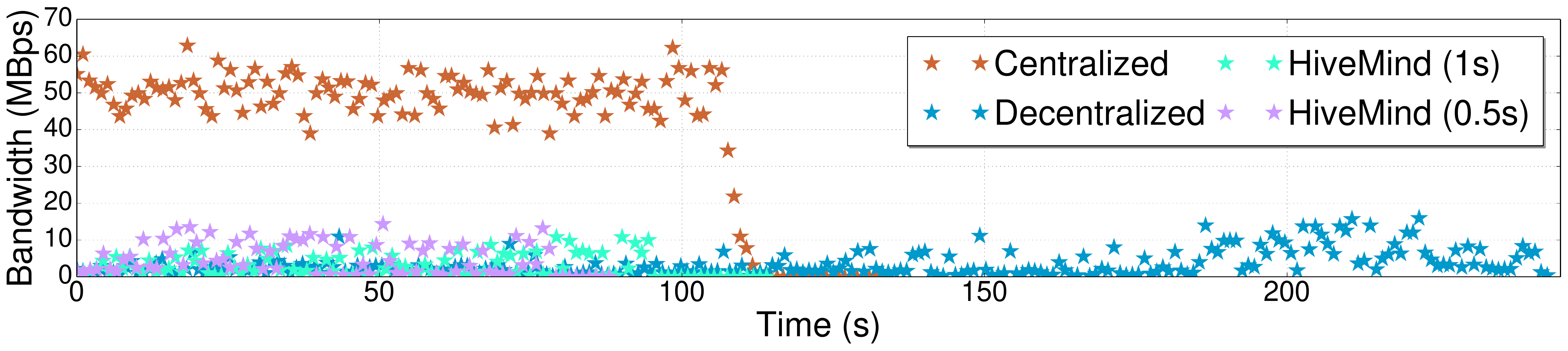} & & \\ 
	\includegraphics[scale=0.144, viewport=330 10 1200 350]{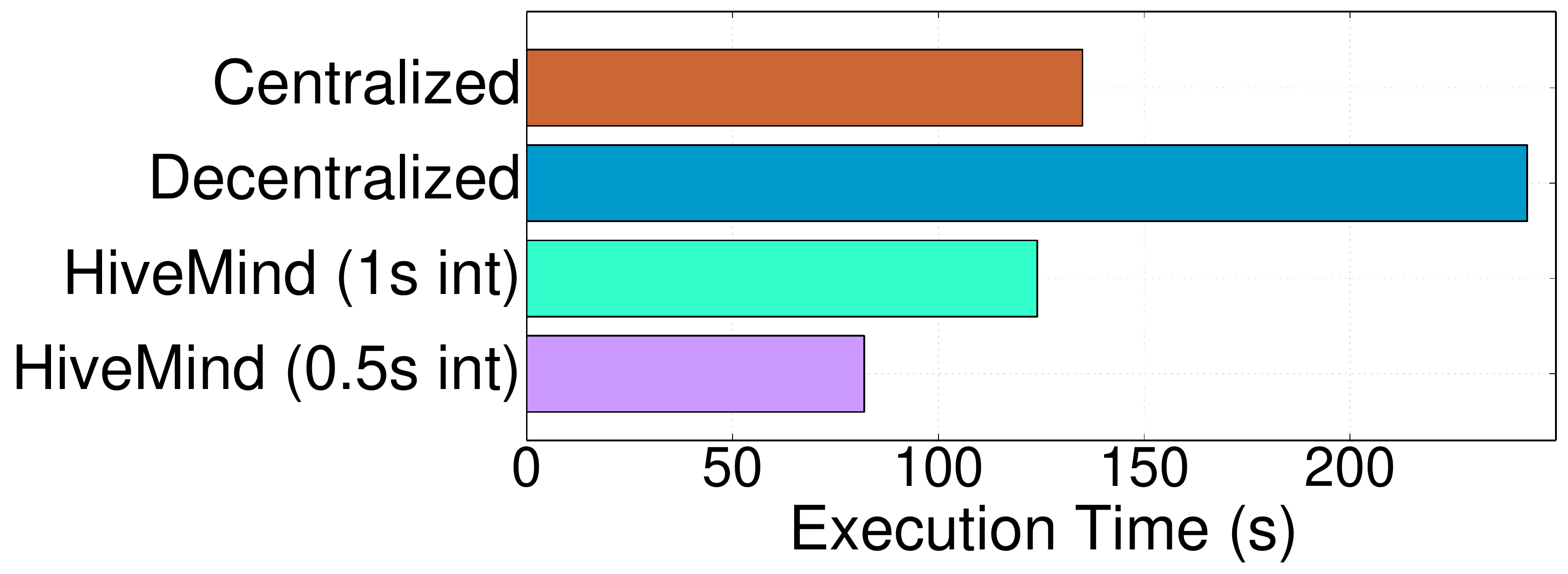} & 
	\includegraphics[scale=0.144, viewport=604 10 1220 330]{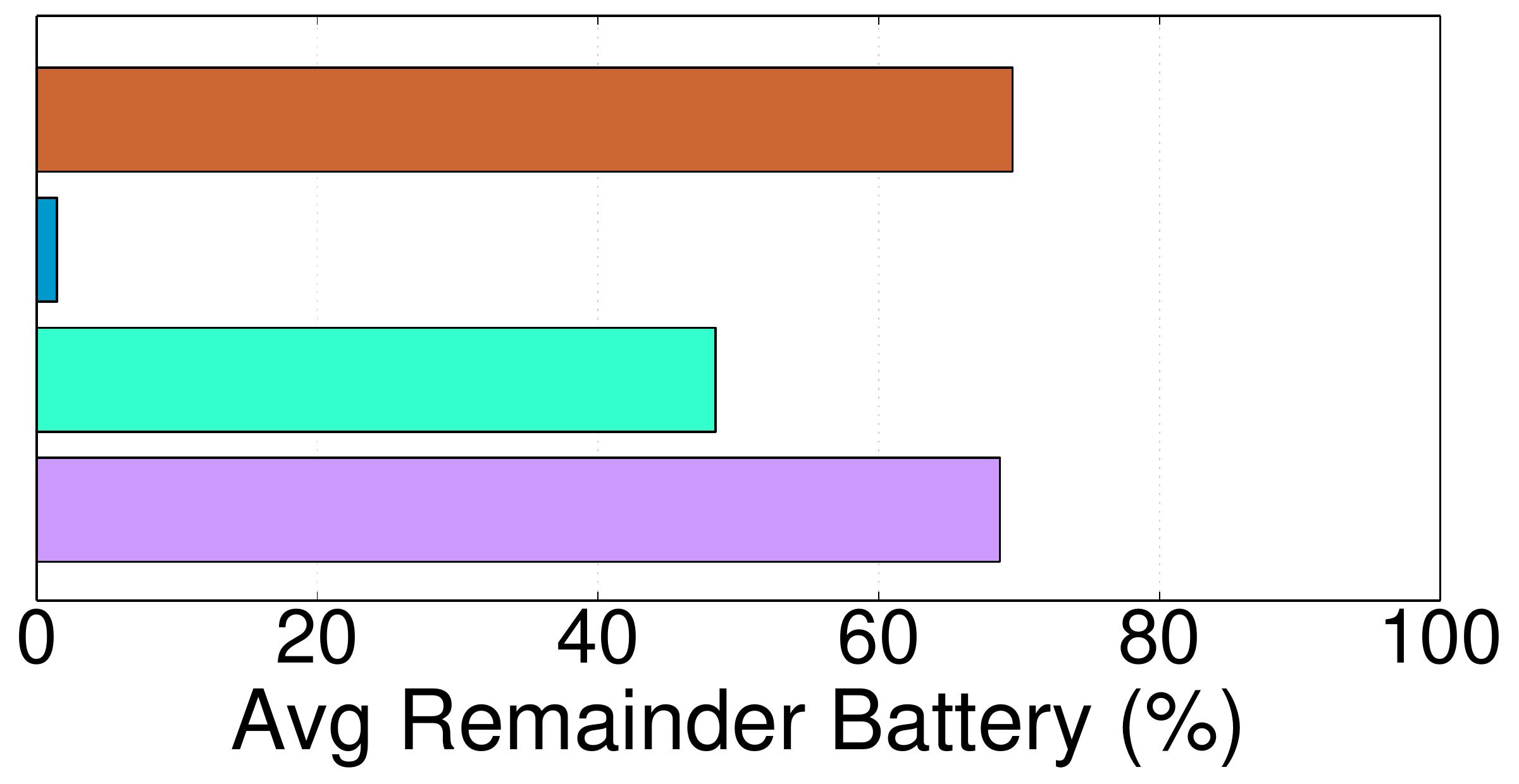} & \\ 
	\includegraphics[scale=0.15, viewport=670 40 900 350]{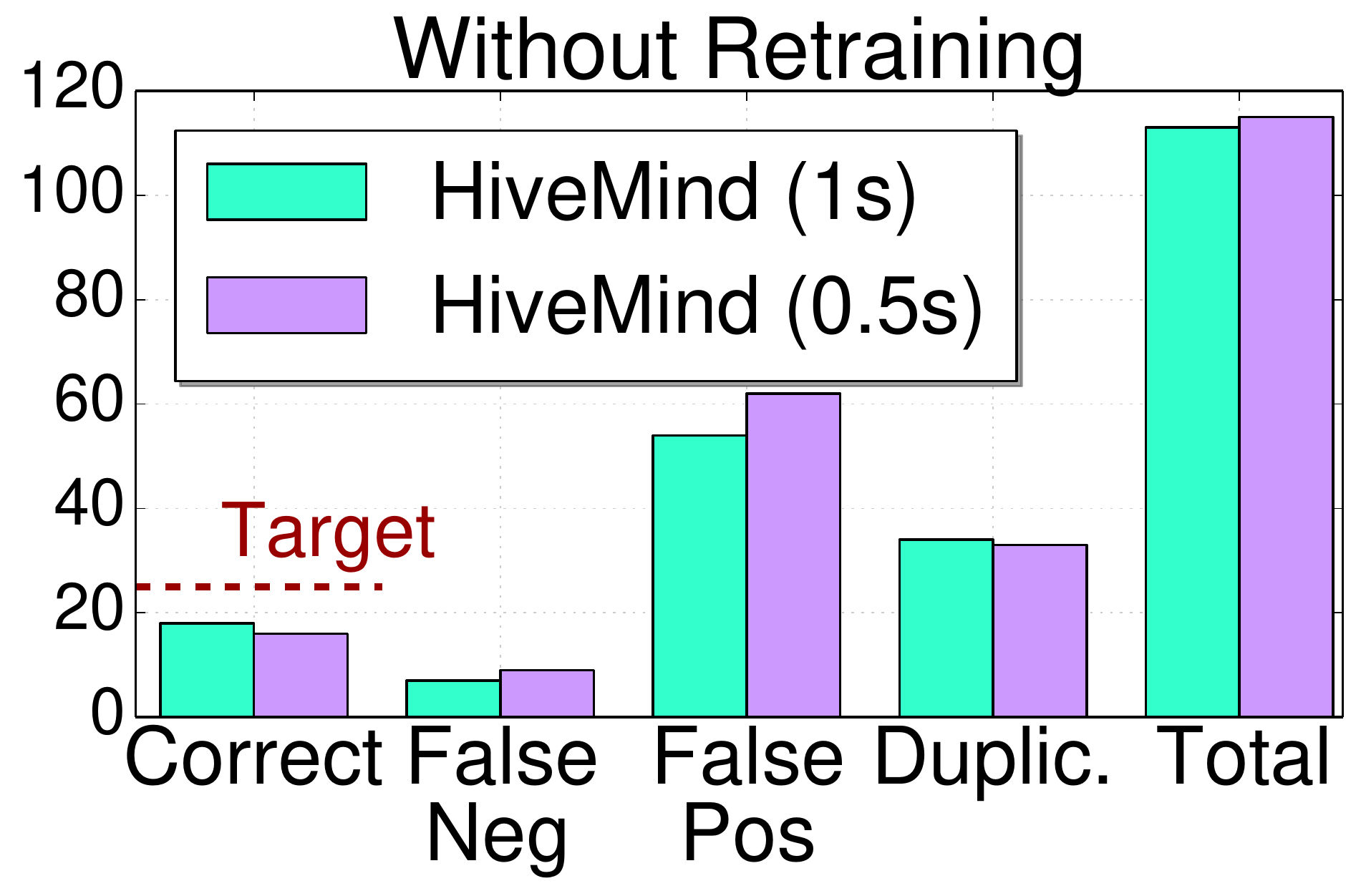} & %accuracy with/without reinforcement learning
	\includegraphics[scale=0.15, viewport=1120 40 1500 350]{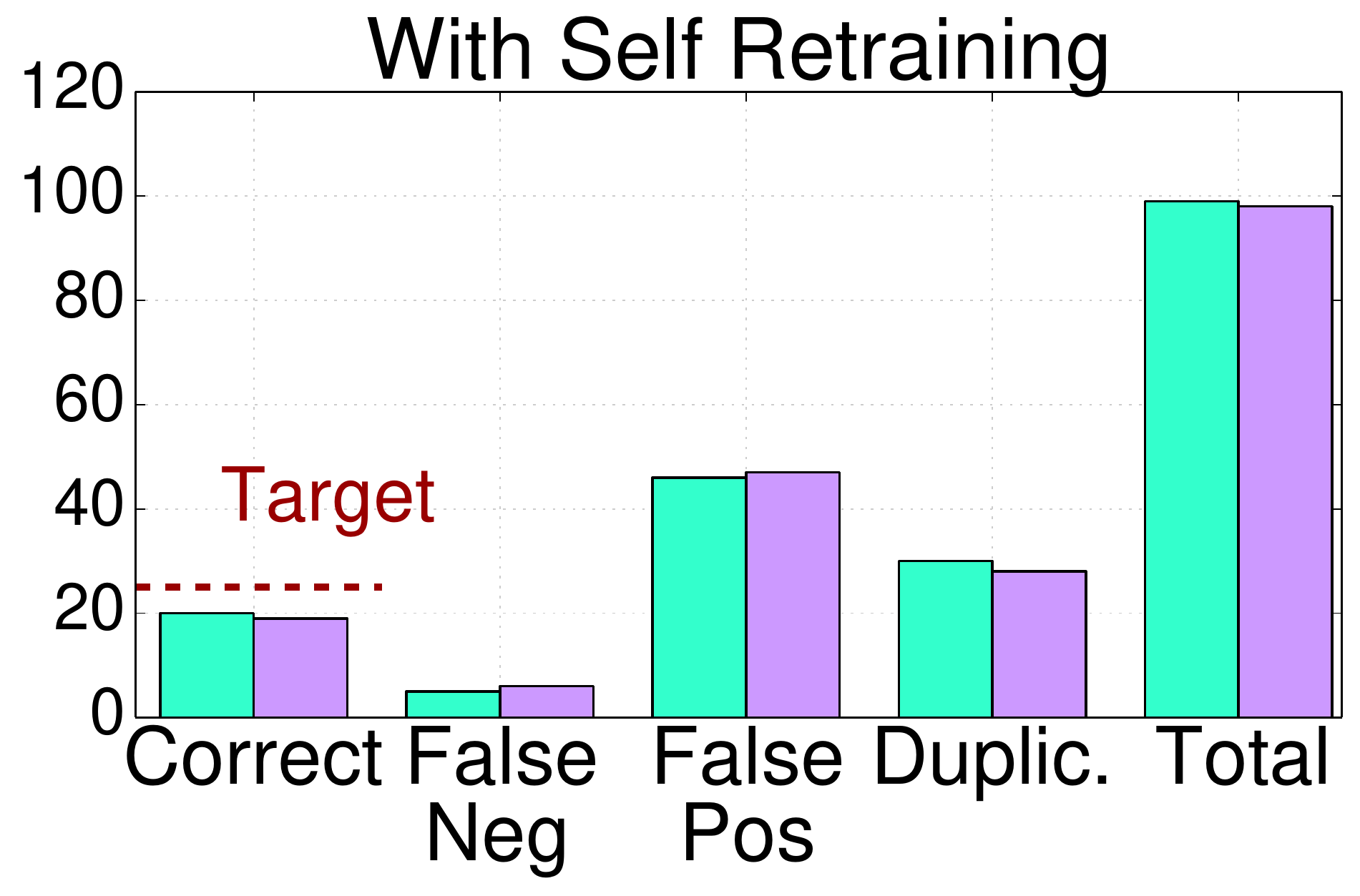} & %accuracy with/without reinforcement learning
	\includegraphics[scale=0.15, viewport=1135 40 1600 330]{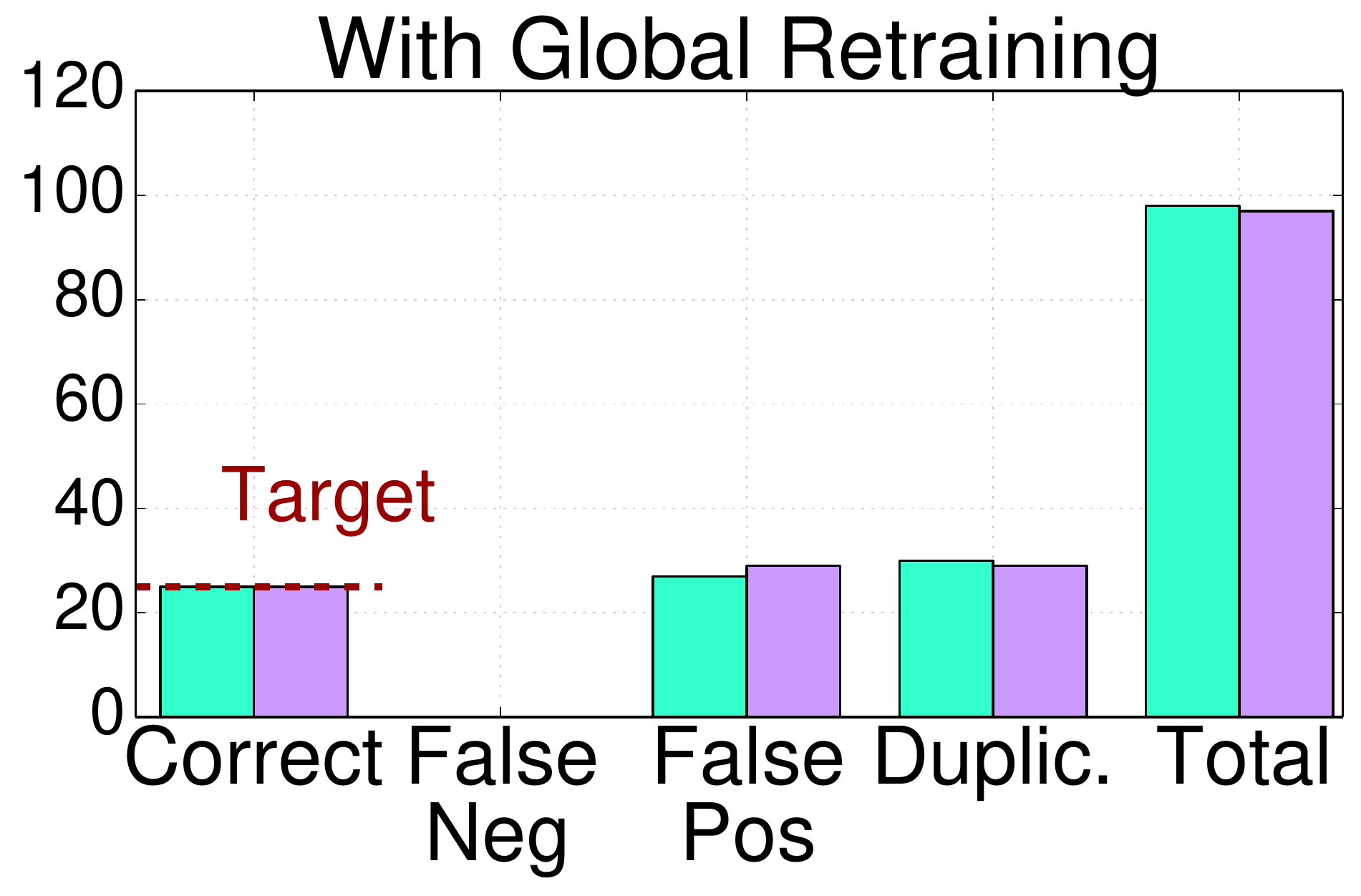} \\ 
\end{tabular}
\caption{\label{fig:hybrid_people_edge} {From top to bottom for the second scenario: (a) latency of tasks running on the drones with HiveMind,
                (b) bandwidth usage across platforms, (c) performance and battery efficiency comparison, and (d) the impact of online
retraining to the accuracy of on-board recognition. }}
\end{minipage}
\end{figure*}

\noindent{\bf{Performance: }}The top rows of Fig.~\ref{fig:hybrid_balls_edge} and~\ref{fig:hybrid_people_edge} show 
the latencies of tasks executing on the edge devices with HiveMind for the two scenarios. %in the hybrid platform. 
On-board recognition now looks for circular objects in the first scenario, and human-shaped objects in the second. % and places bounding boxes around them. %rectangular, larger than 1m less than 2.5m
Photos with such objects are transferred to the cloud, while other images are stored on board. % flash drive. 
%, to verify detection accuracy after the end of the scenario. %not discarded, need to check in the if it was accurate
On-board recognition is considerably faster ($18-117\times$) compared to the decentralized platform, 
where drones had to recognize the precise target objects, 
allowing the devices to consume less battery and maintain more reliable motion control. 
Since HiveMind transfers a lot less data to the cloud, it can also collect images %sensor data 
at a higher rate (twice per second as opposed to once), covering a given area faster. 
Below we also explore a similar policy for the centralized and decentralized platforms. 

Motion and obstacle avoidance takes similar time to the centralized platform, while 
data transfer is significantly reduced, and does not suffer the high queueing latencies 
of the centralized system. Movement again takes longer towards the end of the scenario 
when drones return to their take-off location. The results are consistent across the two scenarios, 
with people-shaped detection being slightly more computationally-intensive, causing 
the scenario to take longer to complete. 

%- latency per task figure

\vspace{0.02in}
\noindent{\bf{Bandwidth utilization: }}The next rows of Fig.~\ref{fig:hybrid_balls_edge} and~\ref{fig:hybrid_people_edge} 
show the network bandwidth utilization for all three platforms. For HiveMind, we show two photo-taking intervals; 
$1s$ at 4m/s speed and $0.5s$ at 6m/s speed. 
We also attempted to do the same with the centralized and decentralized platforms. In the centralized case, bandwidth quickly 
saturates, resulting in high packet drops, once all drones start sending data at peak frequency. In the decentralized system, 
on-board recognition is already stressing the drones' capabilities, therefore requiring 
the process to happen at twice the rate caused their battery to deplete even faster, leaving both scenarios incomplete.  

Even at $1s$ intervals, the centralized platform uses much higher bandwidth than the other platforms in both scenarios. 
The decentralized platform uses the least bandwidth, as only photos with detected objects are transferred to the cloud, with the exception 
of a slightly higher bandwidth usage towards the end of the scenarios, when drones are exchanging their individual results to perform object 
and people disambiguation. 

HiveMind uses more bandwidth than the decentralized platform but less than the centralized system, as it filters out sensor data with 
no objects resembling the target. Unsurprisingly, when the rate of sensor data collection is higher the bandwidth usage is higher too, 
although the whole scenario takes less time to complete. The lower bandwidth usage of HiveMind also means that the platform can scale 
to a larger number of edge devices. For example, while the current wireless router would saturate after 26 drones in the centralized case, 
with HiveMind it can support up to approximately 150 drones. 

\vspace{0.02in}
\noindent{\bf{Battery usage: }}The following rows of Fig.~\ref{fig:hybrid_balls_edge} and~\ref{fig:hybrid_people_edge} show the performance and battery levels 
across platforms and scenarios. HiveMind at $1s$ photo intervals achieves very similar performance to the centralized platform, while consuming more battery 
due to the on-board detection. At $0.5s$ photo intervals both scenarios complete faster, while the reduced total execution time allows the swarm to consume 
less battery than when moving at a slower speed, almost canceling out the effect of on-board recognition, and finishing the scenario with similar battery levels as the centralized 
platform.~\footnote{Speed does not affect battery consumption severely, with most battery depletion happening during take-off, and to keep the drone airborne. } 

\begin{figure*}
	\begin{minipage}{0.48\textwidth}
\centering
\begin{tabular}{cc}
	\includegraphics[scale=0.156, viewport=100 10 1400 290]{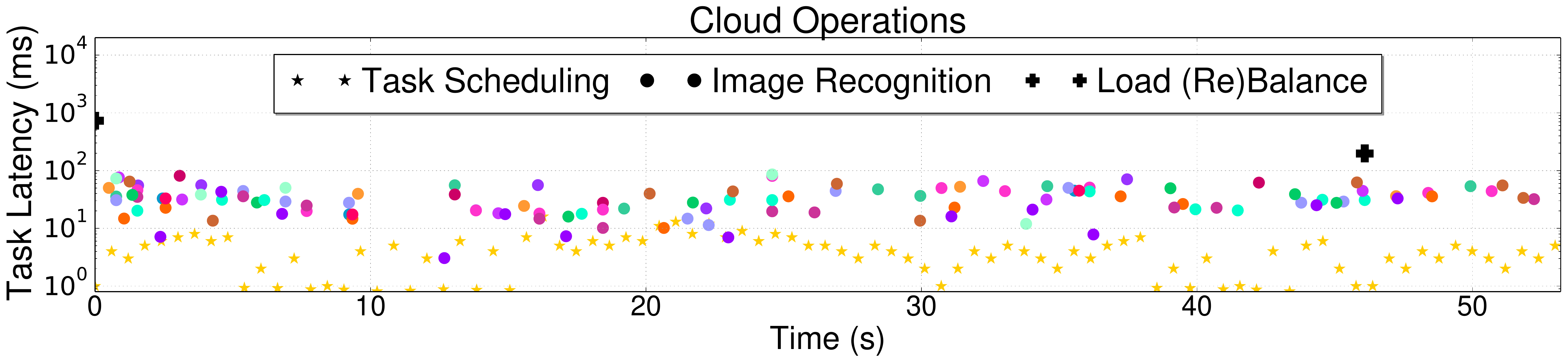} & \\ 
	\includegraphics[scale=0.156, viewport=70 10 1400 290]{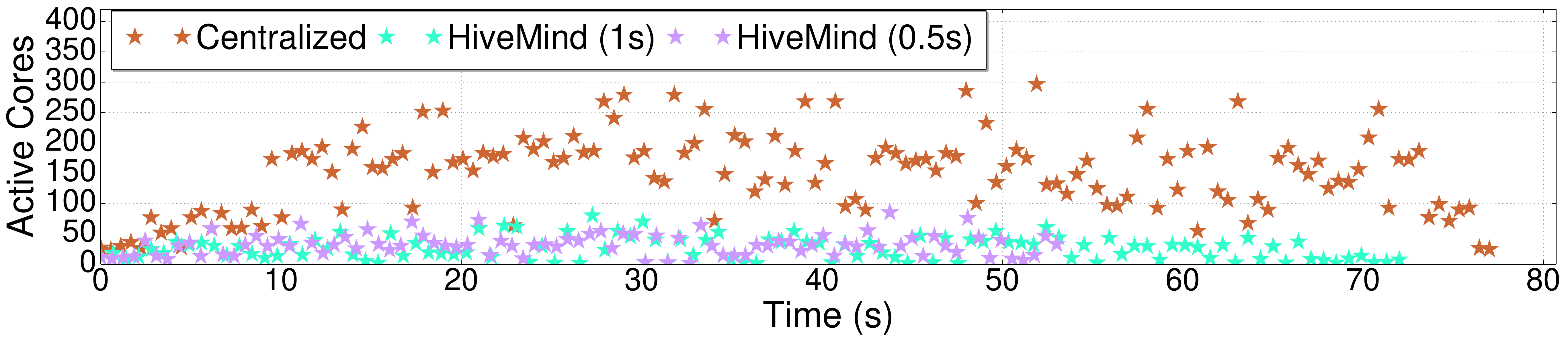} & \\ %#CPUs used
	\includegraphics[scale=0.16, viewport=120 10 1300 340]{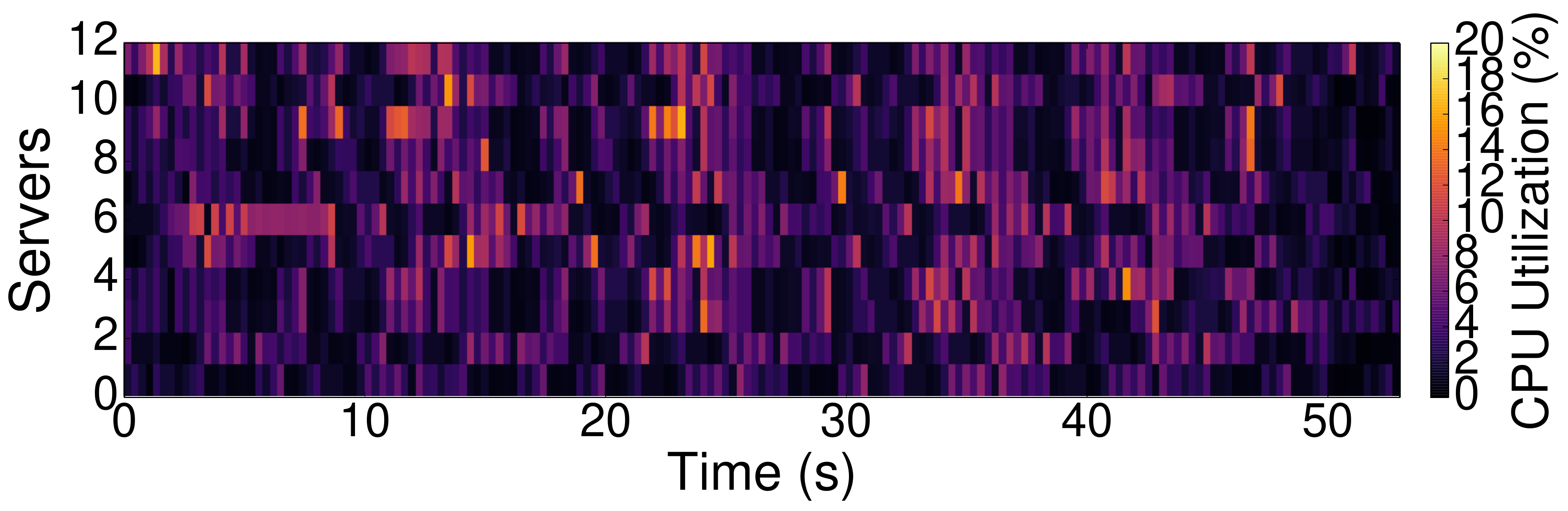} & 
	\includegraphics[scale=0.145, viewport=474 0 1000 370]{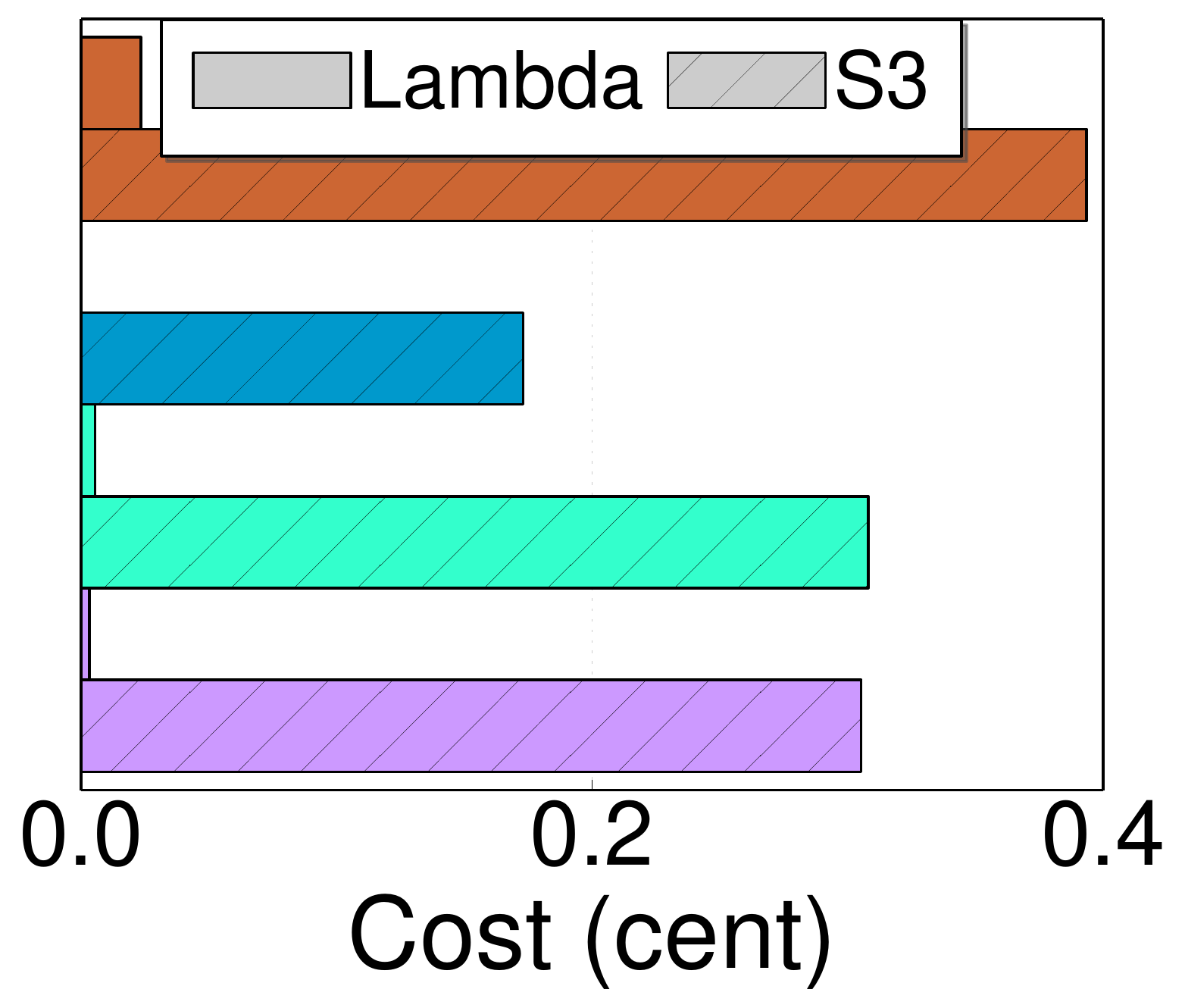} \\
	\includegraphics[scale=0.16, viewport=300 10 1000 60]{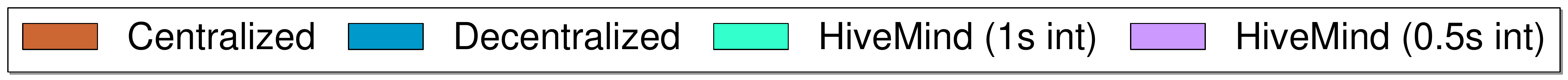} & \\
	\multirow{2}{*}{\includegraphics[scale=0.142, viewport=400 50 1000 340]{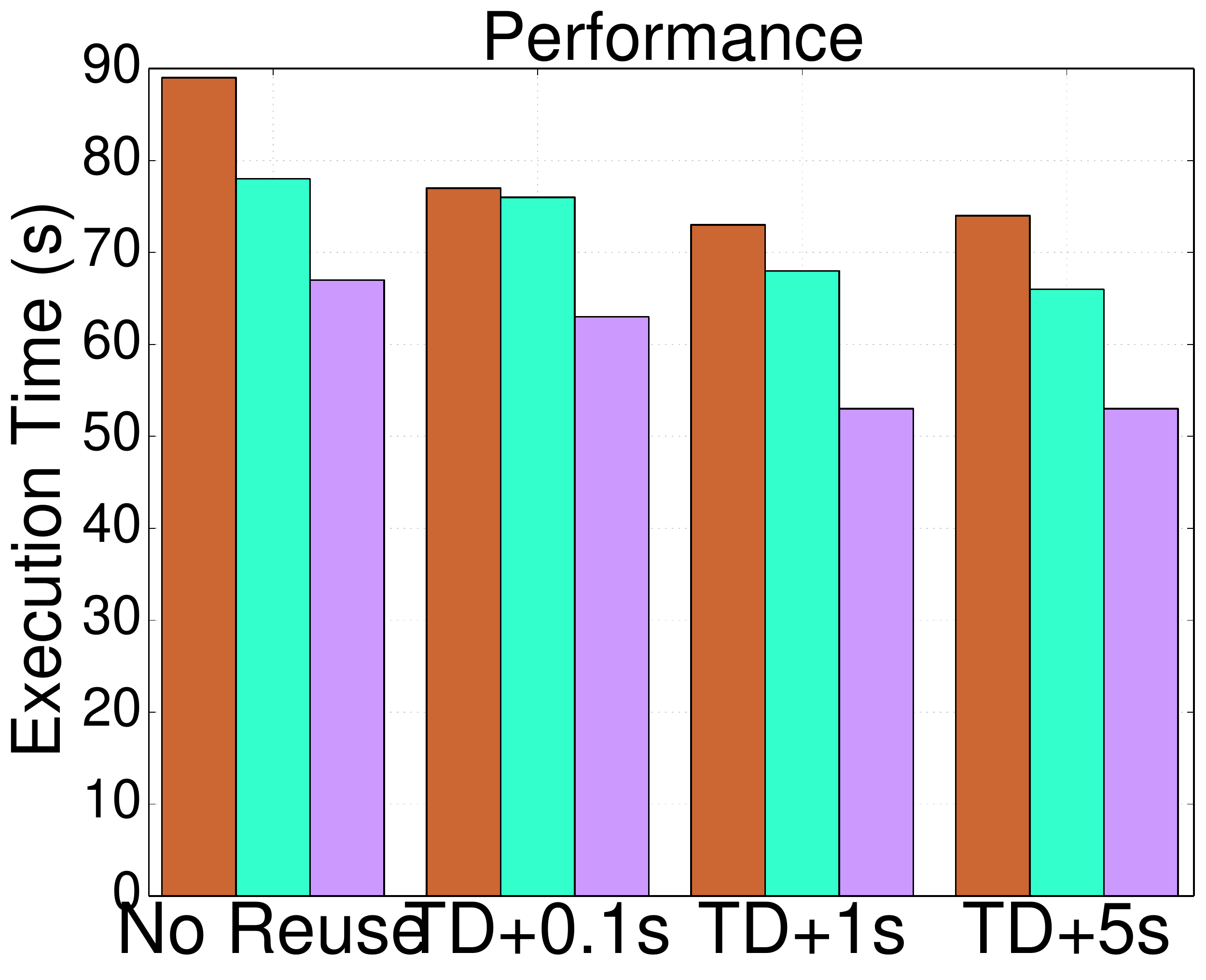}} & %cost and number of unique containers/reused containers
	\includegraphics[scale=0.144, viewport=714 10 1540 330]{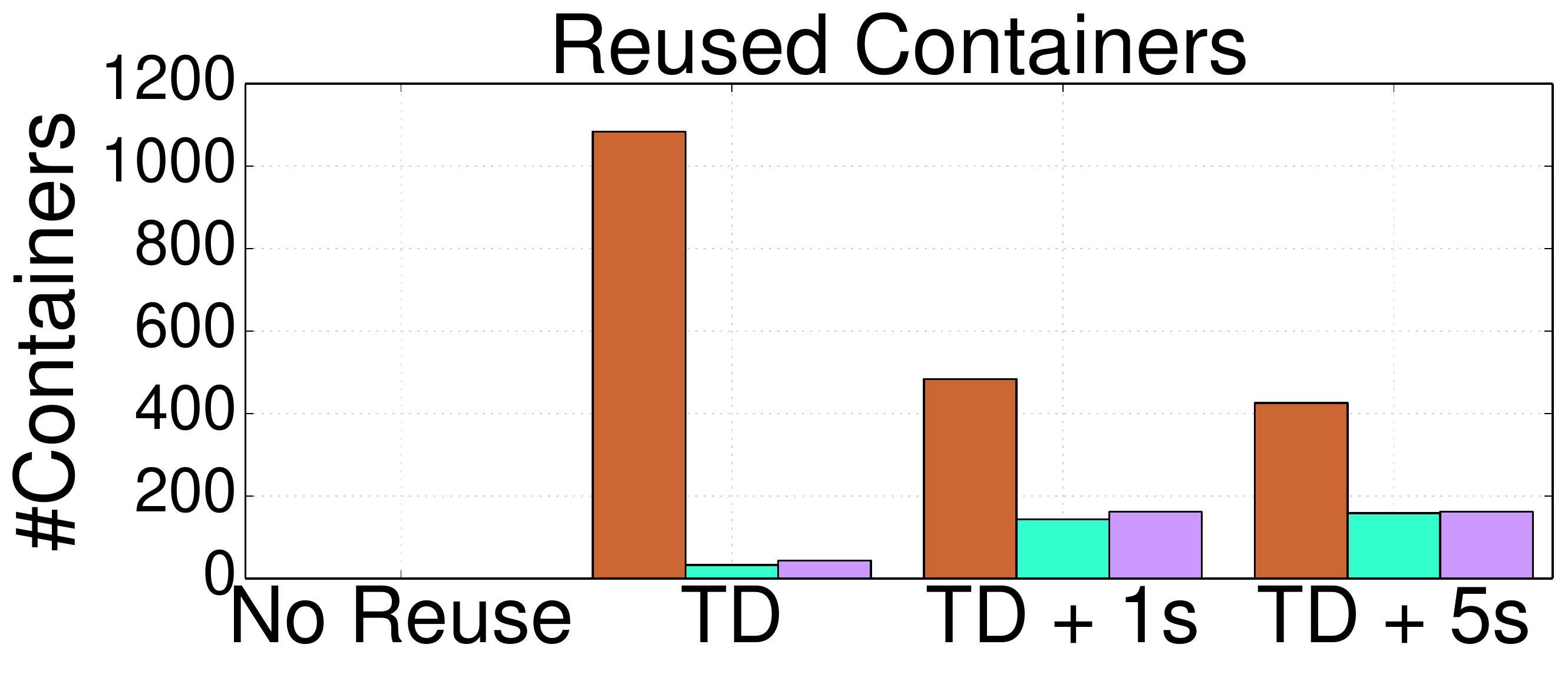} \\ 
	& \includegraphics[scale=0.143, viewport=716 60 1540 304]{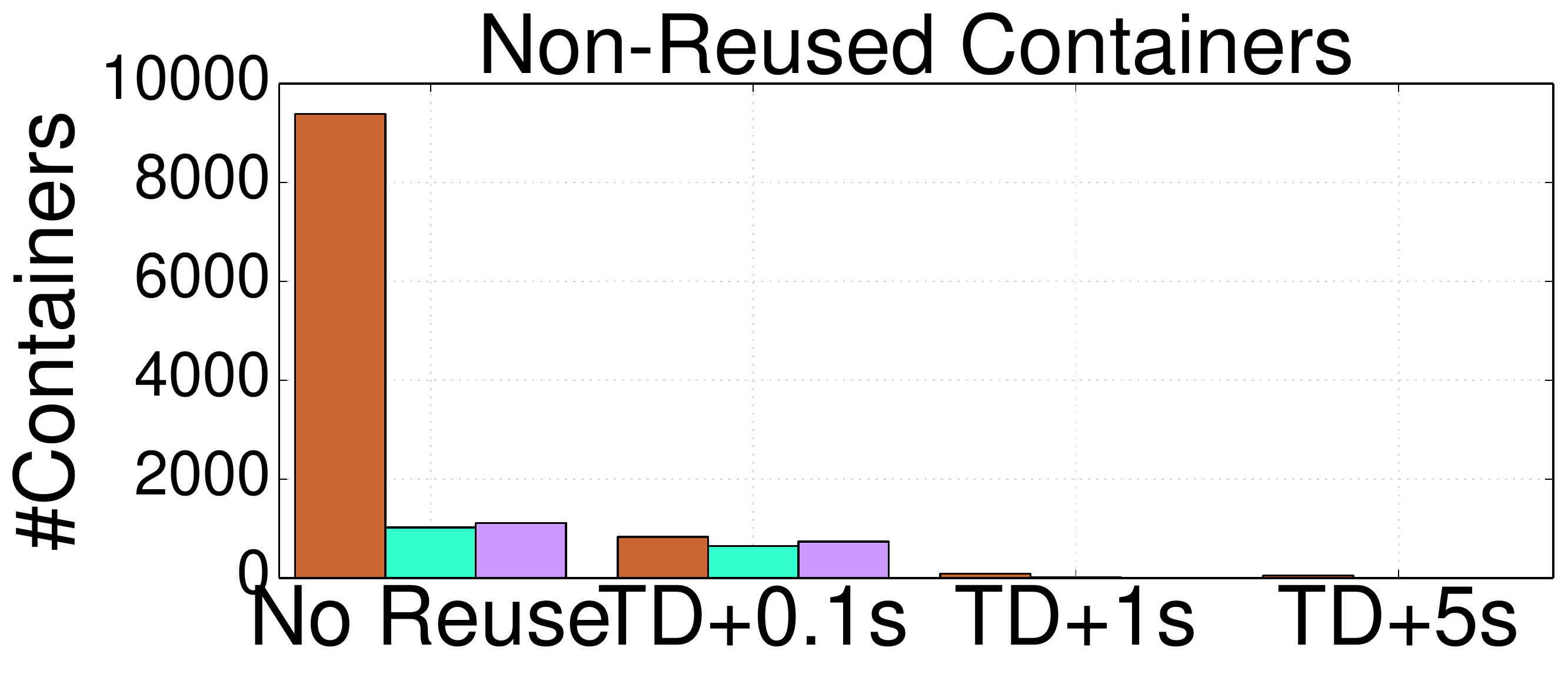} \\ 
\end{tabular}
\caption{\label{fig:hybrid_balls_cloud} {From top to bottom for the first scenario: (a) latencies for cloud tasks with HiveMind, (b) number of active CPUs, (c) CPU utilization per
server and cost for serverless resources, and (d) performance (left) and number of reused and non-reused containers across platforms (right). }}
\end{minipage}
\hspace{0.4cm}
\begin{minipage}{0.48\textwidth}
\centering
\begin{tabular}{cc}
	\includegraphics[scale=0.156, viewport=80 10 1400 290]{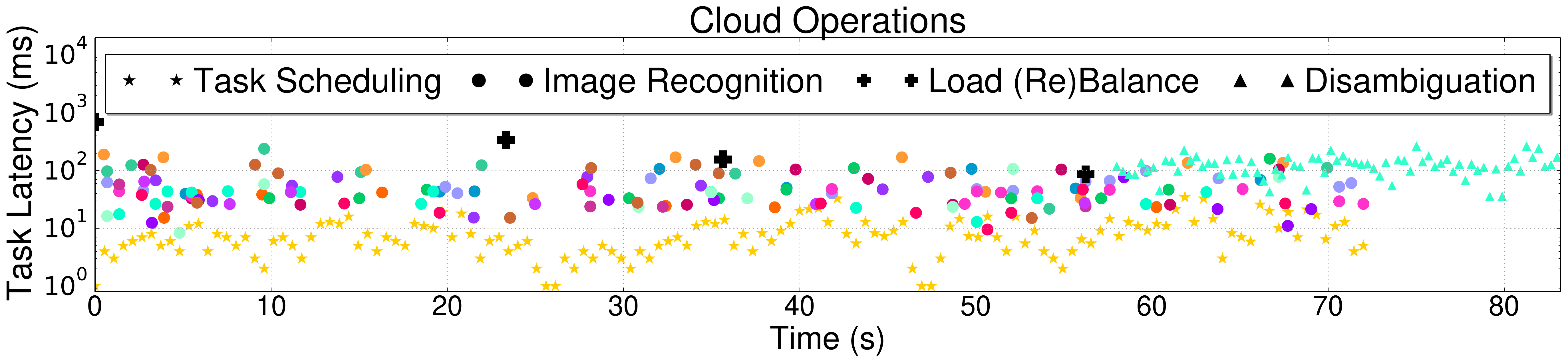} & \\ 
	\includegraphics[scale=0.156, viewport=50 10 1400 290]{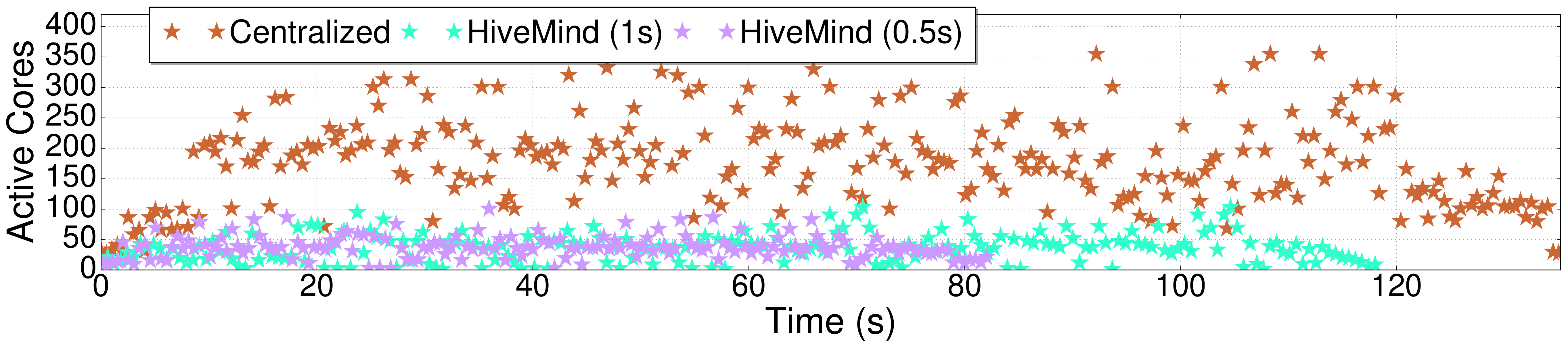} & \\ %#CPUs used 
	\includegraphics[scale=0.16, viewport=100 10 1300 340]{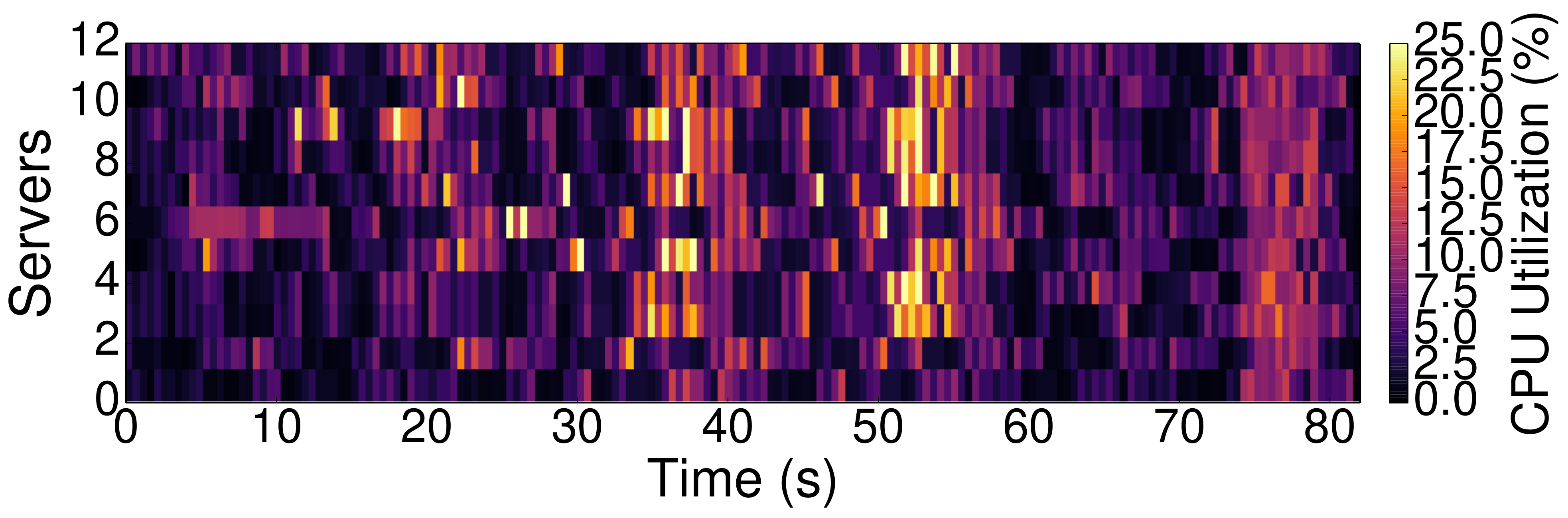} & 
	\includegraphics[scale=0.15, viewport=460 0 1000 370]{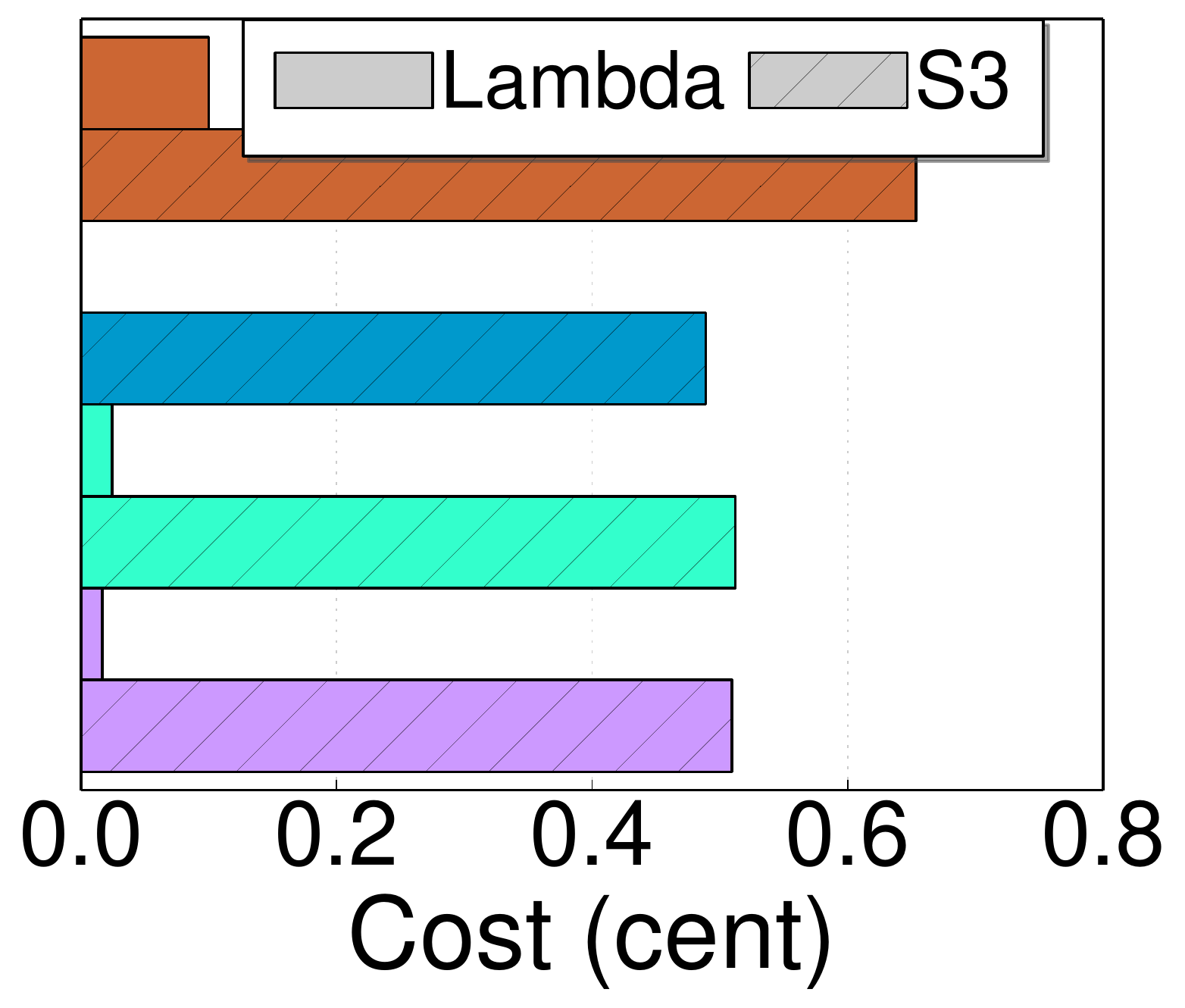} \\
	\includegraphics[scale=0.16, viewport=260 10 1000 60]{HybridPerfLegend4ms.pdf} & \\
	\multirow{2}{*}{\includegraphics[scale=0.142, viewport=400 50 1000 340]{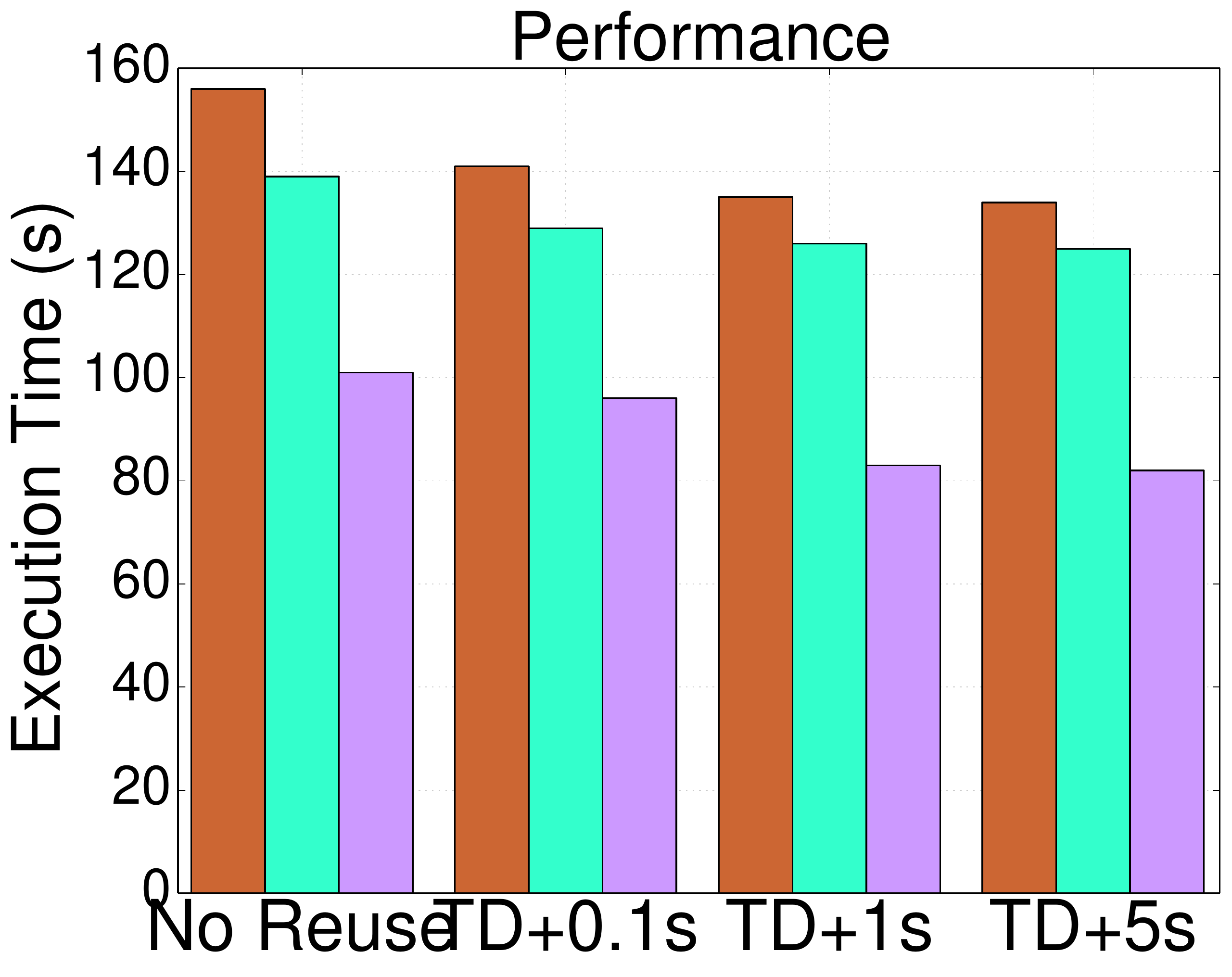}} & %cost and number of unique containers/reused containers
	\includegraphics[scale=0.144, viewport=704 8 1540 328]{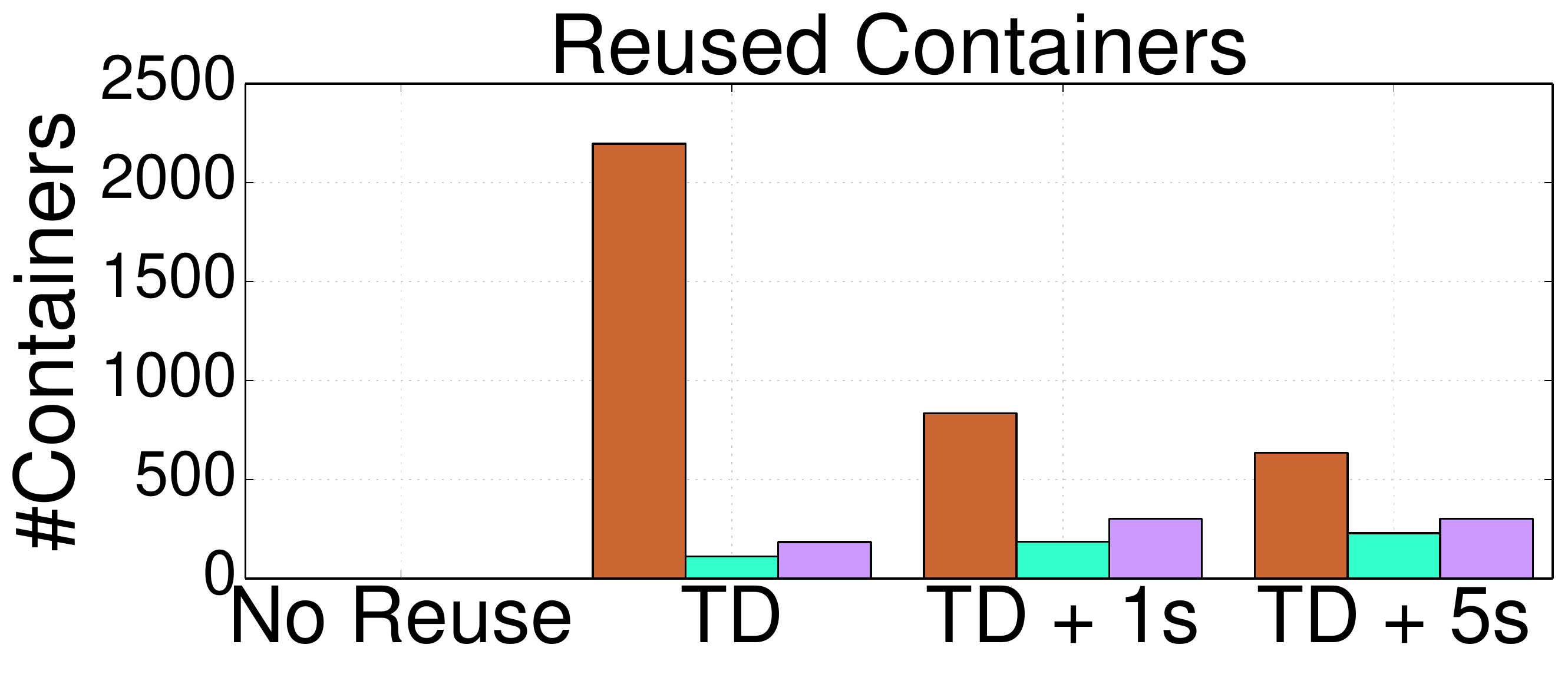} \\ 
	& \includegraphics[scale=0.143, viewport=706 60 1540 304]{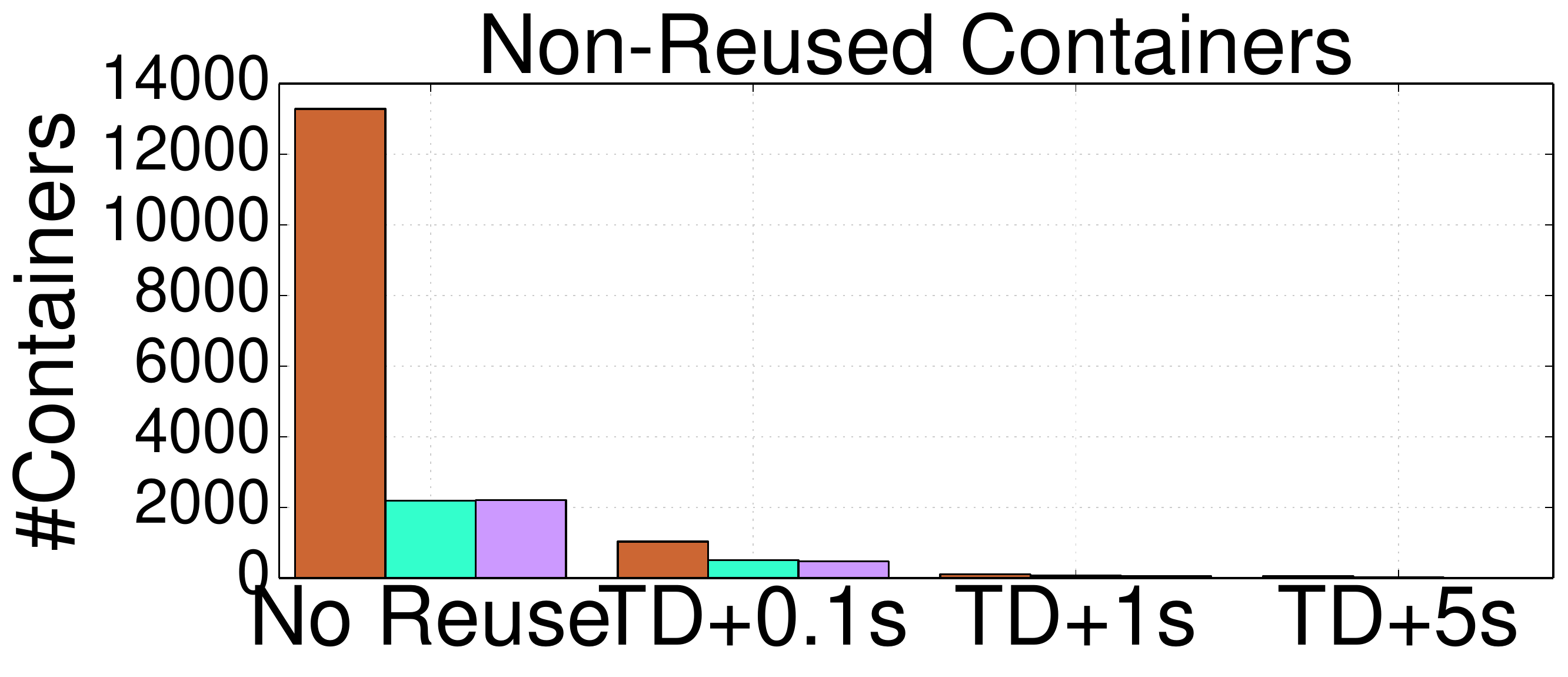} \\ 
\end{tabular}
\caption{\label{fig:hybrid_people_cloud} {From top to bottom for the second scenario: (a) latencies for tasks executing in the cloud with HiveMind, (b) number of active CPUs, (c) CPU utilization per
server and cost for serverless resources, and (d) performance (left) and number of reused and non-reused containers across platforms. }}
\end{minipage}
\end{figure*}

%\vspace{0.02in}
\noindent{\bf{Continuous learning at the edge: }} We now explore the benefit of retraining the on-board image recognition engines. 
We also show the impact of centralized learning on decision quality, compared to decentralized learning at the endpoints. 

The last rows of Fig.~\ref{fig:hybrid_balls_edge} and~\ref{fig:hybrid_people_edge} 
show the accuracy of on-board recognition with HiveMind 
at $1s$ and $0.5s$ photo intervals. We show the number %In each figure, 
of correctly-identified items by drones, any false negatives, i.e., %the number of
the tennis balls and people drones missed respectively, any false %the number of 
positives, i.e., the number of photos transferred to the cloud which did not 
end up containing the target objects, any objects detected multiple times 
(\textit{duplicates}), and finally the total number of recognition 
jobs triggered in the cloud by edge devices. %invocation drones 

The left-most figure shows these statistics when there is no retraining of 
the on-board recognition engines during the scenario's execution. Although 
the drones successfully detect several tennis balls and a few people, they 
also have several false negatives. Since the images for those items/people 
were not transferred to the cloud, there is no way for the centralized system 
to identify them, except by mining all on-board data after the end of the scenario. 
Drones also experience a large number of false positives, by flagging items 
that e.g., have circular shapes but are not tennis balls, which increases 
the total amount of cloud resources used, and data transferred. 

We now enable online retraining, where once a drone starts transferring images 
to the cloud, the cluster uses this drone's images to retrain its on-board 
recognition engine by penalizing false positives, and redeploys the newly-trained 
model to the drone over the network, allowing it to improve as the scenario is 
running.~\footnote{We retrain models in the cloud to avoid 
depleting the on-board resources. }
While this significantly reduces the false positives, as seen in the middle graphs, 
it does not solve the problem of false negatives. To address this, once the scenario 
completes, the cluster mines the on-board storage to flag any false negatives, 
retrains the on-board models, and redeploys them for the next time the scenario executes. 
This results in fewer, although still existing, false negatives, 
as well as fewer total number of cloud recognition jobs, which 
saves cloud resources and network bandwidth. Duplicates are the result of partial overlap 
between images from the same or different drones, so the technique above 
cannot address them, although they are not as impactful 
as false negatives and false positives. 

Finally, we investigate the potential of using the false negatives/positives of the entire swarm to improve 
their on-board models in a collective way, given that their images are transferred to the centralized cluster. 
The right-most graphs of the last row of Fig.~\ref{fig:hybrid_balls_edge} and~\ref{fig:hybrid_people_edge} show 
the impact of global retraining on the drones' detection accuracy. False negatives are completely eliminated 
for both scenarios, while false positives are further reduced. 
This shows that a centralized platform allows edge devices to benefit 
from each other and learn faster than in a fully decentralized system. 

%\vspace{0.02in}
\noindent{\bf{Cloud task performance: }}We now explore metrics related to the centralized cluster. 
The top rows of Fig.~\ref{fig:hybrid_balls_cloud} and~\ref{fig:hybrid_people_cloud} show the latencies for 
cloud image recognition, task scheduling, and load (re)balancing. The results are consistent with 
those for the centralized platform, with the difference that there are much fewer image recognition tasks in 
HiveMind. This also simplifies the job of the serverless scheduler, which now takes an average of $1.8ms$ to schedule 
a new job, compared to $3.2ms$ in the centralized platform. The load balancer had to rebalance the work once for the first 
scenario, when drone \texttt{14}'s battery started draining abnormally quickly, and a couple of times for the second scenario 
to account for drones in highly-populated areas that consumed more battery by needing to transfer more data to the cloud. %had to transfer more data to the cloud. 

%\vspace{0.02in}
\noindent{\bf{Cloud utilization: }}The next two rows in Fig.~\ref{fig:hybrid_balls_cloud} and~\ref{fig:hybrid_people_cloud} show 
the number of active {\smallcapital CPU}s in the cluster for HiveMind 
and the centralized platform, and the {\smallcapital CPU} utilization 
per server for HiveMind with $0.5s$ photo intervals. 
The centralized platform uses $4-6\times$ more {\smallcapital CPU}s than HiveMind, 
since it performs image recognition on all sensor data. Between 
the two variants of HiveMind, collecting data at a higher rate 
also means more cloud resources per unit time, although this 
also results in faster execution. %the scenario finishing faster. 
The second scenario involves more computationally-intensive work, 
resulting in more serverless functions per job, 
therefore the number of {\smallcapital CPU}s is also higher. 

The heatmaps below show that despite 30-75 {\smallcapital CPU}s 
being active at a time with HiveMind, the actual {\smallcapital CPU} 
utilization is low, as each serverless function only keeps a single {\smallcapital CPU} 
occupied for a few msec. This can cause resource underutilization, 
however, we currently primarily focus on optimizing inference latency 
by scheduling new jobs on available, uncontended resources, 
as fast as possible. We plan to explore more resource-efficient 
placement policies in future work. 

%\vspace{0.02in}
\noindent{\bf{Serverless cost: }} The right graphs of the third rows of 
Fig.~\ref{fig:hybrid_balls_cloud} and~\ref{fig:hybrid_people_cloud} show 
the cost of hosting the cluster on a commercial serverless 
framework, specifically {\smallcapital AWS} Lambda~\cite{lambda}, and 
executing each scenario once to completion. 
We use pricing information from the time of submission, and observe that 
the majority of cost for all platforms comes from hosting the training 
and output data on S3, as opposed to processing lambdas. 
The S3 cost for the centralized platforms and HiveMind also includes the 
read and write accesses needed to perform image recognition on the cloud, 
with the overall processing cost for centralized being higher than HiveMind. 

%\vspace{0.03in}
\noindent{\bf{Container reuse: }}Finally we examine the impact of container reuse 
on performance and resource efficiency. 
The last rows of Fig.~\ref{fig:hybrid_balls_cloud} and~\ref{fig:hybrid_people_cloud} 
show the performance and number of reused and uniquely-used cloud containers 
for the centralized platform and HiveMind, when no reuse is allowed, when containers 
remain alive only for $100ms$ after their task completes, and when they remain alive 
for an additional $1s$ and $5s$ after the end of a task's execution. By default, 
each serverless function has its own container, consistent with the allocation policies 
of {\smallcapital AWS} Lambda~\cite{lambda}, Google Functions~\cite{google_functions}, 
and Azure Functions~\cite{azure_functions}. Disallowing container sharing means 
that only a single function can use a container before it is terminated, 
which results in several thousand containers being spawned throughout 
the execution of each scenario, especially for the more computationally-intensive 
people recognition. This causes significant degradation in execution time, as each serverless 
function must sustain the overhead of initializing a new container on 
a new {\smallcapital CPU}. This is especially problematic when all drones are active, and 
unallocated {\smallcapital CPU}s are sparse. 

\begin{wrapfigure}[11]{l}{0.23\textwidth}
\centering
\begin{tabular}{cc}
	\includegraphics[scale=0.13, viewport=60 10 1480 330]{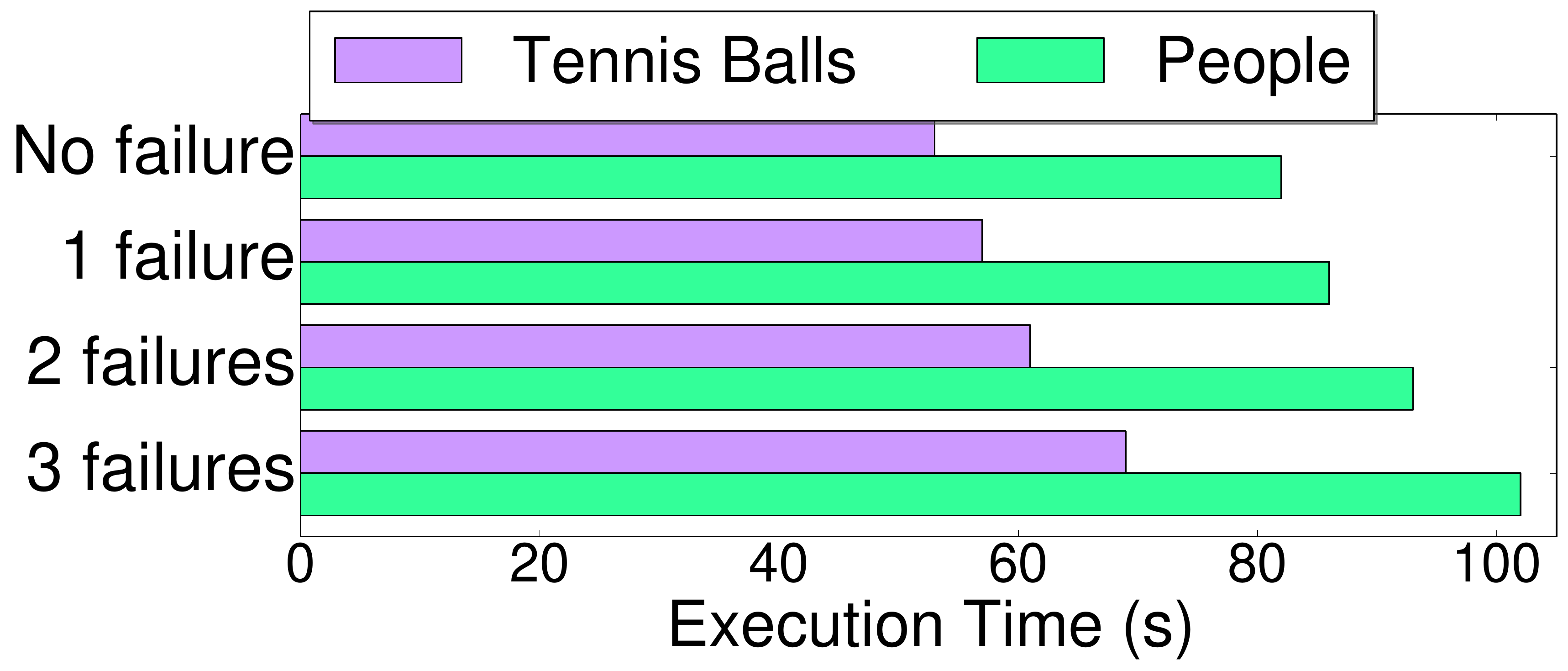} & \\
	\includegraphics[scale=0.13, viewport=60 30 1480 330]{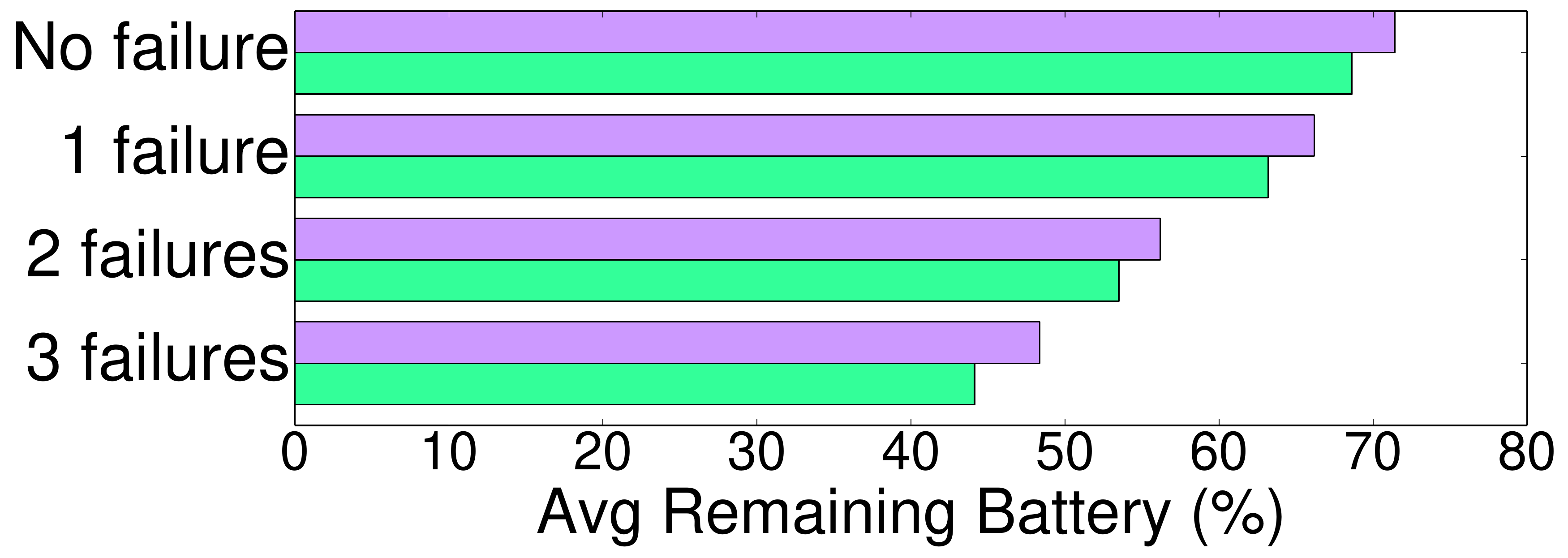} & \\
\end{tabular}
\caption{\label{fig:fault_tolerance} {Performance and battery under drone failures. }}
\end{wrapfigure}
Allowing containers to be reused reduces instantiation overheads, 
but can introduce resource contention, by keeping compute and 
memory allocated without doing useful work. For this experiment we do 
not use sleep states in the containers; adding sleep states would 
free up more resources but delay task starting time. The figure shows 
that keeping containers alive for up to $100ms$ after their task completes 
significantly reduces the uniquely-used containers in the centralized system, 
where there is a high probability that a new job arrives within this interval, 
but does not have a significant impact on the performance and reused containers 
of HiveMind. Keeping containers alive for an additional $1s$ reduces the total 
number of containers in HiveMind, and improves the scenarios' execution time 
by avoiding most instantiation overhead. Keeping containers alive 
for $5s$ almost eliminates uniquely-used containers, although it does not have 
an additional benefit on performance, given that % or number of containers as 
there should already be new data arriving every $1s$ from the drones, 
and that the total number of active tasks is relatively low, such that 
incurring some instantiation overhead does not lead 
to long queueing delays in the serverless scheduler. 
Nonetheless, between the two variants of HiveMind, 
collecting images every $0.5s$ means that there is a higher chance of reusing 
containers than when data arrives at a lower frequency. %Therefore keeping containers alive for longer benefits the $1s$ photo-taking interval more. 
Finally, people recognition takes longer than item recognition, 
which increases the probability of reusing a keep-alive container. 
Unless otherwise specified, we keep containers alive for $1s$ 
past their task's execution. 

\vspace{0.02in}
\noindent{\bf{Fault tolerance: }} %Change results to hybrid
Fig.~\ref{fig:fault_tolerance} shows the performance and average battery at the end of each scenario with HiveMind, excluding failed drones, 
when we force a randomly-selected subset of drones to land and power down at random 
points during execution. 
When a drone fails, the controller redistributes its assigned region to its neighboring drones, 
as discussed in Sec.~\ref{sec:fault_tolerance}. One failure is easily absorbed by the swarm, without 
significant performance and battery degradation. Two failures have a visible impact on battery, 
although the performance impact is limited, 
10.3\% for the first scenario, and 11.8\% for the second. Three failures %Having three devices fail 
lower the battery by the end to 48\% for the first scenario and 43\% for the second, 
and degrade performance by 24\% and 26\% respectively. In all cases HiveMind is able 
to redistribute the load, and complete each scenario. 

\begin{figure}
\centering
\begin{tabular}{cc}
	\includegraphics[scale=0.16, viewport=100 10 700 85]{HybridPerfLegend4ms.pdf} & \\
	\includegraphics[scale=0.186, viewport=150 40 530 410]{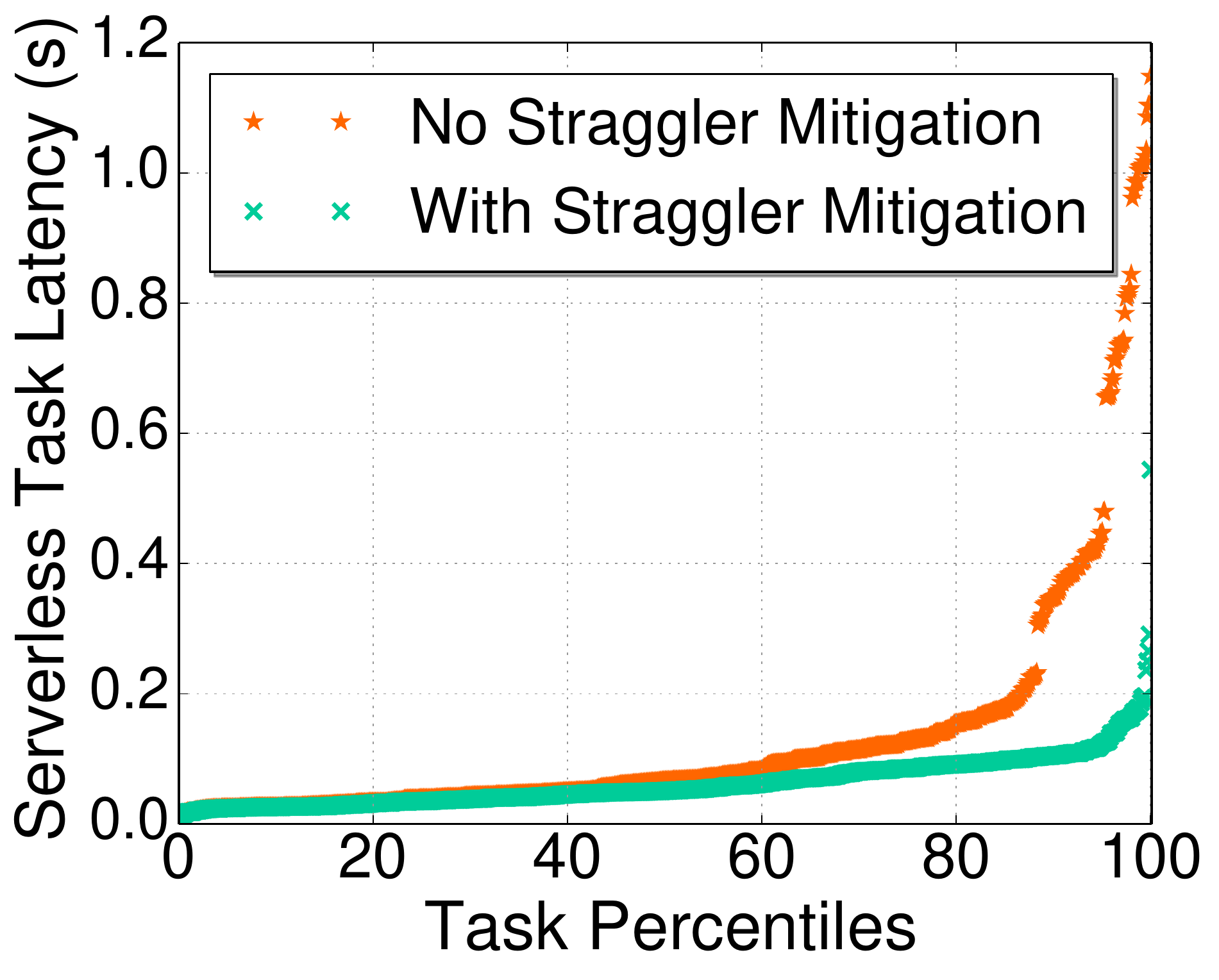} & 
	\includegraphics[scale=0.174, viewport=80 10 560 430]{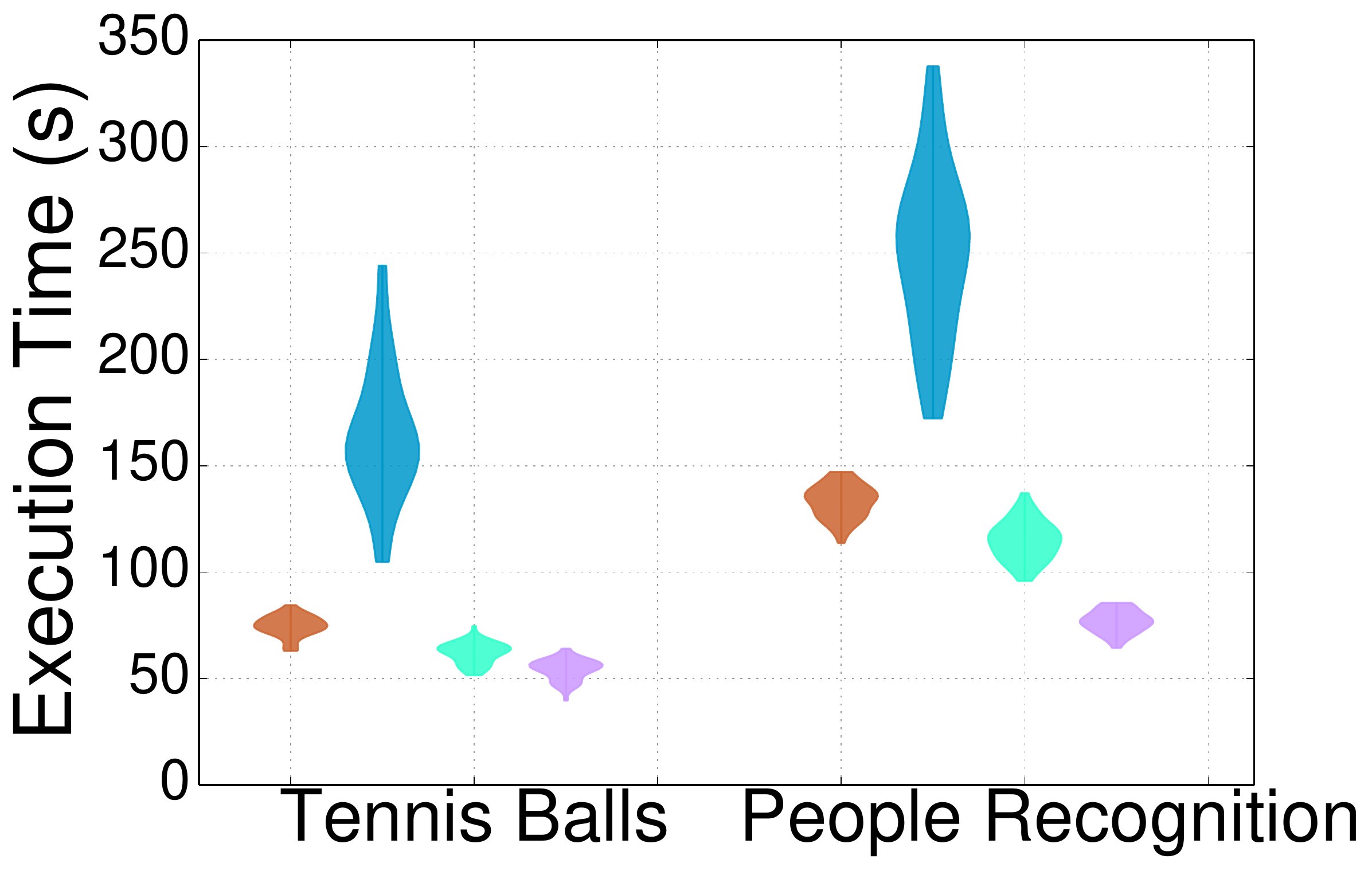} \\
\end{tabular}
	\caption{\label{fig:variance} {(a) Impact of straggler mitigation on the latency of serverless tasks. (b) Performance variance across runs. }}
\end{figure}

%\vspace{0.02in}
\noindent{\bf{Straggler mitigation: }} Fig.~\ref{fig:variance}a shows the impact of 
straggler mitigation on serverless task latency for the second scenario. 
Results are similar for the first scenario. 
By default a small number of serverless tasks 
can take orders of magnitude longer to complete than 
the average, either due to faults or resource contention, delaying execution %the entire scenario's execution 
and keeping resources busy. By detecting stragglers eagerly, 
HiveMind is able to reduce long tails in their execution. % execution of serverless functions. 
The larger the swarm, the more 
tasks are spawned in the backend cluster, and the more critical 
it becomes to handle straggler tasks quickly. 

%\vspace{0.02in}
\noindent{\bf{Performance predictability: }} Finally, Fig.~\ref{fig:variance}b shows violin plots of execution time variability 
for both scenarios across platforms. Across all experiments, we use the same placement for tennis balls, and 
instruct people to follow the same route. 
The centralized system achieves low and predictable execution time, since it leverages 
the ample resources of the backend cluster for most computation. The decentralized system 
has the largest performance variability, especially in the second scenario where computation is more resource-demanding. 
HiveMind experiences slightly higher performance jitter compared to the centralized system, 
since it relies on the drones for the initial image recognition, although its overall execution time 
is better than the fully centralized system, because it reduces both data transfer 
and cloud resource contention. 

\vspace{-0.04in}

\section{Conclusions}
\label{sec:Conclusions}
\vspace{-0.04in}

We have presented HiveMind, a scalable and performant coordination control platform for IoT swarms. 
HiveMind uses a centralized controller to improve decision quality, load balancing, and fault tolerance, 
and leverages a serverless framework to ensure fast and cost-efficient cloud execution. HiveMind additionally
employs lightweight on-board pre-processing to reduce network bandwidth congestion, and scale to larger swarms. 
We evaluated HiveMind on a 16-drone swarm and showed that it achieves better performance, efficiency, and reliability 
than decentralized platforms, and better network efficiency and scalability than fully centralized systems. 
We also showed that HiveMind seamlessly handles failures, 
as well as straggler cloud tasks 
to improve performance predictability.

%\vspace{0.14in}
%\noindent{\large{\bf{Acknowledgements}}}
%\vspace{0.03in}
%\section*{Acknowledgements}

%We sincerely thank all the participants in the user study for their time and effort.
%We also thank Daniel Sanchez, Ed Suh, Grant Ayers, Mingyu Gao, Ana Klimovic, and the rest of the MAST group,
%as well as the anonymous reviewers for their feedback on earlier versions of this manuscript.
%This work was supported by the Stanford Platform Lab, NSF
%grant CNS-1422088, and a John and Norma Balen Sesquicentennial Faculty Fellowship.
%
% The acknowledgments section is defined using the "acks" environment (and NOT an unnumbered section). This ensures
% the proper identification of the section in the article metadata, and the consistent spelling of the heading.

\clearpage
\balance
%
% The next two lines define the bibliography style to be used, and the bibliography file.
%\bibliographystyle{ACM-Reference-Format}
\bibliographystyle{IEEEtran}
\bibliography{references}

\end{document}